\documentclass[twocolumn,showpacs,prb,superscriptaddress, floatfix]{revtex4-2}
\usepackage{graphicx}
\usepackage[dvipsnames]{xcolor}
\usepackage{color}\usepackage{enumerate}
\usepackage{amsmath}\usepackage{amssymb}\usepackage{stmaryrd}
\usepackage{esint}
\usepackage{multirow}
\usepackage{float}
\usepackage{longtable,booktabs}
\usepackage{subfigure}
\usepackage{dsfont}
\usepackage{amssymb}
\usepackage{comment}
\usepackage{bm}
\usepackage[pdftex,colorlinks=true,linkcolor=blue,citecolor=blue]{hyperref}
\usepackage{array}
\usepackage{blindtext}
\usepackage{hyperref}
\usepackage{calc}

\usepackage{mathtools}

\DeclarePairedDelimiterX\braket[2]{\langle}{\rangle}{#1 \delimsize\vert #2}

\newcommand{\be}{\begin{equation}}
\newcommand{\ee}{\end{equation}}

\DeclareFontFamily{U}{mathb}{\hyphenchar\font45}
\DeclareFontShape{U}{mathb}{m}{n}{<5> <6> <7> <8> <9> <10> gen * mathb
<10.95> mathb10 <12> <14.4> <17.28> <20.74> <24.88> mathb12}{}
\DeclareSymbolFont{mathb}{U}{mathb}{m}{n}

\DeclareMathSymbol{\rcirclearrow}{\mathbin}{mathb}{'367}
\makeatletter

\newcommand{\UIUCPHYS}[0]{Department of Physics and Institute of Condensed Matter Theory, University of Illinois at Urbana-Champaign, Urbana, IL 61801, USA}

\begin{document}
\title{Anomalous crystalline-electromagnetic responses in semimetals}

\author{Mark R. Hirsbrunner}\thanks{These authors contributed equally.}\affiliation{\UIUCPHYS}

\author{Oleg Dubinkin}\thanks{These authors contributed equally.}\affiliation{\UIUCPHYS}

\author{Fiona J. Burnell}\affiliation{Department of Physics, University of Minnesota Twin Cities, MN 55455, USA}

\author{Taylor L. Hughes}\affiliation{\UIUCPHYS}

\begin{abstract}
    We present a unifying framework that allows us to study the mixed crystalline-electromagnetic responses of topological semimetals in spatial dimensions up to $D = 3$ through  dimensional augmentation and reduction procedures. We show how this framework illuminates relations between the previously known topological semimetals, and use it to identify a new class of quadrupolar nodal line semimetals for which we construct a lattice tight-binding Hamiltonian. We further utilize this framework to quantify a variety of mixed crystalline-electromagnetic responses, including several that have not previously been explored in existing literature, and show that the corresponding coefficients are universally proportional to weighted momentum-energy multipole moments of the nodal points (or lines) of the semimetal. We introduce lattice gauge fields that couple to the crystal momentum and describe how tools including the gradient expansion procedure, dimensional reduction, compactification, and the Kubo formula can be used to systematically derive these responses and their coefficients. We further substantiate these findings through analytical physical arguments, microscopic calculations, and explicit numerical simulations employing tight-binding models.    
\end{abstract}

\maketitle

\section{Introduction}
Topological responses are a key manifestation of electronic topology in solids. Celebrated examples such as the integer quantum Hall effect~\cite{iqhe,laughlin,tknn} and axion electrodynamics~\cite{wilczek,qi2008} have paved the way for a broader exploration of topological response phenomena in insulating systems. As of now, a wide-variety of phenomena that are directly determined by the electronic topology have been considered, including thermal response~\cite{luttinger195x,chiralcentralcharge}, geometric response~\cite{avron1995,read2008,vignaleXXX,hughesleighfradkin, you2016, Hiroaki2016, teoTopologicalDefectsSymmetryProtected2017a, dislocation_defect_2020, laurila_torsional_2020, dbh2021, gioia2021, chongwang2, jakko, gao_chiral_2021, you_fracton_2022, amitani_torsion_2023, hirsbrunner2023}, and electric multipole response~\cite{zak,vanderbilt1993,bbh}. These responses are robust features of topological insulators (TIs), and topological phases in general, and are often described by a quantized response coefficient, e.g., the integer Hall conductance~\cite{iqhe,laughlin,tknn}, or the quantized magneto-electric polarizability~\cite{qi2008,essin2009,wuarmitage}. 

Interestingly, certain distinctive features of response of topological Weyl or Dirac semimetals can be described by response theories that are closely analogous to those of topologically insulating phases, albeit with coefficients that are determined by the momentum-space
and energy locations of the point or line nodes~\cite{haldaneAHE,burkov2011A, Wan2011, zyuzin2012, RamamurthyPatterns, ramamurthylinenode}.
For point-node semimetals, the relevant response coefficients are momentum-energy vectors determined as a sum of the momentum and energy locations of the point-nodes weighted by their chirality (or by their helicity, for Dirac semimetals), yielding a momentum-energy space dipole.  For example, the low-energy, nodal  contribution to the anomalous Hall effect tensor of a 3D Weyl semimetal is determined by the momentum components of this momentum-energy dipole vector.

The quasi-topological response coefficients of topological semimetals are not strictly quantized since they can be continuously tuned with the nodal momenta. However, the forms of the responses share many features with topological insulators in one lower dimension, or perhaps more precisely, with weak topological insulators in the same dimension~\cite{halperin1987,fu2007topological}. Indeed, topological semimetals and weak topological insulators both require discrete translation symmetry to be protected and both are sensitive to translation defects such as dislocations~\cite{ran2009}. Interestingly, the connection to translation symmetry has motivated recent work which recasts many previously proposed topological responses of these systems as couplings between the electromagnetic gauge field $A_\mu$ and gauge fields for translations $\mathfrak{e}_\mu^a,$ where $\mu$ runs over spacetime indices, and $a$ runs over the spatial directions in which translation symmetry exists. This insight has also led to the development of new response theories that are just beginning to be understood~\cite{gioia2021,jakko,dbh2021,chongwang2, hirsbrunner2023}. 

Motivated by these previous results and our recent related work on higher rank chiral fermions~\cite{dbh2021,zhu2023higher,hirsbrunner2023}, here we study the topological responses of 1D, 2D, and 3D topological semimetals coupled to electromagnetic and strain (translation gauge) fields.
In addition to the well-studied dipole case mentioned above, we also study cases where point-nodes have momentum-energy quadrupole or octupole patterns. Our approach allows us to make clear connections between a wide variety of response theories across dimensions, and clarifies relationships between many of the response theories we discuss. We find that the chirality-weighted momentum-energy \emph{multipole} moments of the semimetals determine new types of quasi-topological responses to electromagnetic fields and strain. We are able to explicitly derive many of these responses from Kubo formula calculations (sometimes combined with dimensional reduction procedures~\cite{qi2008}). Using these results we explicitly study these families of response theories using lattice model realizations.  We also extend our results to the responses of nodal line semimetals (NLSMs) and construct a new type of NLSM with an unusual crossed, cage-like nodal structure. 

Our article is organized as follows. In Sec.\ref{sec:overview} we provide an overview of and intuition about the response theories that will be discussed in more detail, and in model contexts, in later sections. In Sec.~\ref{sec:gradient_exp} we derive a family of effective actions that describe mixed crystalline-electromagnetic responses in various spatial dimensions. From here we proceed in Sec.~\ref{sec:models} by presenting concrete lattice models and explicit numerical calculations that realize and demonstrate the mixed responses in D$=1,2,3$.
We conclude in Sec.~\ref{sec:conclusion} by discussing possible extensions to future work, and potential pathways to experimental observation of some of the described phenomena.

\section{Overview of Response Theories}
\label{sec:overview}
\begin{figure*}[th!]
   \includegraphics[width=0.98\textwidth]{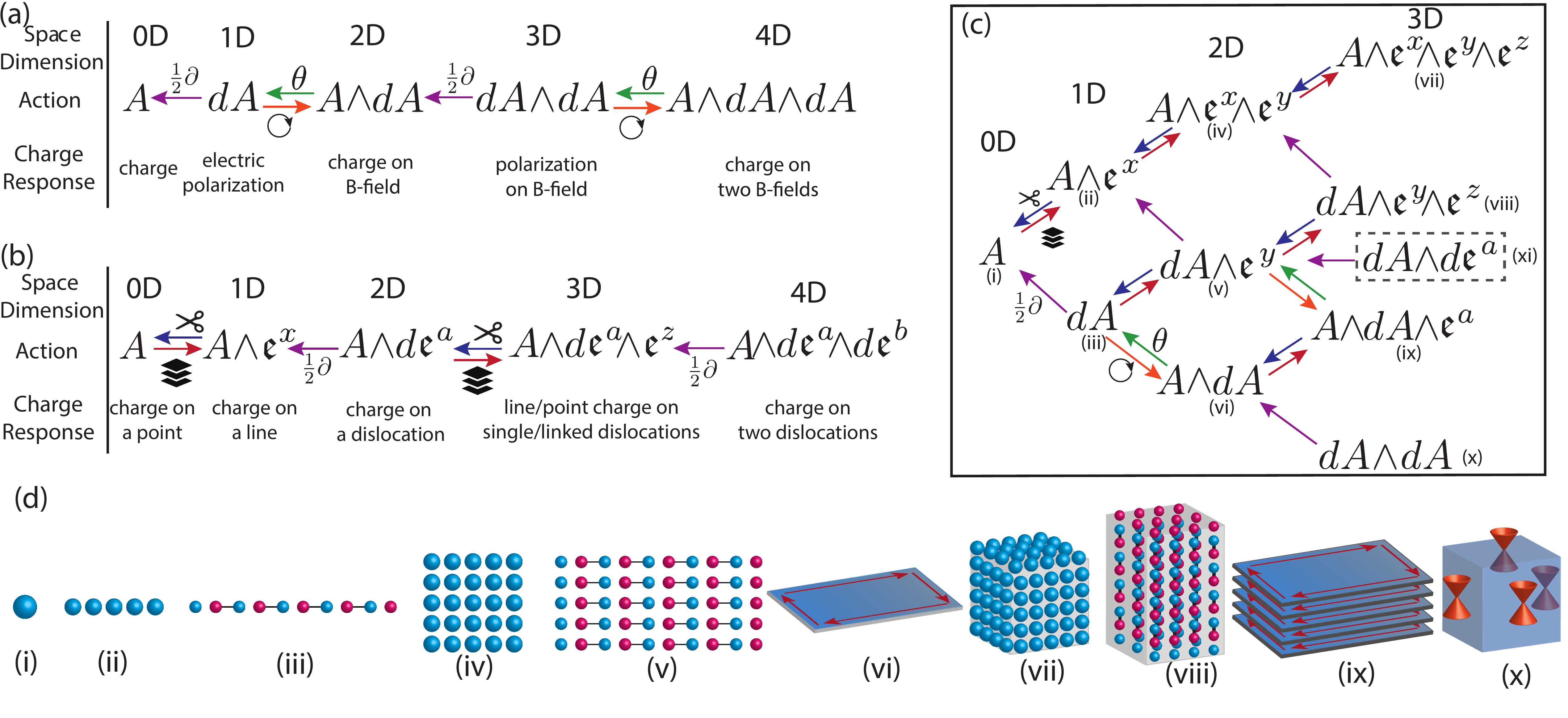}
    \caption{(a) A dimensional hierarchy of theories describing responses of strong topological insulators. The theories are related by dimensional reduction ($\theta$ symbol, green arrow)~\cite{qi2008}, taking the boundary response ($(1/2)\partial$ symbol, purple arrow), or adiabatic pumping ($\rcirclearrow$ symbol, red arrow)~\cite{thoulesspumping}.
    (b) A dimensional hierarchy of insulating systems with mixed crystalline-electromagnetic responses. The theories are related by stacking (layer symbol, dark red arrow) and cutting (scissor symbol, blue arrow).
    (c) A family tree of dimensional hierarchies establishing connections between responses of strong TIs and insulators with mixed crystalline-electromagnetic responses.
    (d) Illustrations representing the nature of the phases constituting the hierarchy depicted in (c).  (i) A single isolated charge. (ii) A line of charges forming a lattice. (iii) An insulating chain having a quantized charge polarization. (iv) A two-dimensional lattice of charges. (v) A two-dimensional weak topological insulator where polarized chains are stacked transverse to their polarization. (vi) A two-dimensional Chern insulator having chiral edge states indicated by red arrows. (vii) A three-dimensional lattice of charges. (viii) A three-dimensional lattice built from a two-dimensional array of polarized chains; alternatively, a stack of two-dimensional weak topological insulators. (ix) A three-dimensional stack of Chern insulators forming a time-reversal breaking weak topological insulator. (x) A three-dimensional strong topological insulator with surface Dirac cones.}
    \label{fig:stiresponse}
\end{figure*}

The systems we consider in this article all exhibit $U(1)$ charge conservation and discrete translation symmetry in at least one spatial direction. In the presence of these symmetries we can consider the responses to background field configurations of the electromagnetic gauge field $A_\mu$ and a collection of translation gauge fields $\mathfrak{e}_\mu^a.$ For example, if the system exhibits translation symmetry in the $x$-direction, then we can consider coupling the system to the field $\mathfrak{e}_\mu^x.$  Our goal is to study low-energy response theories of electrons coupled to translation and electromagnetic gauge fields. 

Since most readers are likely less familiar with the translation gauge fields $\mathfrak{e}_\mu^a$ than the electromagnetic field $A_\mu$, we will briefly review the nature of these fields as they appear in our work. 
In a weakly deformed lattice, $\mathfrak{e}^a$ is given by
\begin{equation} \label{Eq:Edef}
\mathfrak{e}_j^a=\delta_j^a -\frac{\partial u^a}{\partial x^j},\end{equation} where the Kronecker $\delta_j^a$ encodes the fixed reference lattice vectors, $u^a$ is the lattice displacement, and $\frac{\partial u^a}{\partial x^j}$ is the distortion tensor~\cite{landaulifshitzelasticity}.  
The fields $\mathfrak{e}_j^a$ in Eq. (\ref{Eq:Edef}) are reminiscent of gauge fields (see, e.g.,~\cite{KATANAEV1992}): from Eq. (\ref{Eq:Edef}) we immediately see that  line integrals of $\mathfrak{e}$ describe lattice dislocations since $\oint \tfrac{\partial u_a}{\partial x_i} dx^i=b^a$, where $b^a$ is the net Burgers vector of all the dislocations inside the loop~\cite{landaulifshitzelasticity}.  
This points to an analogy with the configurations of the usual electromagnetic field. The analog of magnetic fields derived from $\mathfrak{e}_{\mu}^a$ essentially encodes configurations of dislocations,  each with an amount of flux equal to the corresponding Burgers' vector. Additionally, electric fields are time-dependent strains. In earlier work, e.g., Ref. \onlinecite{hughesleighfradkin}, these fields could have been called frame-fields, but crucially the translation gauge fields encode only the translation/torsional part of the geometric distortion, whereas the frame fields also carry rotational information. 
In keeping with previous literature, here we will call the set of (Abelian) fields $\mathfrak{e}_{\mu}^a$ translation ``gauge" fields, by analogy of their relationship to translation ``fluxes" (i.e. lattice defects).  This language is convenient because, as we will see, actions describing the response to such lattice fluxes are invariant under (vector-charge) gauge transformations of the $\mathfrak{e}_{\mu}^a$ fields. 

A second way in which we will  use the close analogy between $\mathfrak{e}_{\mu}^a$ and electromagnetic gauge fields is through the lattice analog of the usual Aharonov-Bohm effect (holonomy), in which a charged particle encircles a magnetic flux of the gauge field. In the electromagnetic case, a charged particle moving around a magnetic flux generates a $U(1)$ phase factor. For the translation gauge field, taking a particle around a translation magnetic flux having Burgers vector $b^a$ generates a translation operator by the displacement $b^a$. For particles with a fixed translation charge, i.e., a fixed momentum, this generates a momentum-dependent $U(1)$ phase factor. This will lead us to introduce momentum-dependent Peierls factors when performing some lattice calculations. 
To complement this discussion, in Appendix~\ref{app:tp_derivation} we show more explicitly how translation symmetry can be ``gauged" under a teleparallel constraint of the underlying system geometry.  A very similar approach has been used to study the effects of strain on graphene~\cite{Vozmediano2010,Guinea2010,Levy2010} and other semimetallic systems~\cite{Rachel2016,Cortijo2016,Pikulin2016,Grushin2016,Matsushita2020}, where strain can play the role of a valley-dependent magnetic field.

For our purposes there are many ways in which $\mathfrak{e}_{\mu}^a$ can be treated on equal footing with the electromagnetic gauge field. However, there are some important distinctions.   First, the fields $\mathfrak{e}_j^a$ in Eq. (\ref{Eq:Edef}) are not true gauge fields.  This becomes important when considering the possible response actions: while the total charge of a system strictly conserved, momentum conservation is not similarly inviolable (see e.g.~\cite{Pikulin2016,Grushin2016,Matsushita2020} for some interesting physical consequences of this distinction).    Second, responses involving $\mathfrak{e}_\mu^a$ are predicated on the existence of translation symmetry. Thus, if the response is characterized by a boundary effect or a response to a flux/defect, we must be careful to ensure that (at least approximate) translation symmetry is maintained in order to connect the coefficient of the response action to explicit model calculations. Indeed, some responses are not well-defined unless configurations that maintain translation symmetry are used. This is unlike the electromagnetic response for which $U(1)$ charge symmetry is maintained independently of the geometry and gauge field configuration. Other important distinctions have been discussed in recent literature that has begun putting the gauging of discrete spatial symmetries on firmer ground~\cite{thorngren2018,chongwangvishwanath,barkeshli2021}.  One important distinction is that the translation gauge fields correspond to a discrete gauge symmetry $\mathbb{Z}_{N_a}$, where $N_a$ is the number of unit cells in the $a$-th direction.  This discreteness can play an important role in the topological response properties~\cite{gioia2021}, but we will not focus on this aspect in our work.

Using this framework, our goal is to consider the low-energy responses of electrons to the background electromagnetic and translation gauge fields. Given a translationally invariant Bloch Hamiltonian $H$, the response theories we consider can, in principle, be derived from correlation functions of the electromagnetic current
\begin{equation}
    j^\mu=e\frac{\partial H}{\partial k_\mu},
\end{equation} and the crystal momentum current
\begin{equation}
    \mathcal{J}^{\mu}_a = \hbar k_a\frac{\partial H}{\partial k_\mu},\label{eq:mom_current_def}
\end{equation} where the former couples to $A_\mu$ and the latter to $\mathfrak{e}_{\mu}^a$ (see App.~\ref{app:tp_derivation} for more details for the latter).  Indeed, we take exactly this approach in Section~\ref{sec:gradient_exp} to derive response actions for 2D and 3D systems. While our explicit derivations are important for precisely determining the coefficients of the response actions we study, it will be helpful to first motivate the overarching structure that connects a large subset of these response theories. We also note that alternative approaches to determining some of the response actions we discuss have been proposed in Refs.~\onlinecite{gioia2021, chongwang2, jakko}, and where the results overlap with ours, they agree.   

To understand the connections between the response theories we study, it is useful to begin by reviewing the well-known dimensional hierarchy of response theories of strong topological insulators~\cite{qi2008}. We show the general structure in Fig.~\ref{fig:stiresponse}(a), where the response terms are built solely from the electromagnetic gauge field. Furthermore, Chern-Simons and $\theta$-term response actions appear in even and odd spatial dimensions, respectively. There are a number of connections between the theories in different dimensions, and we will now review three of them. First, a Chern-Simons action in D spatial dimensions can be dimensionally reduced to a $\theta$-term action in (D-1)-dimensions by compactifying one spatial direction~\cite{kaluza, klein}. The (D-1)-dimensional system can also represent a TI if the value of $\theta$ is quantized to be $0, \pi$ by a symmetry that protects the (D-1)-dimensional topological insulator~\cite{qi2008}. Second, one can consider the reverse process where quantized adiabatic pumping~\cite{thoulesspumping} in (D-1)-dimensions will convert a $\theta$-term action to a D-dimensional Chern-Simons action. Finally, a $\theta$-term action for a (D-1)-dimensional topological insulator exhibits a half-quantized (D-2)-Chern-Simons response on boundaries where $\theta$ jumps by $\pi.$ These general relationships are summarized in Fig.~\ref{fig:stiresponse}(a) where each type of relationship is color- and symbol-coordinated.

Next we can consider a less familiar set of relationships in Fig.~\ref{fig:stiresponse}(b) between \emph{gapped} theories with mixed crystalline-electromagnetic responses arising from effective actions having both $A_\mu$ and $\mathfrak{e}_\mu^a$ fields.  We emphasize that the precise relationships we refer to in Fig.~\ref{fig:stiresponse}(b) are for gapped systems where the coefficients of the actions are quantized. In contrast, for the majority of this article we will focus on the quasi-topological responses of \emph{gapless} systems which take similar forms, but with non-quantized coefficients. Remarkably, many of the actions we discuss for insulators can be generalized to the non-quantized case. For semimetals, however, the dimensional relationships we point out are more akin to physical guides than a precise prescription for deriving matching coefficients in-between dimensions. 

With this caveat in mind, let us consider the family of theories in Fig.~\ref{fig:stiresponse}(b). In 0D we can consider the response action for a gapped system of electrons,$S[A]=Q\int dt A_0,$ which represents a system with charge $Q=eN_e$ where $N_e$ is the (integer) number of electrons. If we imagine stacking these 0D systems in a discrete, translationally invariant lattice in the $x$-direction, then we will generate a line of charges. Indeed, stacking produces the response for a translation invariant line-charge density which is captured by the next action in the sequence in Fig.~\ref{fig:stiresponse}(b), i.e., $Q\int A\wedge e^x= Q\int dx dt (A_0 \mathfrak{e}_{x}^x-A_x \mathfrak{e}_{0}^x).$ In this action the first term represents the charge density along the line, while the second term represents a current generated if the lattice of charges is moving. The latter consequence becomes manifest in the weakly distorted lattice limit since the current is proportional to the displacement rate: $j\sim \mathfrak{e}_0^x\sim \tfrac{\partial u^x}{\partial t}$. 

We can also imagine a reverse process where we are given a translationally invariant line of charge at integer filling and cut out a single unit cell. Since the system is gapped and translation invariant, this will result in a move in the opposite direction in Fig.~\ref{fig:stiresponse}(b), i.e., from $A\wedge \mathfrak{e}^x$ in 1D to $A$ in 0D with the same \emph{integer} coefficient $Q.$ We can use this example to highlight our caveat about gapped vs. gapless systems mentioned above. That is, while it is reasonable to have a 1D gapless system with non-quantized (i.e., non-integer) charge (per unit cell) described by the 1D action, the cutting procedure will not work properly at non-integer filling since the result will be a 0D point with a fractional charge.

In comparison to the response sequence for strong TIs, we see that stacking is the analog of pumping for the translation gauge field~\footnote{As mentioned, this analogy is precise for gapped systems. For gapless systems the analogy predicts the correct form of the action, but does not uniquely determine the coefficient.}. Indeed, while pumping adds an extra electromagnetic gauge field factor $A,$  stacking adds an extra translation gauge field $\mathfrak{e}^{D+1}$ where D+1 is the stacking direction. As a result, given any action in the strong TI sequence, we can stack copies to get the response action of a primary weak TI (stacks of co-dimension-$1$ strong TIs, e.g., lines stacked into 2D) by adding a wedge product with $\mathfrak{e}^{D+1}.$ We can push the stacking idea further to generate secondary weak TIs (stacks of co-dimension-$2$ strong TIs, e.g., lines stacked into 3D) by a wedge product with $\mathfrak{e}^{D+1}\wedge \mathfrak{e}^{D+2}$ and so on.

The stacking and cutting procedures are not the only relationships between the response theories in Fig.~\ref{fig:stiresponse}(b). Just as in the strong TI sequence, we can find connections between the boundary properties of some D-dimensional systems and the bulk response of a (D-1)-dimensional system. For example, the 2D response action in Fig.~\ref{fig:stiresponse}(b) represents the response of a stack of Su-Schrieffer-Heeger chains (SSH)~\cite{su1979}, each with a quantized polarization of $e/2.$ The boundary of such a 2D system is a line of charge on the edge, albeit with a density of $e/2$ electrons per unit cell on the edge line instead of the integer density we would get by stacking integer-filled 0D points. As such, the boundary of the 2D $A\wedge d\mathfrak{e}^x$ action  represents a line-charge described by the action $A\wedge \mathfrak{e}^x,$ but with a half-integer coefficient.

Now we can combine the dimensional relationships in the sequences of both Fig.~\ref{fig:stiresponse}(a) and (b) to make a family tree of related theories. We show a tree in Fig.~\ref{fig:stiresponse}(c) that includes response actions in 0, 1, 2, and 3 spatial dimensions. In 0D we have only an integer electron charge response that couples to $A_0.$ For 1D, we can either stack charges to form a line of charge (upper branch), or consider an electrically polarized TI (lower branch) where the charge is split in half and moved to opposing ends of the chain while the interior remains neutral (so to speak). In 2D we can stack line charges to get a plane of charge (top branch), stack 1D polarized TIs to get a weak TI (middle branch), or pump charge in a 1D TI to generate a 2D Chern insulator (bottom branch). 

In 3D the set of responses is richer. We can stack plane charges to generate a 3D volume of charges (top branch), stack Chern insulators to get a 3D primary weak TI (second from bottom branch), or stack 2D weak TIs to get a 3D secondary weak TI built from 1D polarized wires (second branch from top). The other well-known possibility is the magneto-electric response for a 3D strong TI~\cite{qi2008,essin2009} (bottom branch). Although it is not shown, this theory is related to a 4D quantum Hall system via pumping (3D to 4D) or dimensional reduction (4D to 3D)~\cite{qi2008}. The final option we consider, which is the middle branch enclosed by a dotted rectangle, is $\int dA\wedge d\mathfrak{e}^a.$ This response theory has not been previously studied in detail. This theory is a total derivative, and yields a gapped boundary with an electric polarization (e.g., a stack of SSH chains on the boundary). This is reminiscent of an electric quadrupole (higher-order) response~\cite{bbh,benalcazar2017}, and we will explore this connection further in Sec.~\ref{sec:3dsemimetalresponses}.

While this discussion has centered on gapped systems, our primary focus is on gapless topological semimetals. Importantly, each of the actions that contains a translation gauge field in the family tree in Fig.~\ref{fig:stiresponse}(c) can also represent a contribution to the response of various types of metals or topological semimetals~\cite{haldaneAHE,burkov2011A, Wan2011, zyuzin2012, RamamurthyPatterns,gioia2021,jakko,dbh2021,chongwang2, hirsbrunner2023}. This is because many semimetals can be generated by translation-invariant stacking of lower dimensional topological phases. Since the momentum $k^a$ in the stacking direction is conserved, one can consider adding up the set of topological response terms for each gapped $k^a.$ A semimetal represents a scenario where the coefficients of these topological terms at each $k^a$ are quantized and have discrete jumps where $k^a$ hits a nodal point. For example, the 2D electric polarization response of a stack of 1D TIs becomes the response of a 2D Dirac semimetal if the wires forming the stack are coupled strongly enough to close the insulating gap~\cite{RamamurthyPatterns}. In the presence of reflection symmetry, each momentum in the stacking direction has a quantized charge polarization that jumps when the momentum hits a gapless 2D Dirac point. Additionally, the 3D response of a stack of Chern insulators becomes the non-quantized anomalous Hall effect response of a time-reversal breaking Weyl semimetal where each fixed-$k$ plane that does not intersect a Weyl point carries a quantized Chern number that jumps at a Weyl point~\cite{haldaneAHE,burkov2011A, Wan2011, zyuzin2012}, and so on.   While many of these response theories have been discussed in detail before, only a few works have highlighted the contributions from the translation gauge fields~\cite{parrikar2,you2016,pikulin2016chiral,gioia2021,jakko,dbh2021,chongwang2, hirsbrunner2023}. As such, a large fraction of our paper will be devoted to both the explicit derivations of the response coefficients of the actions in Fig.~\ref{fig:stiresponse}(c) that have couplings to the translation gauge fields (Sec.~\ref{sec:gradient_exp}), and to the explicit calculations of the physical response phenomena in representative model systems (Sec.~\ref{sec:models}).

\begin{figure}
   \includegraphics[width=0.48\textwidth]{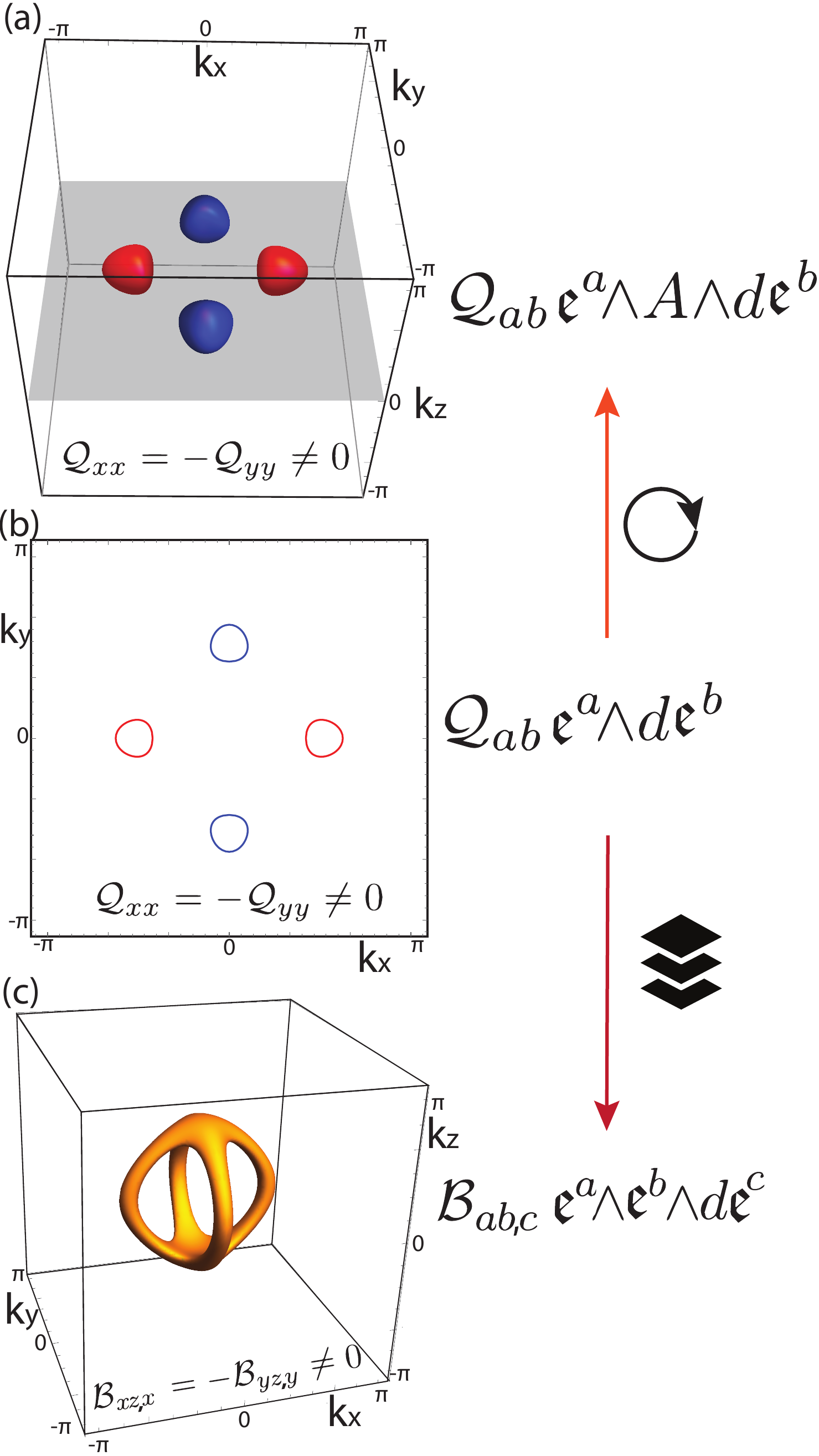}
    \caption{(a) Fermi-surfaces of a 3D time-reversal invariant Weyl semimetal with a quadrupole Weyl node configuration. Red and blue colors denote positive and negative Berry-curvature respectively. The associated action has a coefficient matrix $\mathcal{Q}_{ab}$ which is symmetric and proportional to the Weyl-node quadrupole moment. For this configuration the coefficients $\mathcal{Q}_{xx}$ and $\mathcal{Q}_{yy}$ are non-vanishing. (b) Similar to subfigure (a) except it is the Fermi surfaces for a 2D Dirac semimetal having four Dirac nodes in a quadrupole pattern. The action is described by a symmetric matrix of coefficients $\mathcal{Q}_{ab}.$ (c) the Fermi surface of an unusual cage-like nodal line semimetal built from stacking the Dirac node quadrupole semimetal in subfigure (b). The action has a set of coefficients $\mathcal{B}_{ab,c}$ which is anti-symmetric in $a$ and $b.$ Heuristically the action in (b) can generate the action in (a) by adiabatic pumping, or can generate the action in (c) by stacking.}
    \label{fig:ptiresponse}
\end{figure}

Before we move on to more explicit derivations, we want to motivate three additional response theories we will study that lie outside the family tree in Fig.~\ref{fig:stiresponse}(c). As mentioned above, a remarkable feature of the response actions of point-node semimetals is that their coefficients are determined from the energy-momentum locations of the nodal points. Indeed, for the relevant response actions in Fig.~\ref{fig:stiresponse}(c), the coefficients are obtained as a chirality-weighted momentum dipole moment of the point-nodes (note that Dirac points do not have a chirality, nevertheless there is a signed quantity that plays the same role).
Interestingly, recent work on rank-2 chiral fermions and Weyl semimetals with a chirality-weighted momentum \emph{quadrupole} moment~\cite{dbh2021, gioia2021,chongwang2, hirsbrunner2023} has unveiled a new set of response theories.  This category of theories has actions that include factors of more than one translation gauge field of the same type (e.g., $\mathfrak{e}^a\wedge d\mathfrak{e}^b$, where $a=b$), and as such, does not appear in the family tree in Fig.~\ref{fig:stiresponse}(c). This also implies that the translation gauge field factors in these response theories cannot be obtained by the conventional stacking of lower dimensional systems that we discussed above, since stacking produces wedge products with distinct translation gauge fields. We could also construct related higher dimensional theories (and lower dimensional theories if we considered both space and time translational gauge fields) to form an additional connected tree of theories, but we leave further discussion of those extensions to future work.

To give some explicit examples we show three response theories that follow this pattern in Fig.~\ref{fig:ptiresponse}. Fig.~\ref{fig:ptiresponse}(a) shows the Fermi surface structure of a 3D time-reversal invariant Weyl semimetal having a Weyl node quadrupole moment. The response action of this system is a mixed response between electromagnetic and translation gauge fields, and the inset in the Fermi-surface figure lists which coefficients $\mathcal{Q}_{ab}$ are non-vanishing. Some details of this response were discussed in Refs.~\onlinecite{dbh2021, gioia2021, hirsbrunner2023}, the former of which connected the response to rank-2 chiral fermions on the surface of the 3D Weyl semimetal. Fig.~\ref{fig:ptiresponse}(b) shows the Fermi surface structure of a 2D Dirac semimetal having a Dirac node quadrupole structure.  This response represents a momentum current responding to a translation gauge field (e.g., a strain configuration). Its form shares some similarities with the torsional Hall viscosity~\cite{hughesleighfradkin,parrikar1,parrikar2,bradlyn2015low,bradlynrao}, though a precise connection will be left to future work. Finally, in Fig.~\ref{fig:ptiresponse}(c) we show the Fermi-surface for an unusual nodal line semimetal formed from stacking the Dirac node quadrupole semimetal of Fig.~\ref{fig:ptiresponse}(b). While one might have naively expected two independent Fermi rings, we instead find a new type of Fermi-surface structure where the Fermi lines join at two crossing cap regions to form a cage. 

The symbols on the right-hand-side of Fig.~\ref{fig:ptiresponse} indicate the connections between these theories: (i) the response of the nodal line structure is just a stacked version of the 2D Dirac node quadrupole semimetal response from Fig.~\ref{fig:ptiresponse}(b), and (ii) one can heuristically consider the four-node Weyl response in Fig.~\ref{fig:ptiresponse}(a) to be a dimensional extension of the response in Fig.~\ref{fig:ptiresponse}(b) via pumping.

\section{Effective response actions}
\label{sec:gradient_exp}

Now that we have described the forms of the various response actions of interest, we will spend this section determining their coefficients. All of the response actions in Fig.~\ref{fig:stiresponse}(c) that contain only electromagnetic gauge fields represent insulators, and their coefficients have been studied in detail (e.g., see Ref.~\onlinecite{qi2008}). The actions containing translation gauge fields can represent insulators or gapless systems, and the two can often be distinguished by the values of the coefficients. That is, for insulators we expect the coefficients to be quantized in some units (in even spatial dimensions they are quantized in the presence of some symmetry), while for topological semimetals we expect the coefficients to be a tunable function of the momentum and energy locations of the nodal points or lines. Interestingly, some of the response coefficients for metals/semimetals can take the same values allowed for an insulator, although this would typically require fine-tuning, or extra symmetry. For example, a 1D system can have compensating particle and hole Fermi surfaces such that the total filling is an integer, as one would find in an insulator, yet the system is still gapless. In such a case we will show that the system has additional response terms that have coefficients that are incompatible with a gapped insulator.

Our focus will be on 2D Dirac, 3D Weyl, and 3D nodal line semimetals, and before we begin our derivations it is important to acknowledge a key qualitative difference between these types of topological semimetals. Namely, we recall that 2D topological Dirac semimetals and 3D nodal line semimetals require symmetry (the composite $\mathcal{T}\mathcal{I}$ symmetry) to guarantee the local stability of the gapless points/lines in momentum space. This is inherently different from the case of Weyl semimetals in 3D where Weyl nodes require no extra symmetry to protect them against perturbations. Indeed,  a Weyl node can be gapped out only by bringing another Weyl node of opposite chirality to the same point in the Brillouin zone. A similar story applies to (semi)metallic systems in 1D: each gapless point has a well-defined chirality defined as the sign of the Fermi velocity, and a gap can be opened only after overlapping Fermi points of opposite chiralities.
  
This distinction in symmetry protection is important for the response theories describing Dirac and Weyl semimetals as it reflects the well-known structure of anomalies in even and odd spatial dimensions. Furthermore, it will impact our strategy for deriving the response coefficients for these systems. As an example, the response properties of 2D Dirac semimetals can be determined straightforwardly from the Kubo formula if we first apply a symmetry-breaking perturbation that weakly gaps out the nodes. The resulting insulator response can then be taken to the semimetallic limit if we tune the perturbation to zero. Hence, the effective response action for such systems can be obtained by treating the system as an insulator and applying the Kubo formula, or more generally, a gradient expansion procedure. This method can be applied to 2D and 4D Dirac semimetals, and consequently 3D nodal line semimetals since they are just stacks of 2D Dirac semimetals.  For such semimetals we actually have a choice of what symmetry to break, e.g., inversion or time-reversal. Which one we need to break depends on the nodal configuration and the action we are intending to generate. For example, in the case of a 2D Dirac semimetal with a pair of nodes, breaking time-reversal is well-studied and generates a quantum Hall response via a Chern-Simons term. However, breaking inversion symmetry is relatively less-studied and generates  a mixed Chern-Simons response between an electromagnetic gauge field and a translational gauge field. This is corroborated by the fact that the electromagnetic Chern-Simons action breaks time-reversal, while the mixed Chern-Simons term with these fields breaks inversion.  We will show that the mixed Chern-Simons term has a well-defined limit as the gap closes and inversion symmetry is restored, which leads to a non-trivial response action for the 2D Dirac semimetal.  

Alternatively, the response of isolated chiral gapless points in 1D and 3D can be determined if they are viewed as theories that live on the boundary of a higher dimensional topological insulator or topological semimetal. In the presence of gauge fields, the higher dimensional bulk will generate a current inflow to the boundary to compensate the anomalous response of the gapless boundary modes. From this perspective, we expect that the effective response action of Weyl semimetals in odd spatial dimensions can be obtained by taking the boundary contribution of a higher dimensional system. There are likely other methods that could be applied to derive these response actions in their intrinsic spatial dimension, e.g., via the subtle introduction of an auxiliary $\theta$-field, but we choose our procedure since it reinforces the dimensional relationships discussed in the previous section and requires fewer formal tools.
 
Thus, our strategy for deriving the general form of the coefficients of mixed crystalline-electromagnetic responses is to begin by deriving effective response actions in even spatial dimensions, i.e., 2D and 4D. 
We will do so by identifying gradient expansion contributions (see Appendix~\ref{app:grad_exp} for a brief review) that contain an appropriate effective action constructed out of translational ($\mathfrak{e}^\lambda$) and electromagnetic ($A$) gauge fields.
Then the response of semimetals in odd spatial dimensions can be obtained by looking at the boundary of a response theory defined in one dimension higher. 

\subsection{Effective responses of 2D semimetals}
\label{Sec:SemimetalResponse}
In this subsection we will derive the coefficients of two 2D response actions that contain translation gauge fields, namely response action (v) from Fig.~\ref{fig:stiresponse}(c), and the response action in Fig.~\ref{fig:ptiresponse}(b). We will find that the coefficients of these actions are characterized by the dipole and quadrupole moments of the Berry curvature in the 2D Brillouin Zone, respectively. When we specialize to 2D Dirac semimetals, the distribution of Berry curvature is sharply localized as $\pm \pi$-fluxes at the Dirac nodes. Hence, the coefficients will become proportional to the dipole and quadrupole moment of the distribution of Dirac nodes. 

\subsubsection{Dirac node dipole semimetal}\label{sec:DDSderivation}
Let us start by considering a gapped $\mathcal{T}$-invariant system having broken $\mathcal{I}$ symmetry. Under these conditions the electromagnetic Chern-Simons term, which represents the Hall conductivity, vanishes, and we can consider the mixed linear response of a momentum current responding to an electromagnetic field, or vice-versa. Using the Kubo formula, or applying the gradient expansion procedure described in App.~\ref{app:grad_exp}, we find the contribution to the effective action (when the chemical potential lies in the insulating gap) (See also~\cite{Matsushita2020}):
\begin{equation}
    \begin{split}
        S_{e,A}= -e\int d^3r\ \mathfrak{e}^\alpha_\mu \partial_\nu A_\rho 
         \int \frac{d\omega d^2 k}{(2\pi)^3}k_\alpha\Omega^{(3)}_{\mu\nu\rho}(\omega,k),
    \end{split}
\end{equation}\noindent where 
\begin{equation}
    \Omega^{(3)}_{\mu\nu\rho}(\omega,k)= \text{tr}\left(G_0  \frac{\partial G_{0}^{-1}}{\partial k_\mu} \frac{\partial G_{0}}{\partial k_\nu}\frac{\partial G_{0}^{-1}}{\partial k_\rho} \right), 
\end{equation} and
$G_0(k_\mu)$ is the single-particle Green function. To extract the coefficient of the $\mathfrak{e}^\alpha\wedge dA$ term, we contract $\Omega^{(3)}_{\mu\nu\rho}$ with the totally antisymmetric tensor $\tfrac{1}{3!}\varepsilon^{\mu\nu\rho}$.  This gives the coefficient
\begin{equation}
    c_\alpha=e\frac{\varepsilon^{\mu\nu\rho}}{3!}\int\frac{d\omega d^2 k}{(2\pi)^3}k_\alpha\ \Omega^{(3)}_{\mu\nu\rho}(\omega,k)
    \label{eqn:app_DD_coeff1}
\end{equation}
of the response action
\begin{equation}
    S_{e,A}=c_\alpha\int \mathfrak{e}^\alpha\wedge dA.\label{eqn:responseEdA}
\end{equation}

We note that Eq.~\ref{eqn:app_DD_coeff1} is very similar to the response coefficient of the standard electromagnetic Chern-Simons term apart from the factor of $k_{\alpha}$ in the integrand. As such, assuming $\alpha=x,y$, we can use a well-established result to evaluate the frequency integral to obtain~\cite{Zhong2012}:
\begin{equation}
\begin{split}
    \frac{\varepsilon^{\mu\nu\rho}}{24\pi^2}\int d\omega d^2 k\ \Omega^{(3)}_{\mu\nu\rho}(\omega,k)
    =\frac{1}{2\pi}\int_{BZ} dk_x dk_y\ \mathcal{F}^{xy}(k_x,k_y),
\end{split}
\label{eqn:app_chern}
\end{equation}
where $\mathcal{F}^{xy}$ is the Berry curvature. Hence, we can rewrite $c_{\alpha}$  as an integral over the BZ by substituting this relationship into Eq.~\ref{eqn:app_DD_coeff1} to find:
\begin{equation}
    c_\alpha=\frac{e}{(2\pi)^2}\int_{BZ} dk_x dk_y\ k_\alpha \mathcal{F}^{xy}(k_x,k_y).
\label{eqn:coeffDD}
\end{equation} We have thus arrived at the result that $c_{\alpha}$ is proportional to the $\alpha$-th component of the dipole moment of the distribution of Berry curvature. This coefficient can be non-zero since it is allowed by broken $\mathcal{I}$ and preserved $\mathcal{T}$, i.e., $\mathcal{F}^{xy}(\textbf{k})= -\mathcal{F}^{xy}(-\textbf{k}).$
We also note that $c_\alpha$ is independent of the choice of zone center, and shifts of $k$ in the integrand in general, because the Chern number (Hall conductivity) vanishes in the presence of $\mathcal{T}$.  

In a gapped $\mathcal{T}$-invariant system, restoring $\mathcal{I}$-symmetry forces  $c_{\alpha}$ to vanish, since $\mathcal{F}^{xy}(\textbf{k})=0.$  However, in  gapless systems this need not be the case. To see this, we apply our result from Eq.~\ref{eqn:coeffDD} to a 2D Dirac semimetal by first introducing a weak perturbation $V_{\mathcal{I}}$ which breaks $\mathcal{I}$ and opens up a small gap, and then taking the limit $V_{\mathcal{I}}\rightarrow 0$, in which inversion symmetry is restored.  In the gapped system the Berry curvature  $\mathcal{F}^{xy}$ is distributed smoothly across the entire 2D BZ.
In the gapless limit, however, the Berry curvature distribution will develop sharp peaks of weight $\pi$ localized at the positions of the Dirac points:
\begin{equation}
    \mathcal{F}^{xy}=\sum_{a=1}^{N_D}\pi\chi_a\delta (\textbf{k}-\textbf{k}^a),
\end{equation} where  $a$ runs over all Dirac nodes at momenta $\textbf{k}^a,$ and $\chi_a=\pm 1$ is an integer indicating the sign of the $\pi$-Berry phase around the Fermi surface of the $a$-th Dirac point at a small chemical potential above the node~\cite{RamamurthyPatterns}.  Ultimately, we find the effective response action of a Dirac node dipole semimetal is given by:
\begin{equation}
    S_{DD}=\frac{e\mathcal{P}_\alpha}{4\pi}\int \mathfrak{e}^\alpha\wedge dA,
    \label{eqn:DD_resp}
\end{equation} where \begin{equation}\mathcal{P}_\alpha=\sum_{a=1}^{N_D}\chi_a k^{a}_{\alpha},\label{eqn:berrycurvdipole}
\end{equation} is the dipole moment of the Dirac nodes.

Note that if the Dirac nodes meet at the zone boundary, they can be generically gapped even in the presence of $\mathcal{T}\mathcal{I}$ symmetry. The resulting  insulating phase represents a weak TI having  $\mathcal{P}_\alpha = G_\alpha$, where $G_\alpha$ are the components of a reciprocal lattice vector.   In this case, the action in Eq.~\ref{eqn:DD_resp} describes a stack (i.e., a family of lattice lines/planes corresponding to $G_a$) of 1D polarized TI chains aligned perpendicular to $G_a$. To see this explicitly, take $G_x=\frac{2\pi}{a_x}$, and set $e^\alpha_\beta = \delta^\alpha_\beta$ in Eq.~\ref{eqn:DD_resp} to obtain the action 
\begin{equation}
  \frac{e }{2 } \int \frac{ d x }{a_x} \left( \int d y d t E_y \right )=N_x\frac{e}{2}\int dy dt E_y, \end{equation}
where $N_x$ is the number of unit cells in the $x$-direction. This action is just $N_x$ copies of the usual $\theta$-term action for 1D, electrically-polarized topological insulators $(\theta=\pi)$ parallel to the $y$-direction, stacked along $\hat{x}.$

We have now derived Eq.~\ref{eqn:DD_resp} as a quasi-topological contribution to the response of a 2D Dirac semimetal where the nodes have a dipolar configuration. However, there is another important subtlety that we will now point out. Earlier work has shown that the electromagnetic response of 2D Dirac semimetals with both $\mathcal{T}$ and $\mathcal{I}$ symmetry is an electric polarization proportional to the Dirac-node dipole moment~\cite{RamamurthyPatterns}. Even more recently, connections have been made between mixed translation-electromagnetic responses and the electric polarization~\cite{chongwangvishwanath}. Since we have a clear derivation of the response term we can use our results to understand the precise connection between the electric polarization and the coefficient $c_{\alpha}$ of the $e^{\alpha}\wedge dA$ response action. Using the standard approach of Ref. \onlinecite{vanderbilt1993}, the polarization in 2D is
\begin{equation}
\begin{split}
     P_e^\alpha =\frac{e}{(2\pi)^2}i\int_{BZ} d^2\textbf{k}\langle u_\textbf{k}|\partial_{k_\alpha} u_\textbf{k}\rangle 
\end{split}
\end{equation} where $\mathcal{A}^\alpha(k)=i\langle u_\textbf{k}|\partial_{k_\alpha} u_\textbf{k}\rangle$ is the Berry connection. Hence, we find that the electric polarization $P_{e}^{\alpha}$ is related to $c_\alpha$ by an integration by parts (See Appendix~\ref{app:pol_berry_dipole}):
\begin{equation}\begin{split}
P^{\alpha}_e&=\frac{e}{(2\pi^2)}\varepsilon^{\alpha\beta}\int d^2 k\ k_{\beta} \mathcal{F}^{xy}+\frac{e}{2\pi}W^{\alpha} \\&= \epsilon^{\alpha\beta}c_{\beta}+\frac{e}{2\pi}W^{\alpha},    
\end{split}
\label{eqn:2D_pol_bd+Wy}
\end{equation}\noindent where we have set the lattice constants equal to unity, and the Wilson loop
\begin{equation}
    W^{\alpha}=\oint dk_{\alpha} \mathcal{A}^{\alpha}(k_{\alpha},k_{\beta}=\pi),
\end{equation} is an integral of the Berry connection $\mathcal{A}^{\alpha}$ along the $\alpha$-th momentum direction at a fixed, inversion-invariant transverse momentum $k_{\beta}=\pi$ at the boundary of the BZ. 

From this explicit relationship we can immediately draw some conclusions. First, in the Dirac semimetal limit, we reproduce the result of Ref. \onlinecite{RamamurthyPatterns} where the polarization is proportional to the Dirac node dipole moment: $P^{\alpha}_e=\frac{e}{2(2\pi)}\varepsilon^{\alpha\beta}\mathcal{P}_{\beta}.$ And second, if we have broken inversion symmetry (while $\mathcal{T}$ is still preserved), we see that the polarization and the coefficient $c_{\alpha}$ are not quantized, and \emph{not equal} to each other. This scenario can be found in inversion-breaking insulators with a Berry curvature dipole moment. These insulators will have a charge polarization, and they will also have a mixed translation-electromagnetic response. However, we find from this calculation, and explicit numerical  checks, that they are generically inequivalent.  Ultimately this boils down to the fact that the Wilson loop at the boundary of the BZ requires a symmetry to be quantized, e.g., mirror or inversion. Otherwise, the Wilson loop gives a contribution that distinguishes the polarization and the mixed crystalline-electromagnetic responses. We leave a detailed discussion of this subtle distinction to future work.

To summarize, Eq.~\ref{eqn:DD_resp} captures the generic mixed crystalline-electromagnetic response of the bulk of a 2D system with $\mathcal{T}$-symmetry. In the limit of a Dirac semimetal, the coefficient of the response coincides with the electric polarization of the system. We note that in this limit there will be other non-vanishing response terms since the system is gapless, but Eq.~\ref{eqn:DD_resp} represents a distinct contribution to the total response of the system to electromagnetic and translation gauge fields. We will study an explicit model with this response term in Sec.~\ref{sec:2D_Dirac_dipole}.

\subsubsection{Dirac node quadrupole semimetal}
\label{Sec:2dQuadDirac}
Now we will move on to discuss the response of quadrupole arrangements of 2D Dirac nodes as in Fig.~\ref{fig:ptiresponse}(b). If the Chern number and momentum dipole moment $\mathcal{P}_{\alpha}$ vanish, then our semimetal has a well-defined momentum quadrupole moment, which is independent of the choice of zone center. We now show that such systems are described by the response action:
\begin{equation}
  S_{DQ}=  \frac{\hbar \mathcal{Q}_{\alpha\beta}}{8\pi}\int \mathfrak{e}^{\alpha}\wedge d\mathfrak{e}^{\beta}.
     \label{eqn:QD_resp}
\end{equation}

From the derivation in the previous section we anticipate that, in the limit of a Dirac semimetal band structure, the coefficient $\mathcal{Q}_{\alpha\beta}$ of this response action is related to the momentum quadrupole moment of the Dirac nodes. To confirm this statement let us consider the linear response of a momentum current to a translation gauge field for a gapped system. From the Kubo formula, or gradient expansion, we find a coefficient of the $\mathfrak{e}^{\alpha}\wedge d \mathfrak{e}^{\beta}$ term:
\begin{equation}
    \mathcal{Q}_{\alpha\beta}\equiv \frac{1}{2}\frac{\varepsilon_{\mu\nu\rho}}{3!}\int\frac{d\omega d^2 k}{(2\pi)^3}k_{\alpha} k_{\beta}\ \Omega^{(3)}_{\mu\nu\rho}(\omega, k).
\end{equation} We can use the relationship mentioned in Eq.~\ref{eqn:app_chern} to carry out the frequency integral to obtain the coefficient of Eq.~\ref{eqn:QD_resp}:
\begin{equation}
   \mathcal{Q}_{\alpha\beta}=\frac{1}{\pi}\int_{BZ} dk_x dk_y\ k_{\alpha} k_{\beta}\mathcal{F}^{xy}(k_x,k_y).
\end{equation}

To apply this to the Dirac node quadrupole semimetal shown in Fig.~\ref{fig:ptiresponse}(b),
we evaluate the response by first introducing a symmetry-breaking mass term and then studying the topological response of the resulting gapped system.  In this case the mass term breaks $\mathcal{T}$ but has a vanishing total Chern number. In the example at hand, this can be done by  adding a $k$-independent term that opens a local mass of the same sign for each of the four Dirac points in Fig.~\ref{fig:ptiresponse}b.  Such a mass term preserves $\mathcal{I}$, which in the gapped system automatically guarantees a vanishing dipole moment of the Berry curvature.   This, together with the vanishing Chern number, is necessary so that the momentum quadrupole moment is well-defined, independent of the choice of zone center. For this scenario, in the limit that the perturbative mass goes to zero,
\begin{equation}
    \mathcal{Q}_{\alpha\beta}=\sum_{a=1}^{N_D}\chi_{a} k^{a}_{\alpha}k^{a}_{\beta},\label{eq:momquadchirality}
\end{equation} which is the Dirac node quadrupole moment. In Sec.~\ref{sec:2D_Dirac_qp} we will explicitly study a model with this Berry curvature configuration and a resulting non-vanishing $\mathcal{Q}_{\alpha\beta}.$ We will see that while the Dirac node dipole moment captures the electric polarization (see Appendix~\ref{app:pol_berry_dipole}), the nodal quadrupole moment captures a kind of momentum polarization (see Appendix~\ref{app:quad_pol}) (this time, without the subtlety of the additional Wilson loop contribution discussed above). For comparison, the surface charge theorem relates the bulk electric polarization to a boundary charge, and for the momentum polarization there will be a boundary momentum.

\subsection{Effective responses of 1D (semi)metals}
\label{sec:1dsemimetalderivation}
\begin{figure}[]
   \includegraphics[width=0.48\textwidth]{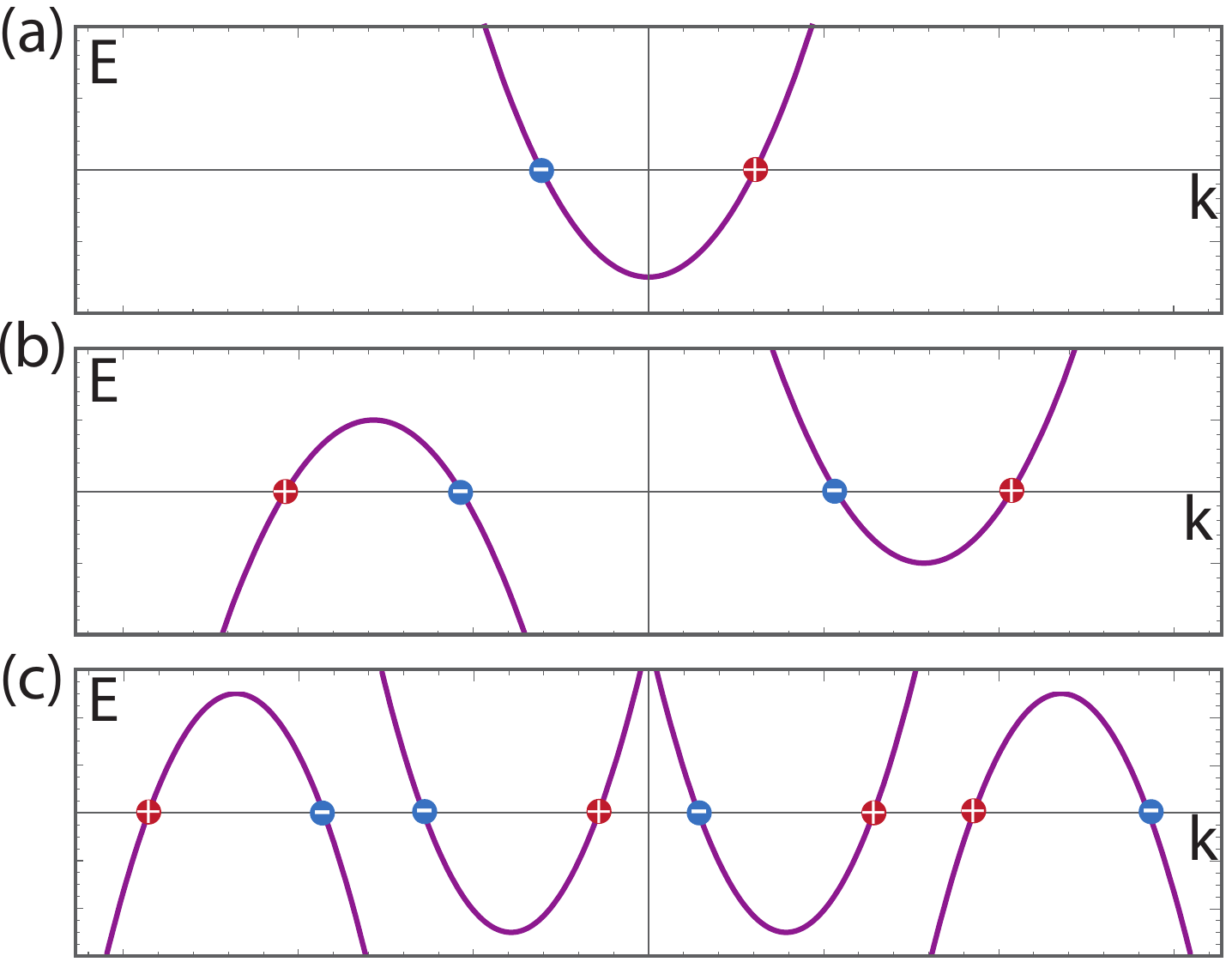}
    \caption{(a) One-dimensional band structure of an ordinary metal. The pair of gapless points is marked by the sign of their respective chiralities, highlighting the momentum-space dipole characterizing the response of the system.
    (b) Band structure of a 1D metal characterized by a momentum quadrupole moment. The system has an integer (vanishing in this case) charge filling, but a non-zero momentum.
    (c) Band structure of a 1D metal characterized by a momentum octupole moment. The system has an integer (vanishing) filling, a vanishing momentum, but a non-vanishing expectation value for the square of the momentum. See Appendix~\ref{app:responsesin1d}.}
    \label{fig:1dresponse}
\end{figure}

Now that we have derived the responses of 2D systems coupled to electromagnetic and translation gauge fields, we will use Figs.~\ref{fig:stiresponse}(b) and \ref{fig:ptiresponse} as guides to generate related responses in 1D and 3D. To get 1D responses we will consider the boundary response of the 2D systems (this subsection), and we will stack the 2D responses to get 3D responses of nodal line semimetals (next subsection).  We note that in the following discussion we will treat translation as a continuous symmetry (as in Appendix~\ref{app:tp_derivation}, as this perspective is useful for obtaining the correct response actions from our diagram calculations). One can see Ref. \onlinecite{gioia2021}, for example, for a discussion that treats the subtleties associated to having a discrete translation symmetry. 

It is well-known that chiral modes in 1D are anomalous, i.e., charge is not conserved when we apply an electric field. In 1D lattice models this anomaly is resolved because of fermion doubling, i.e., for every right-moving chiral mode there is a corresponding left-moving mode that compensates the anomaly. While it is true that the electromagnetic charge anomaly is resolved with such a lattice dispersion, the doubled system  can still be anomalous in a different but related sense if we have translation symmetry (see Ref. \onlinecite{gioia2021} for a similar discussion). 

To be specific, in the presence of translation symmetry we can consider the momentum current in Eq.~\ref{eq:mom_current_def}: $\mathcal{J}_x^\mu=\hbar k_x j^\mu$ where $j^\mu$ is the particle number current. At low energies, current-carrying excitations lie in the vicinity of Fermi points $k^{F,\alpha}_{x}$ and carry corresponding particle currents $j^{\mu}_{(\alpha)}.$ The total contribution to momentum current from these low-lying modes is: 
\begin{equation}
    \mathcal{J}^\mu_x=\sum_{\alpha}\hbar k^{F,\alpha}_{x}j^\mu_{(\alpha)}.
\end{equation}

In the simplest case of a nearest-neighbor lattice model having a single, partially-filled band, we have two Fermi points: $k^F\equiv k^{F,R}_x=-k^{F,L}_x$, with $j^{\mu}_{R} = (\rho_R, v_F \rho_R)$ and $j^{\mu}_{L} = (\rho_L, -v_F \rho_L)$, where $\rho$ is the number density. Interestingly, the momentum current in this scenario is 
\begin{equation}
    \mathcal{J}^\mu_x=\hbar k^F(j^{\mu}_{R}-j^{\mu}_{L}),
\end{equation} which, up to a factor of $\hbar k^F,$ is just the axial current! 

Importantly, even though this lattice model does not have an electromagnetic charge anomaly, $\partial_\mu(e j^\mu_L+e j^\mu_R)=0$, it does have an axial anomaly:
 \begin{equation}
    \partial_\mu( j^\mu_R- j^\mu_L)=\frac{eE^x}{\pi\hbar}.
\end{equation} Taking this point of view, we can reformulate the axial anomaly in this system as a mixed crystalline-electromagnetic anomaly where an electric field $E_x$ violates conservation of the $k_x$ momentum current,
\begin{equation}
    \partial_\mu \mathcal{J}^{\mu}_x=\frac{e\hbar k^F }{\pi\hbar}E^x.\label{eqn:1D_Dmomanomaly}
\end{equation} More generally the anomaly is proportional to the momentum dipole moment of the Fermi points, which replaces a factor of $2k_F$ in Eq.~\ref{eqn:1D_Dmomanomaly} (see App.~\ref{app:responsesin1d}).

There is a conjugate effect that occurs in an applied strain field, which can be implemented as a translation electric field $\mathcal{E}^x_x=\partial_x \mathfrak{e}_{0}^x-\partial_t\mathfrak{e}_x^x.$ Naively such a non-vanishing field will generate violations to the conservation law for the usual  electromagnetic current according to \begin{equation}
    \partial_\mu (ej^{\mu})= \frac{e k^F }{\pi} \mathcal{E}^x_{x},\label{eqn:1D_Dchargeanomaly}
\end{equation} (again see App.~\ref{app:responsesin1d} for a more general expression in terms of the momentum dipole). However, this equation is not quite correct if we have an isolated system with a fixed number of electrons, and hence, we must be careful when considering time-dependent changes to $\mathfrak{e}_x^x$ as we will now describe. 

To gain some intuition for Eq.~\ref{eqn:1D_Dchargeanomaly}, consider increasing the system size by one lattice constant $a$ during a time $T$ by adding an extra site to the system: $\int dx dt\ \mathcal{E}_x^x=a$ (one can also think of threading a dislocation into the hole of a 1D periodic system). From the anomaly equation we would find that the amount of charge in the system changes by $e k^F a/\pi,$ as one would expect for adding a unit cell to a translationally invariant system having a uniform charge density $\rho=e k_F/\pi.$ However, there is a subtlety that we can illustrate by considering a system having a fixed number of electrons $N_e=k^F L_x/\pi,$ which we strain by uniformly increasing the lattice constant. Assuming a uniform system, the anomalous conservation law in this case becomes
\begin{equation}
    \partial_t\rho = \partial_{t}\left(\frac{ek^F}{\pi}\mathfrak{e}_{x}^{x}\right).
\end{equation} Crucially, we note that if we increase the system size with fixed particle number, then $k^F$ will decrease. Indeed, in the small deformation limit the momenta are proportional to $(\mathfrak{e}^x_{x})^{-1}$ since their finite size quantization depends inversely on the system size. Using this result, the conservation law becomes:
\begin{equation}
    \partial_t\rho = \frac{e}{\pi}\left(\mathfrak{e}_{x}^{x}\partial_{t}k^F+k^F\partial_t \mathfrak{e}^{x}_x\right)=\frac{ek^F}{\pi}(-\partial_t \mathfrak{e}_x^x+\partial_t \mathfrak{e}_x^x)=0
\end{equation} where we used $\partial_t (\mathfrak{e}_x^x)^{-1}=-(\mathfrak{e}_x^x)^{-2}\partial_t\mathfrak{e}_{x}^x.$ 

The outcome that $\partial_t \rho=0$ is the result one would expect by stretching the system uniformly while keeping the number of particles fixed. To clarify, at a fixed particle number we know the total charge cannot change, however it perhaps seems counter-intuitive that the \emph{density} does not decrease if we stretch the system. The reason is that the quantity $\rho$ above, which is defined as $\tfrac{\delta S}{\delta A_0},$ is not a scalar density. Indeed, for general geometries the scalar charge density would be defined as
\begin{equation}
    \bar{\rho}=\frac{1}{\mathfrak{e}^x_x}\frac{\delta S}{\delta A_0},
\end{equation} where the $\mathfrak{e}_x^x$ is essentially playing the role of the determinant of a spatial metric. To calculate the total charge we would then use
\begin{equation}
    Q=\int dx\, \mathfrak{e}_x^x \bar{\rho}=\int dx\, \rho.
\end{equation} Indeed, the scalar charge density $\bar{\rho}$ will decrease as the system is stretched since $\partial_t \bar{\rho}\propto \partial_t \mathcal{P}_x$ which decreases as the system size increases at fixed electron number.

The effective response action of the 1D system can be derived as a boundary effective action of an appropriate 2D theory.
In fact, we have already seen such a 2D system when studying the 2D Dirac semimetal with Dirac nodes arranged in a dipolar fashion.
The bulk response for this 2D system with a weak inversion-breaking gap is Eq.~\ref{eqn:DD_resp}. As mentioned above, this bulk theory implies that the system has an electric polarization. From the surface-charge theorem for polarization we expect that the boundary will have a charge density equal the polarization component normal to the boundary. The contribution to the buondary effective action from  Eq.~\ref{eqn:DD_resp} is: 
\begin{equation*}
    S_{\partial}=\frac{e}{4\pi}\mathcal{P}_\alpha \int \mathfrak{e}^\alpha\wedge A.
\end{equation*}
From this we can extract the boundary charge density: $\rho_{\partial}=\frac{e}{2}\frac{\mathcal{P}_{\partial}}{2\pi}\mathfrak{e}_{\partial}^{\partial}$ where $\mathcal{P}_{\partial}$ is the component along the boundary, and  $\mathfrak{e}_\partial^{\partial}$ is the diagonal translation gauge field component along the boundary that is simply equal to unity in non-deformed geometries. 

While the form of this action is what we expect for a 1D metal, the coefficient is half the size it should be. The reason is that on the edge of the 2D Dirac semimetal, the momentum-space projections of the bulk Dirac nodes in the edge BZ represent points where the edge-filling changes by $\pm e/2,$~\cite{RamamurthyPatterns} not $\pm e$  as would be the case for a 1D Fermi-point in a metal. Hence for a metal we expect a result twice as large (we will see a similar result in Sec.~\ref{sec:3dsemimetalresponses} when comparing the boundary response of a 4D system to that of a 3D Weyl semimetal).
Thus the action for the 1D system is 
\begin{equation}
    S_{1D,D}=\frac{e}{2\pi}\mathcal{P}_\alpha \int \mathfrak{e}^\alpha\wedge A.\label{eq:1DD_resp}
\end{equation}
From this form  we can identify $\mathcal{P}_{\alpha}=(-\Delta \mu/\hbar,\Delta k_x)$ such that  $\frac{\mathcal{P}_x}{2\pi}$ is simply the filling fraction of the 1D metal and $\frac{\mathcal{P}_t}{2\pi}$ measures the imbalance of left- and right-moving excitations in the system ($\Delta \mu=\mu_R-\mu_L$).

Introducing a charge current vector
\begin{equation}
    j^\mu=\frac{e}{2\pi}\varepsilon^{\mu\nu}\mathcal{P}_{\nu}=\frac{e}{2\pi}\left(\Delta k_x, \Delta\mu/\hbar \right)^T
    \label{eqn:1D_D_resp}
\end{equation}
we can recast Eq.~\ref{eqn:1D_D_resp} in the most familiar form: $S_{1D,D}=\int dtdx\ j^\mu A_\mu$. Thus, we have now generated the action (ii) from Fig.~\ref{fig:stiresponse}(c). Let us also note that the edge states of the Dirac semimetal can be flat, while the 1D context we mentioned above has a dispersion. However, the key feature of both cases is that as momentum is swept across the 1D BZ (1D surface BZ for the 2D case) the filling of the states changes in discrete jumps at either the Fermi points in 1D, or the (surface-projected) Dirac points in 2D. It is this change in the filling that is captured by the quantity $\mathcal{P}_x,$ and does not depend on the dispersion in a crucial way.

Now that we have this example in mind, we can ask what the analogous 1D boundary system is for the Berry curvature quadrupole action Eq.~\ref{eqn:QD_resp}. We mentioned that this bulk response represents a momentum polarization, which implies that the boundary should have a momentum density parallel to the edge. Indeed, we expect that such a 1D system will have a vanishing Fermi-point dipole moment (i.e., the filling is integer), but a quadrupole moment that is non-vanishing (see Fig.~\ref{fig:1dresponse}(b)). 

From the point of view of the translation gauge fields, such band structures are chiral since either the right movers or left movers carry larger momentum charge.  To see this, consider a 1D Fermi surface with right-movers at momenta $\pm K_F$, and left-movers at momenta $\pm Q_F$.  Let us further restrict our attention to currents for which the net number of right-movers (and of left-movers) is zero, e.g. $\rho_R(K_F) + \rho_R( - K_F) = 0$.  Defining $\delta \rho_R = ( \rho_R(K_F) - \rho_R( - K_F) )$, and $\delta \rho_L =  ( \rho_L(Q_F) - \rho_L( - Q_F) )$, we see that the momentum gauge field couples to 
\begin{equation}
    \mathcal{J}^{\mu}_{x} = K_F \delta \rho_R + Q_F \delta \rho_L \ .
\end{equation}
Thus we see that for $K_F \neq Q_F$ (as in Fig.~\ref{fig:1dresponse}(b)), the momentum gauge field couples differently to right- and left- moving density fluctuations.  In the extreme limit that $Q_F = 0$, the momentum gauge theory is fully chiral.  

More generally, in a 1D system with a Fermi-point quadrupole (c.f. Eq.~\ref{eq:momquadchirality}) $\mathcal{Q}_{xx}=\sum_{a=1}^{N_F}={\rm{sgn}}(v_{Fa})(k_{x}^{(a)})^2$, and fixed electric charge, this chiral coupling leads to  an anomaly in the presence of a non-vanishing translation gauge field:
\begin{equation}
\partial_{\mu}\mathcal{J}^{\mu}_{x}=\frac{\hbar\mathcal{Q}_{xx}}{4\pi}\mathcal{E}_{x}^{x}.    
\label{Eq:ChiralMomentum}
\end{equation} This anomaly implies that if we turn on a translation gauge field (e.g., via strain) then we will generate momentum as shown in App.~\ref{app:responsesin1d}~\footnote{As shown in the Appendix, this anomaly has two contributions. One comes from the low-energy currents that contribute with a factor of $1/2\pi$ and a second from a change of system-size for a ground state carrying a non-vanishing momentum density with a factor of $-1/4\pi$.}.

The response theory describing such a 1D system is similar to that describing the chiral boundary of a Chern-Simons theory.  Indeed, if we start from Eq.~\ref{eqn:QD_resp} and derive the boundary response (and compensate for a similar factor of two as mentioned above in the momentum-dipole case) we arrive at an effective action: 
\begin{equation}
    S=-\frac{\hbar}{4\pi}\int dtdx\ \left(\mathcal{Q}_{xx}\mathfrak{e}^x_x\mathfrak{e}^x_t+\mathcal{Q}_{xt}\mathfrak{e}^x_x\mathfrak{e}^t_t\right).
    \label{eqn:1D_Q_resp}
\end{equation} In this effective action the momentum quadrupole moment of the Fermi points $\mathcal{Q}_{xx}$  encodes the ground state momentum density (see Appendix~\ref{app:responsesin1d}). The quantity $\mathcal{Q}_{xt}$ is the mixed Fermi-point quadrupole moment in momentum and energy, but we leave a detailed discussion of such mixed moments to future work.

The arguments of this section can be extended to look at higher moments of the chirality-weighted Fermi momenta, which are proportional to the ground state expectation values of higher and higher powers of momenta. To describe these properties, and related response phenomena, we can introduce gauge fields $\mathfrak{e}^{abc\ldots}$ that couple to higher monomials of momentum, $k_a k_b k_c\ldots.$ For example, the fields that couple to zero powers or one power of momentum are the electromagnetic $A$ and translation gauge fields $k_x\mathfrak{e}^x$ respectively, and we could introduce a coupling $k_a k_b \mathfrak{e}^{ab}$ to the set of 1-form gauge fields $\mathfrak{e}^{ab},$ e.g., $k_x^2 \mathfrak{e}^{xx}$. We describe the hierarchical anomalies associated to these gauge fields in Appendix~\ref{app:responsesin1d}.

\subsection{Effective responses of 3D nodal line semimetals}
\label{Sec:NLSMResponse}
We can now use our 2D results from Sec.~\ref{Sec:SemimetalResponse} to generate the responses of two types of nodal line semimetals in 3D. To generate the two types we imagine stacking either the action in Eq.~\ref{eqn:DD_resp} or the action in Eq.~\ref{eqn:QD_resp}. The action resulting from the former has been discussed in Refs.~\cite{ramamurthylinenode,chongwang2}; the second is, to the best of our knowledge, new.  From our arguments for gapped systems in Sec.~\ref{sec:overview}, we expect that the form of the actions we obtain from stacking will contain an extra wedge product with the translation gauge field in the stacking direction. To be explicit, suppose we are stacking up 2D semimetals (that are parallel to the $xy$-plane) into the $z$-direction. By stacking decoupled planes of the responses in either Eq.~\ref{eqn:DD_resp} or Eq.~\ref{eqn:QD_resp}, we expect to find
\begin{equation*}
    S=\frac{e\mathcal{P}_{\alpha}}{4\pi a_z}\int \mathfrak{e}^z\wedge\mathfrak{e}^{\alpha}\wedge dA,
\end{equation*}
or
\begin{equation*}
    S=\frac{\hbar \mathcal{Q}_{\alpha\beta}}{8\pi a_z}\int \mathfrak{e}^z\wedge\mathfrak{e}^{\alpha}\wedge d\mathfrak{e}^{\beta},
\end{equation*} respectively, where $\alpha,\beta =x,y.$  The forms of these actions match action (viii) in Fig.~\ref{fig:stiresponse}(c) and the action in Fig.~\ref{fig:ptiresponse}(c) respectively. We note that the stacked, decoupled systems simply inherit the response coefficient of the 2D system.

We want to consider more general configurations of systems with stacked and coupled planes, perhaps stacked in several directions. As we have seen, if the layers we stack are decoupled, then each layer contributes the same amount. This contribution (for a stack in the $z$-direction) is captured by the integral $\tfrac{1}{a_z}\int \mathfrak{e}^z=N_z$ where $N_z$ is the number of layers. However, if the layers are coupled, then each fixed-$k_z$ plane can have a different amount of Dirac node dipole ($\mathcal{P}_\alpha(k_z)$) or Dirac node quadrupole moment ($\mathcal{Q}_{\alpha\beta}(k_z)$) respectively. The total coefficient is then determined by the sum over all values of $k_z.$ One can also have stacks in any direction, not just the $z$-direction. Hence, in this more generic scenario the actions become
\begin{equation}
    S_{DD3}=e\mathcal{B}_{\alpha\beta}\int \mathfrak{e}^\alpha\wedge\mathfrak{e}^{\beta}\wedge dA,
\end{equation} and
\begin{equation}
    S_{DQ3}=\hbar \mathcal{B}_{\alpha\beta,\gamma}\int \mathfrak{e}^\alpha\wedge\mathfrak{e}^{\beta}\wedge d\mathfrak{e}^{\gamma},
\end{equation} with coefficients
 \begin{equation}
 \mathcal{B}_{\alpha\beta}=\frac{1}{4(2\pi)^3}
 \epsilon^{\alpha\beta\sigma}\int d^3 k\  k_{\delta}\mathcal{F}^{\sigma\delta}  
 \label{Eq:BetaCoeff}
\end{equation}
and 
\begin{equation}
\mathcal{B}_{\alpha\beta,\gamma}=\frac{1}{6(2\pi)^3}
\epsilon^{\alpha\beta\sigma}\int d^3 k\  k_{\gamma}k_{\delta}\mathcal{F}^{\sigma\delta}.
\label{Eq:babc}
\end{equation} 
where $\mathcal{F}^{\mu\nu}$ is the Berry curvature of the $k_\mu k_\nu$-plane.  These forms of the coefficients capture scenarios with more complicated nodal line geometries. Indeed, as previously shown in Ref.~\onlinecite{ramamurthylinenode} the coefficient $\mathcal{B}_{\alpha\beta}$ is determined by the line nodes that have non-vanishing area when projected into the $\alpha\beta$-plane. Additionally, for nodal line semimetals with $\mathcal{T}\mathcal{I}$ symmetry the coefficient is proportional to the charge polarization in the direction normal to the $\alpha\beta$-plane~\cite{ramamurthylinenode}. We can see this explicitly by integrating Eq.~\ref{Eq:BetaCoeff} by parts with the same caveats mentioned in Sec.~\ref{sec:DDSderivation} surrounding Eq.~\ref{eqn:2D_pol_bd+Wy}.

Analogously, the coefficient $\mathcal{B}_{\alpha\beta,\gamma}$ can represent a kind of ``momentum"-polarization where the polarization is again normal to the $\alpha\beta$-plane, and the charge that is polarized is the momentum along the $\gamma$-direction. We can see this heuristically by integrating by parts using the derivatives in the $\mathcal{F}^{\sigma\delta}$ to find 
\begin{equation}
\mathcal{B}_{\alpha\beta,\gamma}\sim -\frac{1}{2(2\pi)^3}
\int d^3 k\left(\epsilon^{\alpha\beta\sigma}  k_{\gamma}\mathcal{A}^{\sigma}-\epsilon^{\alpha\beta\gamma}k_i \mathcal{A}^i\right)\label{eq:BabgMomPol}
\end{equation} where we have used the $\sim$ symbol to indicate that there are boundary terms we have dropped that can be important if the line nodes span the Brillouin zone. We can see from this form that the coefficient for the case when $\alpha,\beta,\gamma$ are not all different, e.g. $\mathcal{B}_{xz,x}$, is proportional to the polarization in the $y$-direction (i.e. normal to the $xz$-plane) weighted by the momentum in the $x$-direction.

We note that for $\mathcal{B}_{\alpha\beta}$ to be well-defined, the Chern number in each plane must vanish. In addition to this constraint,  $\mathcal{B}_{\alpha\beta}=0$ is a necessary constraint for $\mathcal{B}_{\alpha\beta,\gamma}$ to be well defined. These hierarchical requirements are analogous to the usual requirements for the ordinary (magnetic) dipole and (magnetic) quadrupole moments of the electromagnetic field  to be independent of the choice of origin. Here the role of the magnetic field distribution is being played by $\mathcal{F}^{\sigma\rho}(k)$, and, for example, the constraint on the vanishing Chern number eliminates the possibility of magnetic monopoles (i.e., Weyl points).

\subsection{Effective responses of 4D semimetals}
Our next goal is to determine the coefficients for the response actions of 3D Weyl point-node semimetals. However, because the Weyl nodes in 3D exhibit an anomaly, the responses are subtle to calculate intrinsically in 3D. Instead, to accomplish our goal we will first carry out more straightforward calculations of the responses of 4D semimetals and then return to 3D either by considering the boundary of a 4D system, or by compactifying and shrinking one dimension of the bulk. Hence, as a step toward 3D semimetals, in this subsection we provide the derivation for effective response actions of semimetals in 4D. 

The first action we consider is of the form
\begin{equation}
    S=c_{\alpha}\int \mathfrak{e}^\alpha\wedge dA\wedge dA,
\end{equation} where for our purposes $\alpha=x,y,z,w.$ Collecting all terms in the gradient expansion that have this field content we obtain:
\begin{equation}
    \begin{split}
        S&= \frac{e^2}{\hbar}\int d^5r\ \mathfrak{e}^\alpha_\mu \partial_\nu A_\rho \partial_\sigma A_\tau \\
        &\times \int \frac{d\omega d^4 k}{(2\pi)^5}k_\alpha \Omega^{(5)}_{\mu\nu\rho\sigma\tau}(\omega,k),
    \end{split}
\end{equation} where
\begin{equation}
    \Omega^{(5)}_{\mu\nu\rho\sigma\tau}(\omega,k)= \text{tr}\left(G_0  \frac{\partial G_{0}^{-1}}{\partial k_\mu} \frac{\partial G_{0}}{\partial k_\nu}\frac{\partial G_{0}^{-1}}{\partial k_\rho} \frac{\partial G_{0}}{\partial k_\sigma}\frac{\partial G_{0}^{-1}}{\partial k_\tau} \right),
\end{equation} and $G_0(\omega,k)$ is the single-particle Green function. To determine the coefficient $c_\alpha$ we project this coefficient onto the totally antisymmetric part and then, just as in Eq.~\ref{eqn:app_chern}, we can carry out the frequency integral~\cite{Zhong2012} to obtain the simpler expression
\begin{equation}
    \begin{split}
        &\int \frac{d\omega d^4 k}{2\pi} \frac{\varepsilon_{\mu\nu\rho\sigma\tau}}{5!} k_\alpha\Omega^{(5)}_{\mu\nu\rho\sigma\tau}(\omega,k)\\
        &=\frac{1}{16}\int_{BZ} d^4\textbf{k}\ k_\alpha \varepsilon_{ijkl} \mathcal{F}^{ij}\mathcal{F}^{kl}.
    \end{split}
\end{equation} Hence, the response coefficient takes the form
\begin{equation}
    \begin{split}
        c_\alpha=\frac{e^2}{\hbar}\frac{1}{16(2\pi)^4}\int_{BZ}d^4k\ k_\alpha\varepsilon_{ijkl} \mathcal{F}^{ij}\mathcal{F}^{kl}=\frac{e^2\mathcal{P}_\alpha}{16\pi^2\hbar},
    \end{split}
\end{equation} where we introduced
\begin{equation}
    \mathcal{P}_\alpha=\frac{1}{16\pi^2}\int_{BZ}d^4\textbf{k}\ k_\alpha\varepsilon_{ijkl} \mathcal{F}^{ij}\mathcal{F}^{kl}.\label{eqn:4dDiracDipole}
\end{equation}

As we see from this calculation, similar to 2D, the 4D response theories can be characterized by the distribution of the quantity $\varepsilon_{ijkl}\mathcal{F}^{ij}\mathcal{F}^{kl}$ across the 4D Brillouin zone. For our focus, let us consider the case where the 4D system is a semimetal with a set of isolated Dirac points (linearly dispersing band touchings where four bands meet).
Without symmetry, these Dirac points are locally unstable in momentum space to the opening of a gap. If we open up an infinitesimally small energy gap, the quantity $\varepsilon_{ijkl}\mathcal{F}^{ij}\mathcal{F}^{kl}$ becomes well-defined across the entire BZ and its distribution takes the following form in the massless limit:
\begin{equation}
\varepsilon_{ijkl}\mathcal{F}^{ij}\mathcal{F}^{kl}=\sum_{a=1}^{N_D}16\pi^2\chi_a\delta(\textbf{k}-\textbf{k}_a).
\end{equation} If we substitute this into Eq.~\ref{eqn:4dDiracDipole} then  we immediately see that $\mathcal{P}_{\alpha}$ becomes the momentum space dipole of the set of 4D Dirac nodes. Let us also comment that if we integrate Eq.~\ref{eqn:4dDiracDipole} by parts we see that $\mathcal{P}_\alpha$ can also be interpreted as a set of magneto-electric polarizabilities~\cite{qi2008,essin2009}. Just as in the case of the polarization of a 2D Dirac semimetal, the integration by parts will generate a boundary term that captures the magneto-electric polarizability coming from the 3D boundaries of the 4D BZ. Hence, the connection between the total magneto-electric polarizability and the mixed translation-electromagnetic response is only exact in the symmetric limit when the boundary term is quantized.

In summary, a 4D response of a system characterized by a dipolar distribution of the $\varepsilon_{ijkl} \mathcal{F}^{ij}\mathcal{F}^{kl}$ quantity reads: 
\begin{equation}
    S=\frac{e^2\mathcal{P}_\alpha}{16\pi^2\hbar}\int \mathfrak{e}^\alpha\wedge dA\wedge dA.\label{eqn:responseEdAdA}
\end{equation}
Similar to 2D, if the dipolar response vanishes we can obtain a momentum quadrupole response coefficient for the action:
\begin{equation}
    S=\frac{e\mathcal{Q}_{\alpha\beta}}{16\pi^2}\int \mathfrak{e}^\alpha\wedge d\mathfrak{e}^\beta\wedge dA,\label{eqn:responseEdEdA}
\end{equation}\noindent where $\mathcal{Q}_{\alpha\beta}$ is a symmetric matrix determined by the momentum space quadrupole moment of the 4D Dirac nodes.
Finally, if both the dipolar and quadrupolar responses vanish we can consider an octupolar distribution that will give the response coefficient for the action:
\begin{equation}
    S=\frac{\hbar\mathcal{O}_{\alpha\beta\gamma}}{48\pi^2}\int \mathfrak{e}^\alpha\wedge d\mathfrak{e}^\beta\wedge d\mathfrak{e}^\gamma,\label{eqn:responseEdEdE}
\end{equation} where $\mathcal{O}_{\alpha\beta\gamma}$ is determined by the momentum space octupole moment of the 4D Dirac nodes. We will leave the discussion of octupolar configurations of Dirac and Weyl nodes to future work. We also mention that, similar to 2D, for these responses to be independent of the choice of BZ origin we require that the second Chern number of the 4D system vanishes.  Alternatively, if the second Chern number is non-vanishing, then the boundary of the system will contain a non-vanishing chirality of Weyl nodes. As such, the anomalous charge response of the chiral boundary will not allow us to uniquely determine the momentum response on the boundary. 

Before moving on to 3D, let us briefly present some physical intuition about the response in Eq.~\ref{eqn:responseEdAdA}. Consider a 4D time-reversal and inversion invariant system having two Dirac nodes separated in the $k_z$-direction. To simplify the discussion, let us also assume the system has mirror symmetry $M_z.$  The assumed symmetries imply that each fixed-$k_z$ volume can be treated as an independent 3D insulator having 3D inversion symmetry, and hence the magneto-electric polarizability of these 3D insulator subspaces is quantized~\cite{qi2008,hughesinversion2011,turner2012inversion}. Now, if we sweep through $k_z$ then each bulk 4D Dirac point crossing changes the magneto-electric polarizability of the fixed-$k_z$ volume by a half-integer (i.e., changes the related axion angle by $\pi$)~\cite{qi2008}. Since the magneto-electric polarizability jumps between its quantized values as we pass through the two bulk Dirac nodes, the $k_z$ Brillouin zone splits into two intervals: (i) an interval with a vanishing magneto-electric polarizability, and (ii) an interval with a non-vanishing quantized magneto-electric polarizability.  Indeed, we could have anticipated this result from the form of the action Eq.~\ref{eqn:responseEdAdA} when $\alpha=z,$ i.e., the action represents stacks of 3D topological insulators that each have a non-vanishing magneto-electric polarizability.

\subsection{Effective responses of 3D semimetals}
\label{sec:3dsemimetalresponses}
From this discussion we see that, in the presence of symmetry, the 4D bulk Dirac node dipole moment determines the magneto-electric polarizability of these 4D topological semimetals via Eq.~\ref{eqn:responseEdAdA}. We want to connect this result to 3D semimetals in two ways. First, we will consider the 3D boundary of the 4D system, and then we will consider the spatial compactification of one spatial dimension. 

Let us begin by considering the boundary response action from Eq.~\ref{eqn:responseEdAdA}. For the model system described at the end of the previous subsection we know the system has a $k_z$-dependent magneto-electric polarizability. Consider a boundary in the fourth spatial direction $w.$ Since the magneto-electric polarizability is changing from inside to outside of the boundary, the boundary itself will have a non-vanishing Hall conductivity. For our example system, each fixed-$k_z$ slice of this boundary will have a Hall conductivity $\sigma_{xy},$ which is quantized, but possibly vanishing. Additionally, since the bulk 4D Dirac nodes are separated in the $k_z$ direction, they will project to gapless points in the 3D surface BZ (on surfaces that have at least one direction perpendicular to the $z$-direction) where the Hall conductivity discretely jumps by $\Delta \sigma_{xy}=\pm\tfrac{e^2}{2h}.$  

From this phenomenology, i.e., discrete Hall conductivity jumps as we sweep through $k_z$ we expect that the boundary response of Eq.~\ref{eqn:responseEdAdA} captures the same response as a Weyl semimetal that has a non-vanishing momentum space dipole moment of the Weyl nodes in the $z$-direction. Indeed the generic boundary contribution from Eq.~\ref{eqn:responseEdAdA} has the form:
\begin{equation}
    S_{WD}=\frac{e^2\mathcal{P}_\alpha}{8\pi^2\hbar}\int \mathfrak{e}^\alpha\wedge dA\wedge A\label{eqn:responseWD}
\end{equation}\noindent 
which was proposed by Ref.~\onlinecite{zyuzin2012} to describe the response of Weyl semimetals, though in the more conventional form using an axion field and without the translation gauge field. Here $\mathcal{P}_\alpha,$ $\alpha=x,y,z$ is the momentum dipole of the Weyl nodes in the $\alpha$-th direction. This action is represented as (ix) in Fig.~\ref{fig:stiresponse}(c). We note that the coefficient in Eq.~\ref{eqn:responseWD} is twice as large as the actual boundary term derived from Eq.~\ref{eqn:responseEdAdA}. This is because when $k_i$ passes through a single Weyl point we have $\epsilon_{ijk}\Delta \sigma_{jk}=\pm \tfrac{e^2}{h},$ where the surface the response of the 4D system has jumps of half the size. This is analogous to the fact that a 1D metal has an integer jump in the filling as we pass through a Fermi point, whereas the surface of a 2D Dirac semimetal has a boundary ``filling" that jumps by a half-integer as we pass through a gapless point in the surface BZ. 

We can repeat this analysis for Eq.~\ref{eqn:responseEdEdA}. The coefficient of this term is proportional to the momentum space quadrupole moment of the nodal points. Unfortunately the phenomenology of this term is not as easy to analyze in 4D because it is not generated from a lower dimensional system in a clear way~\footnote{Even though there is a $\mathfrak{e}^\alpha$ wedge product with a lower-dimensional action, it is not transverse to the lower-dimensional action since $\mathcal{Q}_{\alpha\beta}$ is symmetric. For example, there will be terms where, say, $\mathfrak{e}^x$ couples to $d\mathfrak{e}^x$, which cannot be interpreted as a conventional stacked action.}. By analogy with the previous case, the bulk 4D Dirac nodes will project to a quadrupole of 3D Weyl nodes on the surface. We can extract the form of the 3D action we want by taking the boundary term generated from Eq.~\ref{eqn:responseEdEdA}. Then accounting for the factor of two as in the previous case, we arrive at:
\begin{equation}
    S_{WQ}=\frac{e\mathcal{Q}_{\alpha\beta}}{8\pi^2}\int \mathfrak{e}^\alpha\wedge d\mathfrak{e}^\beta\wedge A.
    \label{Eq:SWQResp}
\end{equation} 
(Note that since $\mathcal{Q}_{\alpha\beta}$ is symmetric, the related contribution of the form $e\mathcal{Q}_{\alpha\beta}/8\pi^2\int \mathfrak{e}^\alpha\wedge \mathfrak{e}^\beta\wedge dA$ vanishes). This action is the same as that shown in Fig.~\ref{fig:ptiresponse}(a). It produces a mixed crystalline-electromagnetic response and represents a rank-2 vector charge response when certain mirror symmetries are preserved~\cite{dbh2021}. Its response coefficient is determined by the momentum space quadrupole moment of the Weyl nodes.

Finally, we come to the action (x) in Fig.~\ref{fig:stiresponse}(c). Let us briefly sketch some salient features of this response, while we leave a detailed discussion to future work.  We can arrive at this action using a formal compactification of the action in Eq.~\ref{eqn:responseEdAdA}~\cite{qi2008}. First we can integrate that action by parts to arrive at 
\begin{equation*}
    \frac{e^2\mathcal{P}_{\alpha}}{16\pi^2\hbar}\int A\wedge d\mathfrak{e}^{\alpha}\wedge dA,
\end{equation*} where we have ignored the boundary term.
We now want to dimensionally reduce the fourth spatial direction $w$, which we accomplish by choosing periodic boundary conditions in $w$ and letting the size of the system in this direction shrink toward zero. In this limit any derivatives with respect to $w$ are (formally in our case) dropped~\footnote{Alternatively we can assume the fields $A_w, e^\alpha_w$ are locked to their ground state values and thus have vanishing derivatives in all directions.}. The resulting non-vanishing contribution is
\begin{equation*}
    \frac{e^2\mathcal{P}_{\alpha}}{8\pi^2\hbar}\oint A_w dw\int  d\mathfrak{e}^{\alpha}\wedge dA,
\end{equation*} where the integral and exterior derivative in the second factor are over only the remaining four spacetime dimensions.
We can now make the definition
\begin{equation}
    \Theta \equiv  2\pi\frac{e}{h}\int A_w dw,
\end{equation} 
 to arrive at action (x) from Fig.~\ref{fig:stiresponse}(c):
\begin{equation}
    \frac{e \mathcal{P}_{\alpha}}{8\pi^2}\int \Theta d\mathfrak{e}^{\alpha}\wedge dA.\label{eqn:responseThetadEdA}
\end{equation}

To illustrate some of the phenomenology of this action let us assume that $\mathcal{P}_z\neq 0.$ Additionally let us assume that we maintain time-reversal and inversion symmetry. As such, $\Theta= 0, \pi$. To begin, we see that the action in Eq.~\ref{eqn:responseThetadEdA} is a total derivative if $\Theta$ and $\mathcal{P}_{\alpha}$ are space-time independent. The resulting pure boundary term is just proportional to the response of a 2D weak TI (or 2D Dirac semimetal), i.e., Eq.~\ref{eqn:responseEdA}. Depending on the symmetry of the surfaces, this implies that we expect the surface to be gapped except for possibly isolated Dirac points. Since the boundary terms appear as $\mathfrak{e}^z\wedge dA$ we expect that surfaces normal to $\hat{x}$ ($\hat{y}$) will harbor a $y$-polarization ($x$-polarization), i.e., the polarization is tangent to the surface. 

Importantly, the sign of the polarization depends on the interpolation of $\Theta$ between its non-trivial bulk value of $\Theta=\pi$ and the trivial vacuum value $\Theta=0$ outside the system. For neighboring surfaces where the effective sign of the polarization changes we anticipate hinge charges where surfaces intersect since the polarizations are converging or diverging from the hinges. Thus, the response of this system is similar to a stack of 2D planes of quadrupole moment having component $q_{xy}\neq 0.$ In this scenario, coupled quadrupole planes could lead to either a higher order weak topological insulator having a quadrupole moment, or a higher order topological semimetal with boundary (and possibly bulk) Dirac nodes~\cite{linHOTSM,wieder2020strong}. To make further progress it would be advantageous to have a microscopic derivation of the coefficient in Eq.~\ref{eqn:responseThetadEdA} intrinsically in 3D. Hence, we will leave further discussion of this action to future work.

\section{Explicit Response Calculations for Lattice Models}
\label{sec:models}

Now that we have completed the derivations of the actions in Figs.~\ref{fig:stiresponse}(c) and \ref{fig:ptiresponse}, we will provide a series of model examples that manifest these responses. Using these models we can numerically calculate the various charge and momentum responses to electromagnetic and translation gauge fields, providing an independent verification of the coefficients derived in the previous section. Some of the models and responses we discuss below have appeared elsewhere in the literature, while others are have not.  We will carry out this analysis in the same order as the previous section, i.e., point-node Dirac semimetals in 2D, nodal line semimetals in 3D, and then point-node Weyl semimetals in 3D. Calculations for 1D systems were carried out analytically in Sec.~\ref{sec:1dsemimetalderivation}, and additional discussion can be found in App.~\ref{app:responsesin1d}.

\subsection{2D Dirac node dipole semimetal and insulator}
\label{sec:2D_Dirac_dipole}
We begin with the time-reversal invariant 2D systems discussed in Sec.~\ref{Sec:SemimetalResponse} that exhibit a mixed crystalline-electromagnetic response.  Since $\mathcal{T}$ is preserved, the usual Chern-Simons, Hall-effect response of the electromagnetic field vanishes. Instead, the response action derived in the Sec.~\ref{sec:2D_Dirac_dipole} takes the form of a mutual Chern-Simons term~\cite{chongwangvishwanath}:
\begin{equation}
    S[\mathfrak{e}^\lambda_\nu,A_\mu]=\frac{e}{4\pi}\mathcal{P}_\lambda \int\mathfrak{e}^\lambda\wedge dA.
   \label{eqn:2D_Dirac_crystalline_resp}
\end{equation}
Unlike the purely electromagnetic polarization response action considered in Ref.~\onlinecite{RamamurthyPatterns}, this formulation of the low-energy response theory also includes bulk electromagnetic responses to the translation gauge fields. For example, by taking a functional derivative with respect to $A_\mu$ we have
\begin{equation}
\begin{split}
    &\rho=-\frac{e}{4\pi}\mathcal{P}_\lambda\varepsilon^{ij}\partial_i\mathfrak{e}^\lambda_{j},\\
    &j^x=\frac{e}{4\pi}\mathcal{P}_\lambda(\partial_t \mathfrak{e}^\lambda_y - \partial_y \mathfrak{e}^\lambda_t),\\
    &j^y=-\frac{e}{4\pi}\mathcal{P}_\lambda(\partial_t \mathfrak{e}^\lambda_x - \partial_x \mathfrak{e}^\lambda_t).
\end{split} 
\label{eqn:CrystalResp}
\end{equation}
We see that the first equation predicts an electric charge density localized on a dislocation in the bulk of the lattice, which is exactly the phenomenology  we expect for a weak topological insulator~\cite{ran2009} or a 2D Dirac semimetal.  The action (\ref{eqn:2D_Dirac_crystalline_resp}) also predicts a bulk momentum response to the electromagnetic field when varied with respect to $\mathfrak{e}^\mu$,
\begin{equation}
\begin{split}
    &\mathcal{J}^t_\lambda=-\frac{e}{4\pi}\mathcal{P}_\lambda B_z,\\
    &\mathcal{J}^i_\lambda=-\frac{e}{4\pi}\mathcal{P}_\lambda\varepsilon^{ij}E_j,
\end{split}\label{eq:DDmomresponse}
\end{equation}
where $E_i$ and $B_i$ are the components of electric and magnetic fields respectively. In the inversion-symmetric limit and in the absence of lattice defects and deformations, for which the crystalline gauge fields reduce to  $\mathfrak{e}^\lambda_\mu=\delta^\lambda_\mu,$ Eq.~(\ref{eqn:CrystalResp}) simply reproduces the boundary charge and current responses of an ordinary 2D Dirac semimetal or weak topological insulator, which harbors a non-vanishing electric polarization. However, as we mentioned in Sec.~\ref{sec:DDSderivation}, and comment further on below, we do not expect the coefficient of this action to match the electric polarization when inversion is strongly broken.

\begin{figure}
    \centering
    \includegraphics[width=0.47\textwidth]{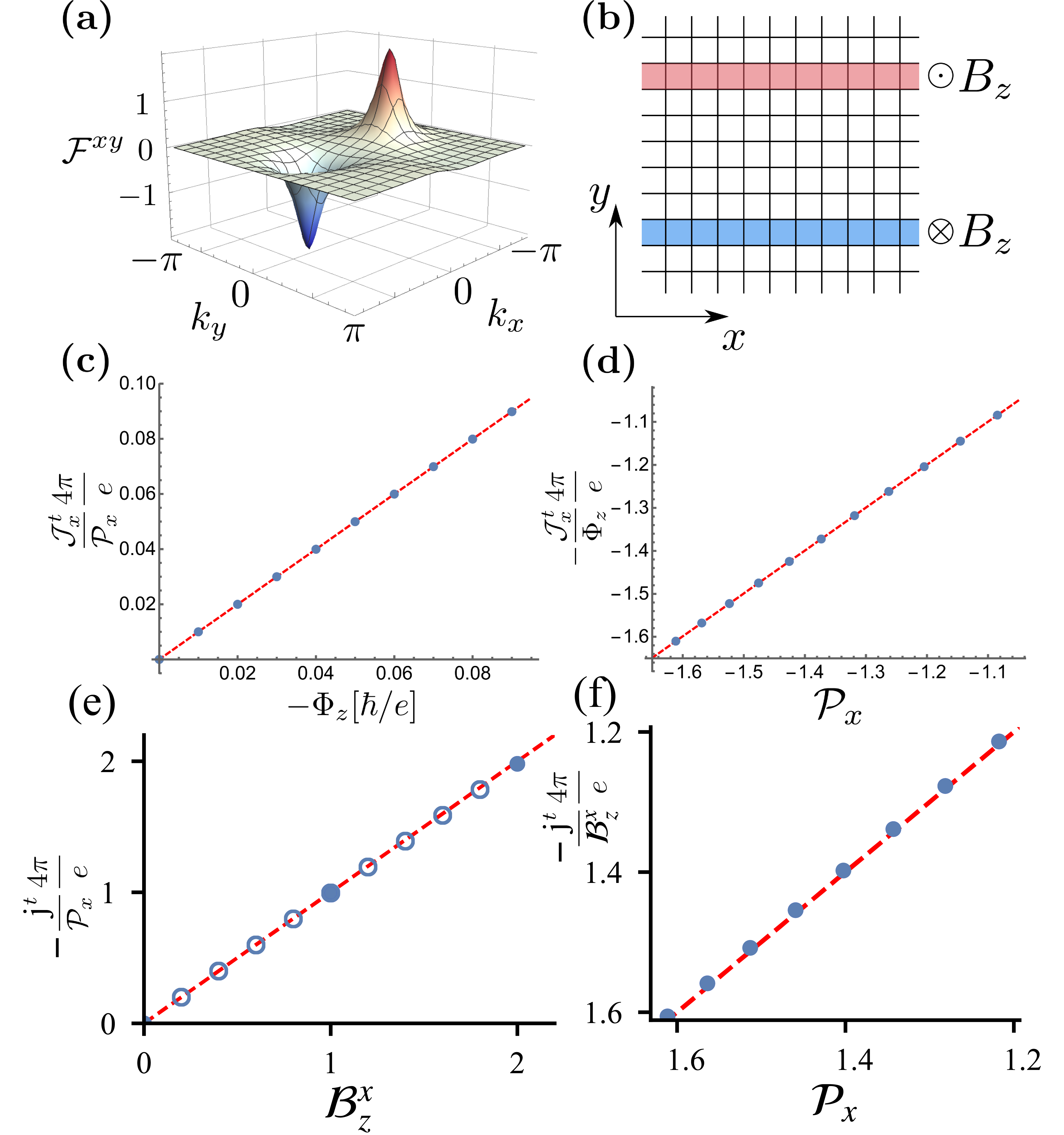}
    \caption{(a) Plot of the Berry curvature across the 2D Brillouin zone for the Dirac node dipole semimetal model (\ref{eqn:2d_ins_dipole}) for $m=1.1$ with an added inversion-breaking perturbation with $V_\mathcal{I}=-0.5$. We use this model to probe the $k_x$ momentum density response. For that we consider a completely periodic system and insert the magnetic flux $\Phi_z$ thorough two lines of plaquettes such that the translational symmetry along the $\hat{x}$-direction is preserved, as shown in panel (b). (c) shows the $k_x$ momentum density localized around one line of plaquettes penetrated by the magnetic field $B_z$ as a function of magnetic flux. (d) shows the $k_x$ momentum density as a function of Berry curvature dipole moment $\mathcal{P}_x$ defined in Eq.~(\ref{eqn:coeffDD})  which we tune in our model by varying the parameter $m$ between $m=1.0$ and $m=1.5$. In (e) and (f) we show analogous calculations for the charge density response to a translation flux with Burgers vector in the $x$-direction as a function of (e) translation flux at fixed Berry curvature dipole, and (f) Berry curvature dipole at fixed translation flux. The open circles in (e) represent Burgers' vector choices that are not integer multiples of a lattice constant. The red dashed lines in (c)-(f) are guides to the eye indicating a slope of 1.}
    \label{fig:2D_berry_dipole}
\end{figure}

While the electric polarization/magnetization responses of Dirac semimetals were discussed in detail in Ref. \onlinecite{RamamurthyPatterns}, the momentum responses in Eq.~\ref{eq:DDmomresponse}, and the charge responses to translation fluxes (i.e., dislocations) in Eq.~\ref{eqn:CrystalResp}  are less familiar. Thus, we will explicitly calculate these responses using a minimal tight-binding model. For simplicity, we employ a two-band Bloch Hamiltonian that can model both 2D Dirac semimetals and weak topological insulators:
\begin{equation}
\begin{split}
    H(\textbf{k})&=V_\mathcal{I}\sigma^x + \sin (k_y a_y) \sigma^y \\
    &+(m-\cos (k_x a_x) - \cos (k_y a_y))\sigma^z.
    \label{eqn:2d_ins_dipole}
\end{split}
\end{equation}
When $V_\mathcal{I} = 0$, $H$ has both inversion symmetry,  $\mathcal{I}=\sigma^z$, and (spinless) time-reversal symmetry, $\mathcal{T}=K$. In this symmetric regime, $m$ can be chosen to produce a semimetal with Dirac points located at, for example, $(k_x,k_y)=(\pm\pi/(2a_x),0),$ when $m=1$. In the semimetal phase, turning on $V_\mathcal{I}\sigma^x$, which breaks inversion while preserving $\mathcal{T}$, generates a mass term that opens a gap at the Dirac points. The signs of the Berry curvature localized near the two now-gapped Dirac points are opposite, as shown in Fig.~\ref{fig:2D_berry_dipole}(a), with the sign at a particular point determined by the sign of the perturbation $V_\mathcal{I}$. Hence the total Berry curvature of the occupied band integrated over the entire BZ, equivalent to the Chern number, is zero, and hence the Berry curvature dipole is well-defined.

To confirm our analytic calculations of the response coefficients we will first calculate the momentum density localized around an out-of-plane magnetic flux $\Phi_z$ using the tight-binding model Eq. (\ref{eqn:2d_ins_dipole}). 
In order to determine the $k_x$ momentum density in the lattice model, we must introduce magnetic flux in a fashion that preserves translation symmetry in the $\hat{x}$-direction. We show the configuration that we employ in Fig.~\ref{fig:2D_berry_dipole}(b). This configuration keeps the crystal momentum $k_x$ as a good quantum number and allows us to compute the value of $\mathcal{J}^t_x$ as the probability density of the occupied single particle states weighted by their momentum $\hbar k_x$. The results of the numerical calculations are presented in Fig.~\ref{fig:2D_berry_dipole}(c,d), where we study how the excess $k_x$ momentum density bound to magnetic flux behaves as a function of both the magnetic flux $\Phi_z$ at fixed Berry curvature dipole $\mathcal{P}_x,$ and and as a function of $\mathcal{P}_x$ at fixed $\Phi_x.$  Our numerical results match our analytic calculations precisely. 

We can interpret this result by noting that the momentum current in Eq.~\ref{eq:DDmomresponse} can be obtained in the semiclassical limit by considering the momentum current carried by electron wavepackets subject to an anomalous velocity~\cite{chang1995berry,chang1996berry}. The equation of motion of an electron wavepacket with momentum $\mathbf{k}$ formed from a single band is
\begin{equation}
    v^i(\mathbf{k}) = \frac{\partial\mathcal{E}}{\hbar\partial k_i} + \frac{e}{\hbar}\epsilon^{ij}E_j\mathcal{F}^{xy}(\mathbf{k}),
\end{equation}
where $v^i(\mathbf{k})$ is the wavepacket velocity, $\mathcal{E}(\mathbf{k})$ is the energy spectrum of the band, $E_j$ is the electric field, and $\frac{e}{\hbar}\epsilon^{ij}E_j\mathcal{F}^{xy}(\mathbf{k})$ is the anomalous velocity. The momentum current of the occupied states is obtained by adding up the contributions $\hbar k_\mu v^i(\mathbf{k})$ in the BZ and contains a term arising from the anomalous velocity given by
\begin{equation}
    \begin{aligned}
    \mathcal{J}^i_\lambda &= -\frac{e}{(2\pi)^2}\epsilon^{ij}E_j\int dk_x dk_y \, k_\lambda \mathcal{F}^{xy}(k_x, k_y) \\
    &= -\frac{e}{4\pi}\mathcal{P}_\lambda\epsilon^{ij}E_j.
    \end{aligned}
\end{equation}

We can also numerically probe our response equations by studying the charge response to the deformation of the lattice.  To do so, we introduce a translation flux to rows of plaquettes located near $y=N_y/4$ and $y=3N_y/4$, analogous to the magnetic flux configuration we just considered. This effectively inserts two rows of dislocations such that if one encircles a plaquette containing translation flux, the Burgers vector is in the $x$-direction. This effectively creates opposite translational magnetic fields $\mathcal{B}^x_z=\partial_{x}\mathfrak{e}_y^x-\partial_y \mathfrak{e}_x^x$ penetrating the two rows of plaquettes. Again, we choose this geometry since it is compatible with translation symmetry in the $x$-direction. In our lattice model we insert the translation flux by explicitly adding generalized Peierls' factors that are $k_x$-dependent, i.e., $\exp\left(i k_x\int \mathfrak{e}^x_i dx^i\right)$ such that the colored regions in Fig.~\ref{fig:2D_berry_dipole}(b) contain non-vanishing translation flux. The resulting electron charge density localized on the translation magnetic flux has a dependence on both the $\mathcal{B}^x_z$ field strength and the Berry curvature dipole moment $\mathcal{P}_x$ as shown in Fig.~\ref{fig:2D_berry_dipole}(e),(f). This again matches the expectation from our analytic response equations.  

We emphasize that the effective action (\ref{eqn:CrystalResp}) describes the \emph{mutual} \emph{bulk} response between the electromagnetic and the momentum currents in semimetallic and insulating systems with vanishing Chern number. We showed in Sec.~\ref{sec:DDSderivation} that one must be careful when comparing this response to the charge polarization. In particular, our numerics show that, even in the presence of significant inversion-breaking, the bulk momentum density response to a magnetic flux tracks the value of the coefficient $c_{\alpha}$ from Eq.~\ref{eqn:coeffDD} as demonstrated in Fig.~\ref{fig:2D_berry_dipole} (d). In contrast, as shown in Sec.~\ref{sec:DDSderivation}, the expression for the electric polarization, Eq.~\ref{eqn:2D_pol_bd+Wy}, contains an additional term that is proportional to the value of a Wilson loop along the boundary of the BZ. This value is not quantized when inversion symmetry is broken, and, for large values of $V_{\mathcal{I}}$, this contribution becomes significant enough that the polarization response clearly deviates from the result one would expect from a naive interpretation of Eq.~\ref{eqn:CrystalResp}. However, the mutual response between the electromagnetic and translation gauge fields described by this action remains valid. This subtlety is not the focus of our current article, so we leave further discussions to future work.

\subsection{2D Dirac quadrupole semimetal}
\label{sec:2D_Dirac_qp}

Next, we consider the class of 2D semimetallic phases characterized by the quadrupole moment of the Berry curvature introduced in Section~\ref{Sec:2dQuadDirac}. We know from Section~\ref{Sec:2dQuadDirac} that the low-energy effective response action for this system takes the form:
\begin{equation}
    S=\frac{\hbar}{8\pi}\mathcal{Q}_{\alpha\beta}\int \mathfrak{e}^\alpha \wedge d\mathfrak{e}^\beta.
    \label{eqn:2D_Dirac_quadr_cryst_resp}
\end{equation}
This action generates a momentum current response 
\begin{equation}
    \mathcal{J}^\mu_\alpha = -\frac{\hbar}{4\pi} \mathcal{Q}_{\alpha\beta}\varepsilon^{\mu\nu\sigma}\partial_{\nu}\mathfrak{e}^\beta_\sigma
    \label{eqn:2D_quad_curr_response}
\end{equation}  These currents describe both a bulk momentum polarization (e.g., yielding momentum on the boundary where $\mathcal{Q}_{\alpha\beta}$ changes), and a bulk energy-momentum response to translation gauge fields. We note that this response is exactly analogous to that of the Dirac node dipole semimetal discussed above if we replace the electromagnetic field with a translation gauge field.

To illustrate and explicitly confirm the responses numerically we use the following 2-band square lattice Bloch Hamiltonian with next-nearest-neighbor hopping terms:
\begin{equation}
\begin{split}
    H(\textbf{k})&=V_\mathcal{T} \sigma^x + \sin (k_x a) \sin (k_y a)\sigma^y\\
    &+ (m-\cos( k_x a) -\cos (k_y a))\sigma^z.
    \label{eqn:2D_BC_quad_ham}
\end{split}
\end{equation}
This model has an inversion symmetry (i.e., $C_2^z$ symmetry) that is realized trivially on-site with $\mathcal{I}=\mathbb{I}$, mirror symmetry along the $k_x = k_y$ axis, and, when $V_\mathcal{T}=0$, time-reversal symmetry $\mathcal{T}=\sigma^z K$. This model can be tuned to a semimetal phase as well, for example, setting $m=1$ we find four gapless Dirac points located at $(k_x,k_y)=(\pm \pi/2a,0)$ and $(k_x,k_y)=(0,\pm\pi/2a)$. 

\begin{figure*}
    \centering
    \includegraphics[width=0.92\textwidth]{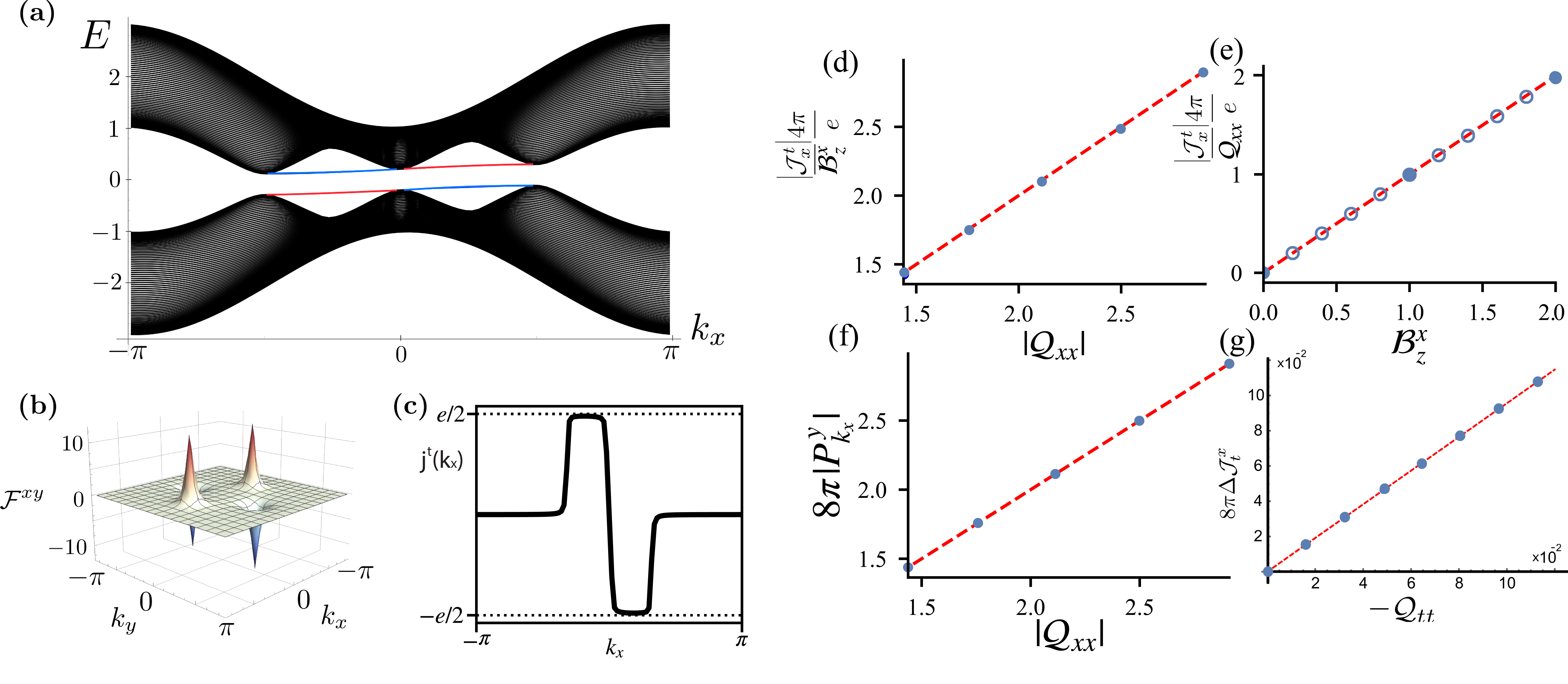}
    \caption{(a) Spectrum of the 2D Dirac node quadrupole semimetal (\ref{eqn:2D_BC_quad_ham}) in a ribbon geometry ($y$-direction open, $x$-direction periodic) for $m=1$, the $\mathcal{T}$-breaking perturbation set to $V_\mathcal{T}=-0.2,$ and the energy tilt in Eq.~\ref{eq:quadtilt}   $\epsilon=0.1$. At half filling, the ground state of the model is momentum-polarized: occupied states localized near $y=1$, which are indicated by the blue color, carry a positive value of the $k_x$ momentum, while the occupied states near $y=N_y$ have a negative value of $k_x$. (b) Berry curvature distribution across the Brillouin zone for a small gapping perturbation $V_\mathcal{T}=-0.2.$ (c) The boundary charge distribution as a function of momentum. (d), (e) $k_x$ momentum bound to a row of dislocations (c.f. Fig.~\ref{fig:2D_berry_dipole}(b)) as a function of $\mathcal{Q}_{xx}$ at fixed $\mathcal{B}_z^x$ in (d) and as a function of $\mathcal{B}_z^x$ at fixed $\mathcal{Q}_{xx}$ in (e). (f) Plot of momentum polarization $P^y_{k_x}$ obtained from computing $k_x$-momentum bound to an edge normal to $\hat{y}.$ (g) As a consequence of non-zero $\epsilon$ we see that the velocities of single-particle states in (a)localized on opposite edges have the same sign, while the energy and $k_x$ momentum charges are exact opposite. This leads to boundary energy currents as illustrated in panel (g) as a function of $\mathcal{Q}_{tt}.$ }
    \label{fig:2D_ins_quadrupole}
\end{figure*}

To confirm the response action is correct, we first need to calculate the Dirac-node quadrupole moment. To see that the Berry curvature quadrupole moment is well-defined, we first note that the choice of $V_{\mathcal{T}}$ as a mass perturbation forces $\mathcal{P}_\alpha$ to vanish. We also need the Chern number to vanish, which is guaranteed by the mirror symmetry. With these symmetries, the Berry curvature peaks at Dirac points that are related by inversion symmetry have the same sign, while the peaks related by mirror symmetry carry opposite signs, resulting in a quadrupolar distribution of the Berry curvature, as in Fig.~\ref{fig:2D_ins_quadrupole}(b). Since the Chern number and $\mathcal{P}_\alpha$ both vanish, the quadrupolar distribution is well-defined and signals the presence of a well-defined elastic response in this model (see also Ref.~\onlinecite{bradlynrao}). The diagonal elements of the Dirac-node quadrupole moment of our model are equal and opposite, $\mathcal{Q}_{xx}=-\mathcal{Q}_{yy},$ and the off-diagonal elements are zero. Since the sign of the Berry curvature flux for 2D Dirac points with $\mathcal{TI}$-symmetry is ambiguous, we once again treat our system in the insulating regime with non-zero $V_\mathcal{T}$ first and then recover the semimetallic case by taking the limit $V_\mathcal{T}\to 0$.

Using this model, let us first focus on the momentum polarization response and highlight the difference with the 2D Dirac node dipole semimetal case from Section~\ref{sec:2D_Dirac_dipole}. If the bulk has a momentum polarization we expect translation-symmetric edges to have a bound momentum density. We will first make a general argument for the existence of the boundary momentum and then confirm the results numerically for our model. Let us suppose our system has a boundary normal to the $y$-direction. We expect such a boundary will carry $k_x$ momentum if $\mathcal{Q}_{xx}\neq 0.$ To show this, let us make a gauge transformation on the fields in Eq.~\ref{eqn:2D_Dirac_quadr_cryst_resp}: $\mathfrak{e}_\mu^a\to \mathfrak{e}_\mu^a +\partial_\mu \lambda^a$ for some vector function $\lambda^a.$ Since there is a boundary, the response action is not gauge invariant and we find the variation $\delta_{\lambda} S= -\frac{\hbar\mathcal{Q}_{ab}}{8\pi}\lambda^a (\partial_0 \mathfrak{e}_x^b-\partial_x \mathfrak{e}_0^b).$ Our system has no translation-twisting of the boundaries, i.e., $\mathfrak{e}_{x}^{y}=\mathfrak{e}_{y}^{x}=0$, so we find the variation reduces to $\delta_{\lambda} S=-\frac{\hbar\mathcal{Q}_{xx}}{8\pi}\lambda^x (\partial_0 \mathfrak{e}_x^x-\partial_x \mathfrak{e}_0^x).$ This variation can be canceled by adding an action of the form Eq.~\ref{eqn:1D_Q_resp}. That is, we expect to have 1D degrees of freedom on the boundary that harbor a non-vanishing $k_x$-momentum density captured by an effective 1D  quadrupole moment $\mathcal{Q}_{xx}$ that matches the value of the 2D quadrupole moment. Interestingly, we note that the coefficient of Eq.~\ref{eqn:1D_Q_resp} is twice that of the variation we need to cancel. Hence, the edge of our system has a fractional momentum density, i.e., a 1D system with the same $\mathcal{Q}_{xx}$ would have twice as much momentum. This is analogous to the fractional boundary charge density one finds from the half-quantized electric charge polarization.

We confirm this response numerically by studying the model (\ref{eqn:2D_BC_quad_ham}) on a lattice in a ribbon geometry that is open in the $\hat{y}$-direction and periodic in $\hat{x}$. Figure~\ref{fig:2D_ins_quadrupole} (a) shows the resulting band structure, for which a gap is opened by a non-vanishing $V_\mathcal{T}$ and the occupied states have two symmetrically positioned sets of flat band states: one in an interval having $k_x<0$ and the other in an interval having $k_x>0$. The occupied boundary states with $k_x<0$ (red) are localized near the top ($y=N_y$) boundary, while the occupied boundary states with $k_x>0$ (blue) are localized near the bottom ($y=1$) boundary. At half filling we find that the excess/deficit charge near the boundary depends on $k_x$ as shown in Fig.~\ref{fig:2D_ins_quadrupole}(c). We see that the states at positive and negative $k_x$ are imbalanced, indicating a non-vanishing $k_x$ momentum density on the edge. Indeed, each state between the Dirac nodes contributes an amount to the total edge momentum equal to $k_x$ weighted by a factor of $\pm 1/2$, since the particle density on the edge at each $k_x$ in this range is $\pm 1/2.$ Because states at opposite $k_x$ have opposite excess/deficit probability density, the total sum is non-vanishing and depends on $\mathcal{Q}_{xx}$ as shown in Fig.~\ref{fig:2D_ins_quadrupole}(f). We find that the bulk momentum polarization $P^y_{k_x}=\frac{\hbar \mathcal{Q}_{xx}}{8\pi}$ matches the numerically calculated boundary momentum density, as expected for a generalized surface charge theorem~\footnote{We comment that even though the Chern-Simons term for the translation gauge fields shares some properties with the electromagnetic Chern-Simons term, there is a key distinction: The translation gauge fields have a constant background. This allows the Dirac node quadrupole system to have a static momentum polarization, whereas the electromagnetic Chern Simons term in a Chern insulator would predict generating an electric polarization as one tunes the vector potential.}.
To further probe the response equations, we subject the Dirac node quadrupole semimetal to the same linear array of dislocations employed in the previous subsection (c.f. Fig.~\ref{fig:2D_berry_dipole}(b)). From Eq.~\ref{eqn:2D_quad_curr_response}, we expect to find momentum density localized on dislocations. Since our geometry preserves translation in the $\hat{x}$-direction, we can compute the amount of $k_x$ momentum bound to dislocations, similar to how we computed the amount of charge bound to dislocations in the previous subsection. We show our results in Fig.~\ref{fig:2D_ins_quadrupole}(d)(e) where we first plot momentum density as a function of $\mathcal{Q}_{xx}$ for fixed translation flux $\mathcal{B}_z^x,$ and then plot momentum density as a function of $\mathcal{B}_z^x$ for fixed $\mathcal{Q}_{xx}.$ Both results match the analytic value from the response action.

Finally, let us briefly consider a case when the mixed energy-momentum  quadrupole moment $\mathcal{Q}_{ta}$ is non-vanishing. In this scenario the effective action (\ref{eqn:2D_Dirac_quadr_cryst_resp}) implies the existence of a bulk orbital \emph{momentum magnetization} of 
\begin{equation}
    M^z_{k_\mu}=-\frac{\hbar}{8\pi}\mathcal{Q}_{t\mu},
    \label{Eq:MomMag}
\end{equation} that will manifest as boundary momentum currents, even in equilibrium (note we assume $\mathfrak{e}_t^t=1)$.  
To generate a non-vanishing $\mathcal{Q}_{t\mu}$ in our model (\ref{eqn:2D_BC_quad_ham}), we turn on an additional perturbation
\begin{equation}
    \Delta H(\textbf{k})=\epsilon \sin(k_x) \mathbb{I}_{2\times 2}. \label{eq:quadtilt}
\end{equation}
When $m=1$ and $V_\mathcal{T}\to 0_-$, this induces $\mathcal{Q}_{tx}=-\pi \epsilon$ and $\mathcal{Q}_{tt}=\epsilon^2$, leading to momentum $k_x$ magnetization, $M^z_{k_x}=-\frac{\hbar}{8\pi}\mathcal{Q}_{tx}$, and bulk energy magnetization, $M^z_{k_t}=-\frac{\hbar}{8\pi}\mathcal{Q}_{tt}$, following from Eq.~(\ref{Eq:MomMag}). 
In Fig.~\ref{fig:2D_ins_quadrupole}(g) we plot the boundary energy current response $\Delta\mathcal{J}^x_t$ as a function of $\mathcal{Q}_{tt}.$ We calculate this quantity by summing the particle current $\frac{1}{\hbar}\frac{\partial H}{\partial k_x}$ weighted by the energy $\epsilon (k)$ of each state. The slope of the plot confirms the coefficients predicted in Eq.~(\ref{Eq:MomMag}).

\subsection{3D nodal line dipole semimetal}
\label{sec:3D_dipole_NLSM}
\begin{figure}
    \centering
    \includegraphics[width=0.47\textwidth]{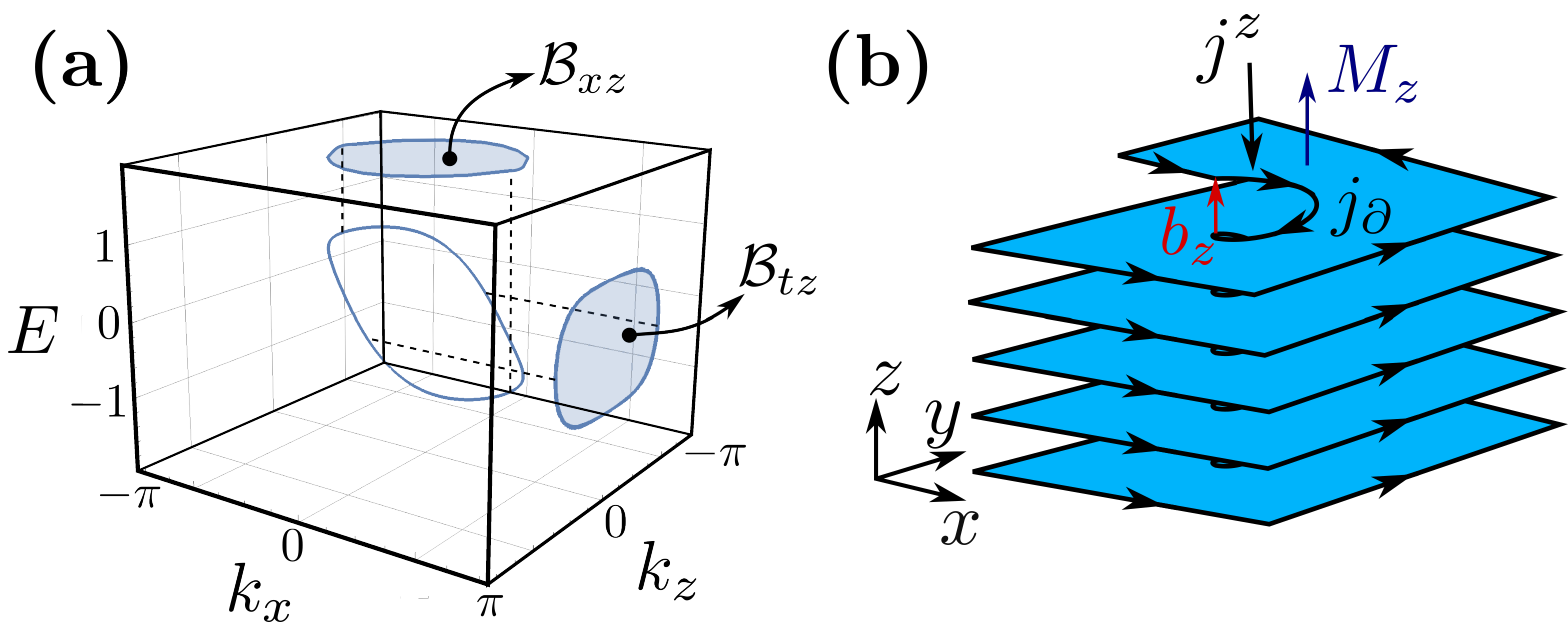}
    \caption{(a) Fermi line of a 3D NLSM (\ref{eqn:3D_dipole_NLSM_ham}) with $V_\mathcal{I}=0$, $m=2$ that is tiled in the energy-momentum space $\{k_z,k_x, E\}$ by the perturbation (\ref{eqn:Btz_inducing_term}) where we set $\epsilon=1$. The projections of this curve onto the $\{k_x,k_z\}$ and $\{k_z,E\}$ planes give the exact values of the $\mathcal{B}_{xz}$ and $\mathcal{B}_{tz}$ coefficients respectively. (b) A screw dislocation characterized by a Burgers' vector $b^z=a_z$ creates an internal boundary carrying a current circulating around the magnetization vector $M_z$. Note that the currents' direction is perpendicular to the Burgers' vector and the Magnetization vector $M_z$, as predicted by Eq.~\ref{eqn:NLSM_screw_current}.}
    \label{fig:dipole_NLSM}
\end{figure}

Heuristically we can consider nodal 3D semimetals as arising from stacks of 2D Dirac node dipole semimetals. Furthermore, similar to the 2D case, with inversion symmetry the bulk response action
\begin{equation}
    S[\mathfrak{e}^\lambda,A]=e\mathcal{B}_{\mu\nu}\int \mathfrak{e}^\mu\wedge\mathfrak{e}^\nu\wedge dA
    \label{eqn:dipole_NLSM_cryst}
\end{equation} can be interpreted as
a charge magnetization $M_i$ and electric polarization $P_e^i$: 
\begin{equation}
    e\mathcal{B}_{ta}=M^i \mathfrak{e}_i^a,\quad e\mathcal{B}_{ab}=\varepsilon_{ijk}P^k_e \mathfrak{e}_{i}^a \mathfrak{e}_{j}^b
\end{equation} where we have taken functional derivatives of Eq.~\ref{eqn:dipole_NLSM_cryst} with respect to the magnetic and electric fields respectively, and  used $\mathfrak{e}_t^t=1.$ For an unmodified geometry we recover the results of Ref.~\onlinecite{ramamurthylinenode}, i.e.,
\begin{equation}
    e\mathcal{B}_{ta}=M^a,\,\,\, e\mathcal{B}_{ab}=\varepsilon_{abk}P^k_e.
\end{equation} Microscopically, the coefficient $\mathcal{B}_{ab},$ where $a,b=1,2,3,$ is proportional to the area of the line nodes that project onto surfaces normal to the $ab$-plane as illustrated in Fig.~\ref{fig:dipole_NLSM}(a).

The bulk action also implies a non-vanishing  momentum response to electromagnetic fields:
\begin{equation}
\mathcal{J}^\mu_\lambda=2e\mathcal{B}_{\lambda\eta}\varepsilon^{\mu\nu\rho\sigma}\mathfrak{e}^\eta_\nu\partial_\rho A_\sigma,
  \label{Eq:NLSMResponse}
\end{equation}  and a conjugate electromagnetic response to translation gauge fields:
\begin{equation}
j^\mu=2e\mathcal{B}_{\lambda\eta}\varepsilon^{\mu\nu\rho\sigma}\mathfrak{e}^\lambda_\nu\partial_\rho\mathfrak{e}^\eta_\sigma.\label{eqn:linenodecurrentresp}\end{equation}

To illustrate how these responses manifest in an explicit model, we can construct a Hamiltonian for a 3D nodal line dipole semimetal by
stacking copies of the 2D Dirac node dipole semimetal in Eq.~(\ref{eqn:2d_ins_dipole}) in the $\hat{z}$-direction.  When there is no hopping between the 2D layers, such a system will have two lines of gapless states spanning the BZ along the $k_z$ direction, located at $(k_x,k_y)=(\pm K,0)$ (for our model). 
Adding hopping terms in the $\hat{z}$-direction leads to a Bloch Hamiltonian:
\begin{equation}
\begin{split}
    H(\textbf{k})&=V_\mathcal{I} \sigma^x + \sin (k_y a_y)\sigma^y\\
    &+ (m-\cos (k_x a_x) - \cos(k_y a_y) - \cos(k_z a_z))\sigma^z.
    \label{eqn:3D_dipole_NLSM_ham}
\end{split}
\end{equation}
Taking $V_\mathcal{I} \to 0$ and $m=2$, we find a single loop of gapless states located in the $k_y=0$ plane, described by the equation $\cos (k_x a_x) + \cos(k_z a_z)=1$. 
The stack of 2D Dirac node dipole semimetals will naturally endow the 3D nodal line system with electric polarization (and/or magnetization). Correspondingly, this model has a single non-zero component of the antisymmetric tensor $\mathcal{B}_{xz}$ defined in Eq.~(\ref{Eq:BetaCoeff}), which encodes a charge polarization in the $\hat{y}$-direction. 
From Eq.~\ref{Eq:NLSMResponse}, a non-vanishing $\mathcal{B}_{xz}$ also implies a $k_x$ momentum line-density localized on a magnetic flux tube oriented in the $\hat{z}$-direction:
\begin{equation}
\mathcal{J}^t_x=2e\mathcal{B}_{xz}\varepsilon^{tzij}\mathfrak{e}^z_z B_z=2e\mathcal{B}_{xz}B^z ,
\end{equation} 
similar to a stack of decoupled 2D Dirac semimetallic layers (in the last equality we replaced $\mathfrak{e}_z^z=1)$. This is the 3D analog of the response shown in Figs.~\ref{fig:2D_berry_dipole}(c) and (d) for the 2D Dirac semimetal.

We can see an example of a charge response if we tilt the nodal line to introduce a non-zero value of $\mathcal{B}_{tz}$ as illustrated in Fig.~\ref{fig:dipole_NLSM}(a). In our model we can tilt the node by adding an extra dispersion 
\begin{equation}
    \Delta H(\textbf{k})=\epsilon\sin(k_x a_x)\mathbb{I}_{2\times 2},
    \label{eqn:Btz_inducing_term}
\end{equation}
to the Hamiltonian. This term breaks $\mathcal{T}$ and induces a net magnetization $M_z= e\mathcal{B}_{tz}$, setting up the corresponding circulating boundary currents in the system~\cite{ramamurthylinenode}.

Now, when $\mathcal{B}_{tz}$ is non-vanishing, Eq.~\ref{eqn:linenodecurrentresp} implies that a screw dislocation with Burgers vector $b^z \hat{z}$ hosts a bound electromagnetic current. Indeed, if we assume the screw dislocation is located at $(x,y)=(0,0)$ and runs along the $z$-axis we find
\begin{equation}
j^z=-2e\mathcal{B}_{tz}\varepsilon^{tzjk}\mathfrak{e}^t_t\partial_j\mathfrak{e}^z_k=-2e\mathcal{B}_{tz}b^z\delta(x)\delta(y),
    \label{eqn:NLSM_screw_current}
\end{equation} where we used $\mathfrak{e}_t^t=1$ and $\nabla \times \mathfrak{e}^z=b^z \delta(x)\delta(y).$

We can illustrate the origin of this current by considering the magnetization $M_z$ (and associated boundary currents) induced by $\mathcal{B}_{tz}.$  A screw dislocation with Burgers vector $b^z\hat{z}$ can be constructed by cutting a seam into layers normal to $\hat{z}$ and re-gluing them along the seams with neighboring layers above or below. When cut, the boundary current associated to $M_z$ will appear, and after re-gluing this current will be routed vertically along the screw-dislocation line, i.e., along the $z$-direction as shown in Fig.~\ref{fig:dipole_NLSM} (b). The magnetization $M_z$ gives rise to a surface bound current $j_\partial=M_z$ circulating around the $\hat{z}$-axis in each layer.  
The effective number of current loops winding around the dislocation line per unit length is equal to the Burgers vector $b^z$.  Thus the total current in the $\hat{z}$-direction is:
\begin{equation}
    j^z=-b^z j_\partial =-2e\mathcal{B}_{tz} b^z,
\end{equation}
which reproduces the result obtained from the response action. Furthermore, we can understand the sign of the current from Fig.~\ref{fig:dipole_NLSM}(b) where we see that the current on the dislocation has an opposite orientation to the current generated by $M_z.$ Another interesting consequence of Eq. (\ref{eqn:dipole_NLSM_cryst}) is the topological piezoelectric effect discussed in Ref.~\cite{Matsushita2020}.

\subsection{3D nodal line quadrupole semimetal}

In Sec.~\ref{Sec:NLSMResponse}, we derived the effective response action: 
\begin{equation*}
S[\mathfrak{e}^\lambda]=\hbar\mathcal{B}_{\lambda\eta, \alpha}\int \mathfrak{e}^\lambda\wedge\mathfrak{e}^\eta\wedge d\mathfrak{e}^\alpha.
  \end{equation*}
for the nodal line quadrupole semimetal. The response action implies the energy-momentum currents:
\begin{equation}
\mathcal{J}^\mu_\lambda=2\hbar\left(\mathcal{B}_{\lambda\eta,\alpha}-\mathcal{B}_{\eta\alpha,\lambda}\right)\varepsilon^{\mu\nu\rho\sigma} \mathfrak{e}^\eta_\nu\partial_\rho\mathfrak{e}^\alpha_\sigma, 
\end{equation} where we have used that $\mathcal{B}_{\lambda\eta,\alpha}$ is  anti-symmetric under exchange of the first two indices.

In analogy with the 2D Dirac node dipole and Dirac node quadrupole semimetals, we expect that most of the responses from the Dirac nodal line dipole semimetal in Sec.~\ref{sec:3D_dipole_NLSM} can be translated to describe some of the responses of this action if we replace charge currents and densities with momentum currents and densities etc. Indeed, we showed in Eq.~\ref{eq:BabgMomPol} that when $\lambda$ and $\eta$ are both spatial indices,
$\mathcal{B}_{\lambda \eta,\lambda}$ implies a momentum polarization in a direction perpendicular to $\lambda$ and $\eta,$ and carrying momentum parallel to $\lambda$. By analogy, the mixed temporal-spatial components $\mathcal{B}_{i t,j}$ describe a momentum magnetization in the $i$-th direction carrying momentum in the $j$-th direction. The momentum magnetization is further responsible for generating bound-currents on screw-dislocations, i.e., the momentum magnetization will have circulating boundary momentum currents and a momentum current along screw dislocations similar to the charge bound currents on dislocations shown in Section~\ref{sec:3D_dipole_NLSM}. 

\begin{figure}
    \centering
    \includegraphics[width=0.4\textwidth]{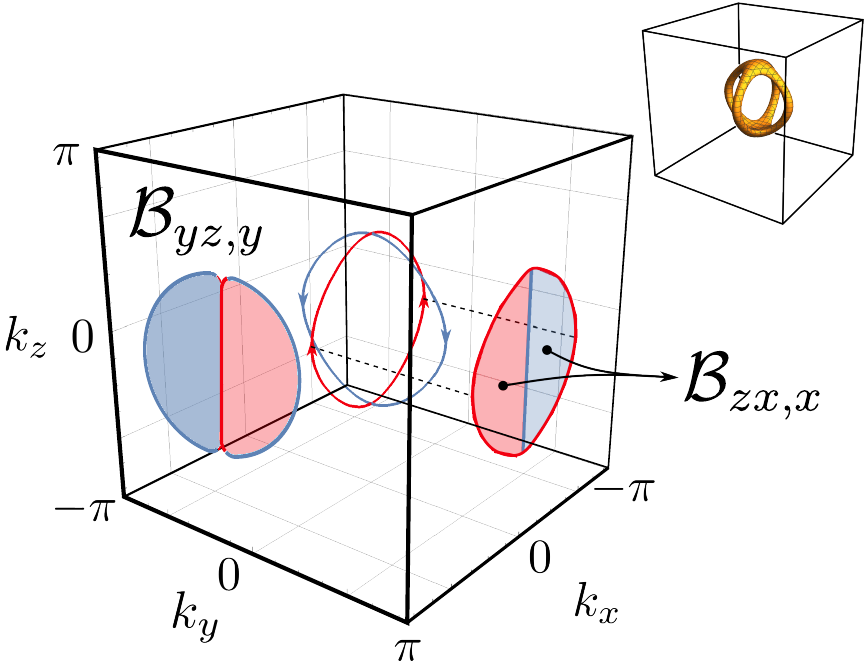}
    \caption{Fermi Lines of the model (\ref{eqn:3D_quad_NLSM}) with $m=2$ and $V_\mathcal{T}\to 0_-$. Resolving this structure as a pair of loops with \emph{fixed orientation} we can project them onto the $k_x k_z $ or $k_y k_z$ surfaces to determine the momentum polarization. The colored regions of the projected nodes indicate flat drumhead states that would appear in open boundary conditions on one boundary (red) or the opposing boundary (blue). By looking at the relative positions of the two areas bounded by the projected loops in the surface BZ, we see that one surface will have one sign of the $k_x$ or $k_y$ momentum, and the other surface will have the other. For example, for the $k_x k_z$ surface BZ the the projections indicate a dipole moment of $k_x$ momentum polarized along the $y$ direction captured by the response coefficient $\mathcal{B}_{zx,x}$.  Inset: Cage-like nodal Fermi surface in the model (\ref{eqn:3D_quad_NLSM}) with $E_{F}=0.2$.}
    \label{fig:3D_NLSM_quadrupole}
\end{figure}

To be more explicit, we can illustrate the momentum polarization in a model by showing the analog of the surface charge theorem, i.e., momentum polarization will yield surface momentum densities. To obtain a Hamiltonian for the nodal line quadrupole semimetal, we begin by stacking 2D Dirac node quadrupole semimetals (see Fig.~\ref{fig:2D_ins_quadrupole} (b)) along the $\hat{z}$-direction. When the planes are completely decoupled, this construction produces a set of four straight Fermi lines stretching in the $k_z$-direction. If we couple the two-dimensional planes, then we arrive at the following Bloch Hamiltonian:
\begin{equation}
\begin{split}
    H(\textbf{k})&=V_\mathcal{T} \sigma^x + \sin (k_x a) \sin (k_y a)\sigma^y\\
    &+ (m-\cos( k_x a) -\cos (k_y a)-\cos(k_z a_z))\sigma^z.
    \label{eqn:3D_quad_NLSM}
\end{split}
\end{equation}
For a wide range of parameters this model has a pair of nodal line loops that form a cage structure as shown in Figs.~\ref{fig:ptiresponse} and \ref{fig:3D_NLSM_quadrupole} with $m=2$ and  $V_{\mathcal{T}}=0.$ In general, the local gaplessness of the nodal loops can be protected by the product $\mathcal{TI}.$ The cage structure created by the joined, intersecting loops can be split apart by, for example, breaking mirror symmetry along the $k_x=k_y$ axis while preserving $\mathcal{TI}$. However, even in this case the nodal loops still produce a non-vanishing contribution to the response coefficient $\mathcal{B}_{\alpha\beta,\gamma}.$  Hence, the response is more general than the specific cage-like nodal configuration. Calculating the response coefficient for the action in the limit $V_\mathcal{T} \rightarrow 0_-$, we find that $\mathcal{B}_{xz, x}=-\mathcal{B}_{zx, x}, $ $\mathcal{B}_{yz,y}=-\mathcal{B}_{yz, y}  $ are non-vanishing, as shown in Fig.~\ref{fig:3D_NLSM_quadrupole}.

Using this model we can illustrate the origin of the boundary momentum resulting from the bulk momentum polarization. The discussion is analogous to the calculation of the boundary momentum of the 2D Dirac node quadrupole semimetal in Sec.~\ref{sec:2D_Dirac_qp}. Indeed, the analogy is clear since the cage nodal structure is just arising from a family of 2D Dirac node quadrupoles parameterized by $k_z.$ To specify an unambiguous momentum polarization we turn on a small $\mathcal{T}$-breaking perturbation $V_\mathcal{T}.$ After doing this, and as shown in Fig.~\ref{fig:3D_NLSM_quadrupole}, we see that the two nodal loop segments that lie in the $k_y=0$ plane (one for $k_x>0$ and one for $k_x<0$) carry the same Berry flux in the $k_z$-direction (red arrows in Fig.~\ref{fig:3D_NLSM_quadrupole}). Similarly, the two loop segments in the $k_x = 0$ plane carry the same Berry flux (blue arrows), which is opposite to that carried by the $k_y =0$ segments. Consequently, the loop segment in the $k_y=0$, $k_x>0$ half-plane must connect with a loop segment in the $k_x=0$ plane in order to form a closed nodal loop with a consistent helicity/flux sign. 

To clarify the consequences of this nodal configuration let us consider the $k_x k_z$ plane in Fig.~\ref{fig:3D_NLSM_quadrupole}. We can calculate a Berry-Zak phase~\cite{zak} in the $k_y$ direction parameterized by $(k_x, k_z),$ and for our model we find a Berry phase of magnitude $\pi$ inside the projected nodal region in the $k_x k_z$ plane. When $V_\mathcal{T}$ is turned on, the sign of the $\pi$ Berry-Zak phases are no longer ambiguous, and are opposite for the projected areas at $k_x>0$ and $k_x<0.$ If we calculate the total polarization in the $y$-direction when summed over all $k_x$ and $k_z$ it will vanish. However, the polarization weighted by the $k_x$ momentum will be non-zero. The occupied drumhead surface states in the $k_x k_z$ surface-BZ (see Fig.~\ref{fig:3D_NLSM_quadrupole} and c.f. Fig.~\ref{fig:2D_ins_quadrupole}(a,b,c)) will have an imbalanced $k_x$ momentum, but, when combined with the bulk charge density, a vanishing charge (c.f. Fig.~\ref{fig:2D_ins_quadrupole}(c)).  This is a reflection of the surface charge theorem for a vanishing charge polarization, and non-vanishing momentum polarization. We numerically calculated the magnitude of the bound surface momentum, finding it to be in agreement with the value predicted by the response action, $2\hbar\mathcal{B}_{xz,x}.$ We see from this picture that to have a non-zero response $\mathcal{B}_{xz,x}$, we want two oppositely oriented nodal loops with identical, non-vanishing areas when projected in the $k_x k_z$-plane, but positioned so that the sums of all $k_x$ inside each nodal loop are different from each other, e.g., in our model they are opposite values.

As an additional explicit example of a non-vanishing response allowed in our model we can consider the momentum density
\begin{equation}
\mathcal{J}^0_x = 2\hbar \mathcal{B}_{xz,x}\epsilon^{ijk}(2\mathfrak{e}_i^z\partial_j \mathfrak{e}_k^x-\mathfrak{e}_i^x\partial_j \mathfrak{e}_k^z)
\end{equation}
generated by a geometric deformation. To generate a non-vanishing response let us consider an $xz$-planar interface. Since we must preserve translation symmetry along $x$ to calculate $k_x$ momentum, and we want to preserve translation in $z$ for convenience, we have the following terms:
\begin{equation*}
    \mathcal{J}^0_x=2\hbar \mathcal{B}_{xz,x}\left(2\mathfrak{e}_x^z\partial_y \mathfrak{e}_z^x-2\mathfrak{e}_z^z\partial_y \mathfrak{e}_x^x-\mathfrak{e}_x^x\partial_y \mathfrak{e}_z^z+\mathfrak{e}_z^x\partial_y \mathfrak{e}_x^z\right).
\end{equation*} If we cut the system at $y=0$, both sides of the interface will carry a surface $k_x$-momentum density $\mathcal{J}^0_{x,surf}=\pm 2\hbar \mathcal{B}_{xz,x},$ since the system has a $k_x$ momentum polarization along $\hat{y}$ with this magnitude. Since each interface carries an opposite sign of the momentum density, if we glue them back together there will be no momentum at the interface.  Now, for $y>0$ let us perturb away from the background translation gauge field configuration to 
 $\mathfrak{e}_i^a=(1+\epsilon^a)\delta_{i}^a$ where $\epsilon^a=(\epsilon^x,\, 0,\, \epsilon^z)$ is a small deformation.    The momentum density response to leading order in $\epsilon^a$ is 
\begin{equation}
\mathcal{J}^0_x=2\hbar\mathcal{B}_{xz,x}\left[-2\epsilon^x\delta(y)-\epsilon^z\delta(y)\right],\label{eq:quadnodallineinterface}
\end{equation} which we see is localized at the interface $y=0.$ 

We can interpret this response by noting that changing $\mathfrak{e}_x^x$ or $\mathfrak{e}_z^z$ effectively changes the area of one side of the interface ($y>0$) relative to the other ($y<0$). Since the total $k_x$ momentum on both sides of the interface should be unchanged by this deformation (we maintain translation symmetry in $x$ during the process), then increasing the area for $y>0$ must lower the momentum \emph{density}. Indeed, the surface $k_x$ momentum density on $\hat{y}$ surfaces must be inversely proportional to $L_x$ and $L_z.$ Finally, since we are considering $k_x$-momentum density, the quantization of which depends on $L_x^{-1}$, $\mathcal{J}^0_x$ actually depends on $L_x^{-2}$, hence the difference between the coefficients of $\epsilon^x$ and $\epsilon^z$ in Eq.~\ref{eq:quadnodallineinterface}.

\subsection{3D Weyl node dipole semimetal}

\begin{figure*}
    \centering
    \includegraphics[width=0.90\textwidth]{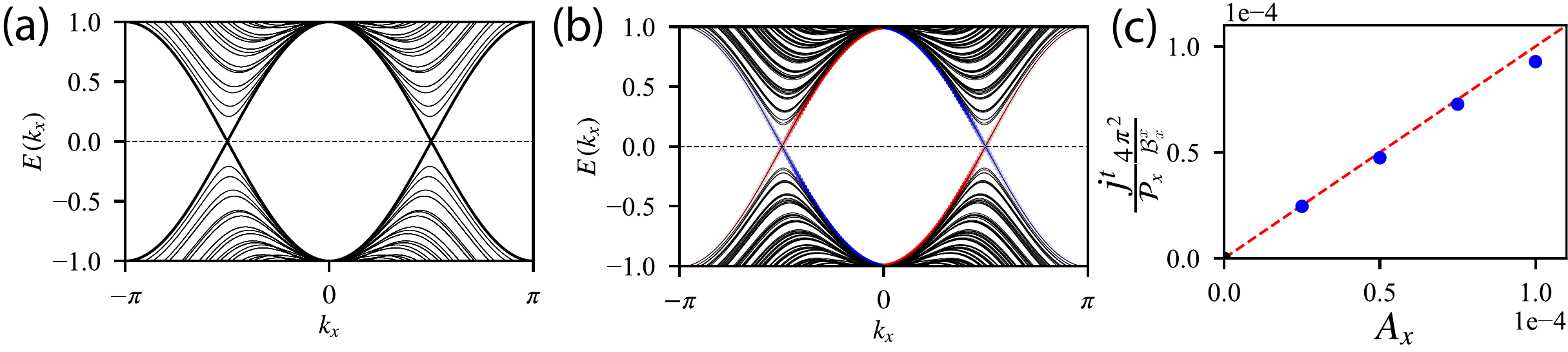}
    \caption{(a) The bulk spectrum of a Weyl semimetal with two nodes on the $k_x$-axis (b) The spectrum of the same Weyl semimetal with periodic boundary conditions and two screw dislocations with opposite Burgers vectors threaded along the $x$-direction. Red and blue coloration indicates on which dislocation the chiral modes are localized. Each dislocation has a net positive (red) or negative (blue) chirality.  (c) Numerical calculation of the charge density bound to a screw dislocation as $A_x$ is tuned. }
    \label{fig:weyldipoledislocation}
\end{figure*}

The electromagnetic and geometric response of time-reversal breaking 3D  Weyl semimetals have been discussed extensively in the literature~\cite{zyuzin2012, parrikar2,RamamurthyPatterns,BurkovReview18,dbh2021,gioia2021,Hiroaki2016,Kodama2019,Huang19,Huang20a,Huang20b,Liang20,Nissinen20,pikulin2016chiral,you2016,dislocation_defect_2020,laurila_torsional_2020,chongwang2,jakko,gao_chiral_2021,amitani_torsion_2023,nissinen2018tetrads,nissinen2019elasticity,ferreiros2019mixed,chu2023chiral}. Here we focus on a few particular consequences of the mixed crystalline-electromagnetic response and the matching between the response field theory and microscopic lattice model calculations. Recall that the response action for a 3D Weyl semimetal with a non-vanishing Weyl-node dipole moment $\mathcal{P}_\lambda$ is
\begin{equation}
    S[\mathfrak{e}^\lambda_\nu,A_\mu]=\frac{e^2\mathcal{P}_\lambda}{8\pi^2\hbar}\int \mathfrak{e}^\lambda\wedge A\wedge dA.
    \label{eqn:3D_dipole_resp_cryst}
\end{equation} This response implies the following bulk electromagnetic and momentum currents: 
\begin{gather}
j^\mu = -\frac{e^2\mathcal{P}_\lambda}{4\pi^2\hbar}\varepsilon^{\mu\nu\rho\sigma}\mathfrak{e}^\lambda_\nu\partial_\rho A_\sigma+\frac{e^2\mathcal{P}_\lambda}{8\pi^2\hbar}\varepsilon^{\mu\nu\rho\sigma}A_\nu\partial_\rho \mathfrak{e}^\lambda_\sigma \ ,
\label{Eq:WeylDipoleCurrents} \\
\mathcal{J}^\mu_\lambda = \frac{e^2\mathcal{P}_\lambda}{8\pi^2\hbar}\epsilon^{\mu\nu\rho\sigma}A_\nu \partial_\rho A_\sigma.\label{Eq:WeylDipoleMomCurrents}
\end{gather} In the presence of dislocations the translational flux is non-vanishing, and hence the bulk electromagnetic current is anomalous: 
\begin{equation} \label{Eq:WeylScrew}
\begin{split}
        \partial_\mu j^\mu=-\frac{e^2\mathcal{P}_\lambda}{8\pi^2\hbar} \varepsilon^{\mu\nu\sigma\rho} \partial_\mu \mathfrak{e}^\lambda_\nu  \partial_\sigma A_\rho.
\end{split}
\end{equation} This reflects the fact that the action Eq.~\ref{eqn:3D_dipole_resp_cryst} is not gauge-invariant in the presence of dislocations. Indeed, in our explicit tight-binding model calculations below we find the spectrum on a single screw dislocation line contains a pair of chiral modes of the same chirality (one near each bulk Weyl node momentum as shown in Fig.~\ref{fig:weyldipoledislocation}(b)). These modes are responsible for the anomalous current on dislocation lines, as was first described by Ref.~\onlinecite{ran2009}.

To verify the electromagnetic response to the applied crystalline gauge field we consider a simple 2-band model of a 3D Weyl semimetal with a pair of gapless nodes:
\begin{equation}
\begin{split}
    H(\textbf{k})&=\sin (k_z a_z) \sigma^x + \sin (k_y a_y) \sigma^y \\
    &+ (2-m-\cos (k_x a_x) - \cos (k_y a_y) - \cos (k_z a_z))\sigma^z. 
\end{split}
\end{equation}
The Weyl node with the positive chirality $\chi=+1$ is located at $\textbf{k}=(\arccos(-m),\,0,\,0)$ and the node with $\chi=-1$ is at $\textbf{k}=(-\arccos(-m),\,0,\,0)$. The Weyl node dipole moment therefore has only one non-zero component $\mathcal{P}_x=2\arccos(-m)$ and the resulting response action is
\begin{equation}
     S[\mathfrak{e}^\lambda_\nu,A_\mu]=\frac{e^2\mathcal{P}_x}{8\pi^2\hbar}\int d^4 x \epsilon^{\mu\nu\rho\sigma}\mathfrak{e}^x_\mu A_\nu \partial_\rho A_\sigma.
\end{equation} 

\begin{figure}
    \centering
    \includegraphics[width=0.45\textwidth]{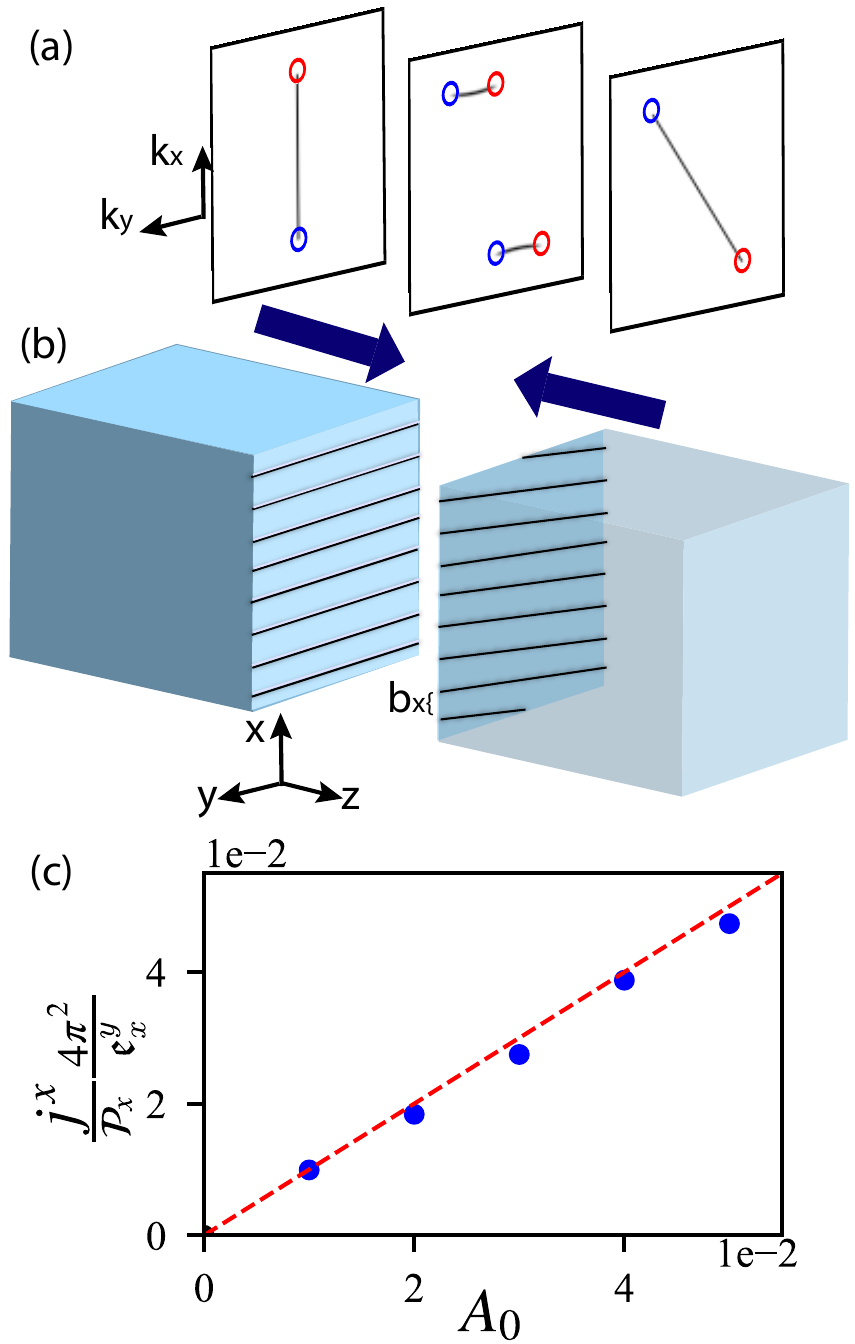}
    \caption{(a) The three panels show numerically calculated Fermi arcs in (left) the surface BZ with un-deformed geometry, (right) the surface BZ with $\mathfrak{e}_y^x$ non-vanishing, and (center) the arcs localized at the interface formed by gluing the two sides of the interface together. The colored circles in the first and third panels represent the surface BZ projections of the bulk Weyl nodes on either side of the interface. The color is a guide to show the connectivity/orientation of the Fermi arcs, not the chirality of the bulk nodes. On both sides of the interface the bulk nodes have the same chirality, but since they are effectively projected onto surfaces having opposite normal vectors they generate Fermi arcs having opposite chirality.  (b) Illustrations of (left) the un-deformed geometry and (right) the deformed geometry with $\mathfrak{e}_y^x$ non-vanishing. (c) The numerically calculated current localized at the interface between un-deformed and deformed geometries as a function of the chemical potential shift $A_0$.}
    \label{fig:twistedinterface}
\end{figure}

Let us first consider the response arising from the constant background translation fields $\mathfrak{e}_x^x=1$ and $\mathfrak{e}_y^x=b^x/L_y$, which describe a twist such that a particle traversing the lattice in the $y$-direction translates by $b^x$ in the $x$-direction. We note that such a configuration is volume preserving since $\det({\bf{e}})=1$, where the matrix ${\bf{e}}$ has matrix elements $e_{ij}=\mathfrak{e}_i^j.$ When $b^x=0$ the response action is
\begin{equation*}
\frac{e^2\mathcal{P}_x}{8\pi^2\hbar}\int \mathfrak{e}_{x}^x dx\int dt dy dz \epsilon^{x\nu\rho\sigma} A_\nu \partial_\rho A_\sigma.
\end{equation*}
Using the relation $\int dx \mathfrak{e}_x^x=L_x$ we find an anomalous Hall effect in the $yz$-plane such that $\sigma_{yz}=\frac{e^2}{h}\frac{\mathcal{P}_x L_x}{2\pi},$ which is the standard result~\cite{zyuzin2012,Wan2011}. Now, if we turn on $b^x$ we will still have the same $\sigma_{yz},$ but we will also have the additional term
\begin{equation*}
    \frac{e^2\mathcal{P}_x}{8\pi^2\hbar}\int \mathfrak{e}_y^x dy\int dt dx dz \epsilon^{y\nu\rho\sigma} A_\nu \partial_\rho A_\sigma.
\end{equation*} Because of the different index on the $\epsilon$-symbol, this term represents an anomalous Hall effect in the $xz$-plane with $\sigma_{zx}=\frac{e^2}{h}\frac{\mathcal{P}_x b_x}{2\pi}.$ We can find a simple interpretation for this effect: when we turn on $\mathfrak{e}_y^x$, the minimal coupling $k_x\to k_x, k_y\to k_y+k_x\mathfrak{e}_y^x$ shifts the bulk Weyl nodes, $(\pm \mathcal{P}_x/2,\,0,\,0) \to (\pm \mathcal{P}_x/2,\,\pm \mathcal{P}_x b_x/(2L_y),\,0).$ Hence an effective $\mathcal{P}_y=\frac{\mathcal{P}_x b_x}{L_y}$ is generated when the Weyl momenta are sheared. Indeed, we expect that, at least for uniform, traceless translation gauge field deformations, the response phenomena can be simply interpreted as  transformations of the Weyl node dipole $\mathcal{P}_i\to \mathfrak{e}^j_i\mathcal{P}_j.$ We show an explicit example of this in the first and third surface-BZ panels of Fig.~\ref{fig:twistedinterface}(a) where the bulk nodes and their connected Fermi arcs have been rotated in the deformed geometry relative to the undeformed geometry. We note that if the deformation is not volume-preserving, then we must be careful when considering what is held fixed while volume is changing in order to interpret the resulting phenomena.

In addition to these cases of fixed background translation fields, let us consider varying those fields in space. We are interested in the electromagnetic response to applied translational \emph{magnetic} fields $\mathcal{B}_i^a=\epsilon^{ijk}\partial_j \mathfrak{e}_k^a$.  Since the nodes in our model are separated in $k_x$, we will consider geometries where the Burgers vector of the translation magnetic field also points along the $x$-direction, $\mathcal{B}_i^x\neq 0$. 

First let us consider a system containing a domain wall as a function of $z$, such that at $z=0$ the field $\mathfrak{e}_y^x$ jumps from 0 to $b_x/L_y.$ For $z<0$ we have bulk Weyl nodes that project onto the $z$-surface at  $(\pm \mathcal{P}_x/2,0),$  while for $z>0$ the bulk Weyl nodes have been transformed and sit at  $(\pm \mathcal{P}_x/2,\pm \mathcal{P}_x b_x/(2L_y)).$ We show the numerically calculated Fermi arcs for our un-deformed and deformed models in the left and right surface BZ panels of Fig.~\ref{fig:twistedinterface}(a).

Now let us glue the $z<0$ and $z>0$ sides to each other to make a domain-wall interface. We schematically illustrate the interface geometry in Fig.~\ref{fig:twistedinterface}(b). Since the normal vector on each side of the interface is opposite, we expect the Fermi arcs for $z<0$ to have the opposite chirality to their corresponding arcs for $z>0.$  Indeed, as shown in the center surface BZ panel of Fig.~\ref{fig:twistedinterface}(a), the Fermi arcs on both sides can hybridize because of their opposite chiralities and form new arcs in the 2D subsystem of the interface. These new Fermi arcs encode the fact that the Hall conductivity $\sigma_{xz}$ is varying at this interface. These effects are all manifestations of the fact that the Weyl node dipole moment $\mathcal{P}_i$ is changing at the interface, and hence we expect Fermi arcs to be trapped generically at the interfaces of this type. We note that a similar strain geometry, and the corresponding Weyl node configuration, was discussed in~\cite{Grushin2016}.

From Eq.~\ref{Eq:WeylDipoleCurrents} we see that applying a uniform, non-vanishing $A_0$ to the system described above should generate a charge current in the $x$-direction. We can see the microscopic origin of this current as follows. If we increase $A_0,$ each linearly-dispersing point on the Fermi arc will have an excess charge density $\delta n({\bf{k}})=\frac{e A_0}{2\pi\hbar |v_F({\bf{k}})|}$ where $v_F({\bf{k}})$ is the Fermi velocity at the Fermi arc located at ${\bf{k}}$ in the surface-BZ. Hence, the contribution to the current of such a point on the Fermi arc is $j^x({\bf{k}})=e v_{F}({\bf{k}}) \delta n({\bf{k}}).$    For our model and geometry, the contributions to the $j^x$ current that are linear in the deformations of $\mathfrak{e}_i^a$ arise from the Fermi arcs stretching between $(K,0)\to(K,K\mathfrak{e}_y^x)$ and $(-K,0)\to(-K,-K\mathfrak{e}_y^x).$ Each of these arcs has a fixed value $k_x=\pm K$ and each arc has an opposite Fermi velocity. Hence 
\begin{equation*}
\begin{gathered}
j^x =e v_F(K,k_y) \delta n \frac{K\mathfrak{e}_y^x}{2\pi} \\
+ e v_F(-K,k_y) \delta n \frac{K\mathfrak{e}_y^x}{2\pi}
=\frac{e^2 \mathcal{P}_x\mathfrak{e}_y^x A_0}{4\pi^2 \hbar}{\rm{sgn}}(v_F),
\end{gathered}
\end{equation*}
 where $K\mathfrak{e}_y^x/2\pi$ counts the density of states on the Fermi arc in the $k_y$ direction, ${\rm{sgn}}(v_F)$ is sign of the velocity on the $k_x=+K$ arc, and $\mathcal{P}_{x}=2K^2$ is the un-deformed value. This result matches the prediction from the response theory and matches the numerical results in Fig.~\ref{fig:twistedinterface}(c)~\footnote{While the coefficient of the response in Eq.~\ref{Eq:WeylDipoleCurrents} is half the size of our numerical and analytic result, our calculations inherently determine the \emph{covariant} anomaly of the interface Fermi arc states which receives inflow from the bulk term in Eq.~\ref{Eq:WeylDipoleCurrents}, i.e., inflow from a boundary term of the same magnitude, hence doubling the result~\cite{callanharvey,naculich1988,stone2012,parrikar1}.}.

We can also study a system with a pair of screw dislocation lines. We explicitly insert two screw dislocations at positions $(y,z)=(N_y/4,0)$ and $(y,z)=(3N_y/4,0)$, running parallel to the $\hat{x}$-axis with Burgers vectors $b^x=+1$ and $b^x=-1$, respectively. In Fig.~\ref{fig:weyldipoledislocation}(a) we show the energy spectrum of a Weyl semimetal with Weyl nodes on the $k_x$-axis with periodic boundary conditions and no dislocations. In Fig.~\ref{fig:weyldipoledislocation}(b) we show the spectrum of the same system after two screw dislocations have been inserted as described above. The blue/red coloration indicates on which dislocation the states are localized. We see that near each Weyl point the right-moving modes are on the red dislocation while the left-moving modes are on the blue dislocation,  as described by Eq. (\ref{Eq:WeylDipoleCurrents}). Hence, each dislocation has a net chirality. 

To test the response equation we apply a non-vanishing $A_x$ and numerically calculate the charge density localized on a single dislocation. We can carry out a microscopic calculation of the charge bound to a dislocation as a function of $A_x.$ Let us assume a nodal configuration with a positive node at ${\bf{k}}=(\mathcal{P}_x/2,0,0)$ and a negative node at $(-\mathcal{P}_x/2,0,0).$ In the presence of a dislocation having Burgers vector $b^x,$ each $k_y k_z$-plane sees an effective magnetic flux $\Phi(k_x)=\frac{b^x k_x}{2\pi}\Phi_0$, where $\Phi_0=h/e.$ Hence each $k_y k_z$-plane having a non-vanishing Chern number will contribute to the charge as
\begin{equation}
\Delta Q=\frac{e L_x}{2\pi}\int_{BZ}C(k_x) \frac{k_x b^x}{2\pi}dk_x=0,
\end{equation} where $C(k_x)$ is the Chern number of each $k_y k_z$-plane parameterized by $k_x.$ If we turn on a non-vanishing $A_x$ ($k_x\to k_x+\tfrac{e}{\hbar}A_x)$ and re-calculate the bound charge we find
\begin{equation}
\begin{aligned}
\Delta Q\vert_{A_x} &= - \frac{L_x}{2\pi}\int_{-\tfrac{\mathcal{P}_x}{2}-\tfrac{e}{\hbar}A_x}^{\tfrac{\mathcal{P}_x}{2}-\tfrac{e}{\hbar}A_x}\frac{k_x b^x}{2\pi}dk_x \\
&=\frac{e^2\mathcal{P}_x b^x L_x}{4\pi^2\hbar}A_x.
\end{aligned}
\end{equation} This result is exactly what is found in our numerics shown in Fig.~\ref{fig:weyldipoledislocation}(c). Both of these results match the analytic prediction in Eq.~\ref{Eq:WeylDipoleCurrents} after including an extra factor of two which takes into account the bulk and boundary inflow to the boundary~\cite{callanharvey,naculich1988,stone2012,parrikar1}.

\subsection{3D Weyl node quadrupole semimetal}

\begin{figure*}
    \centering
    \includegraphics[width=0.95\textwidth]{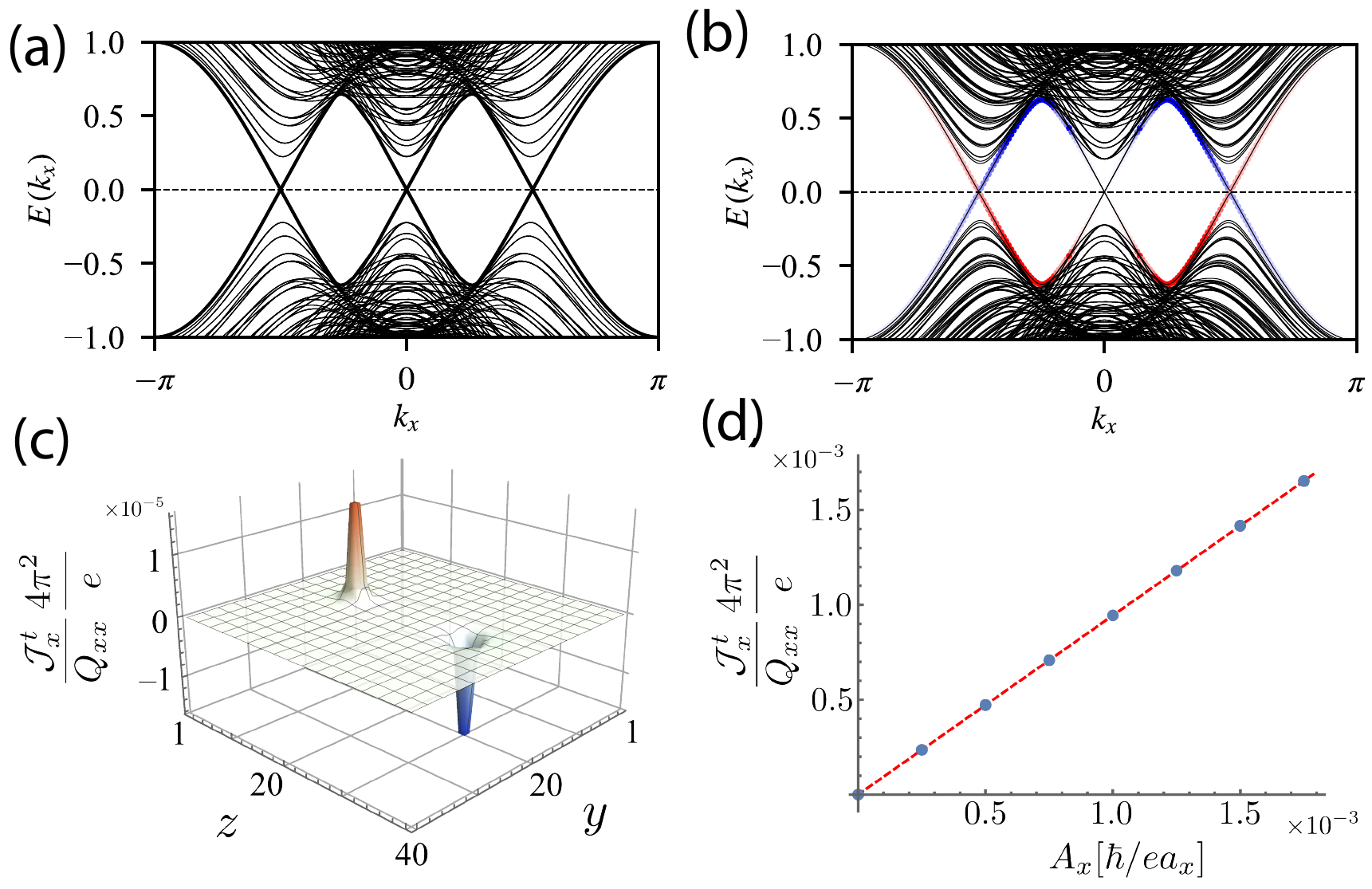}
    \caption{(a) The bulk spectrum of a Weyl semimetal with two nodes of one chirality on the $k_x$-axis and two nodes of the opposite chirality on the $k_y$-axis. (b) The spectrum of the same Weyl semimetal with periodic boundary conditions and two screw dislocations with opposite Burgers vectors threaded along the $x$-direction. Red and blue coloration indicates on which dislocation the chiral modes are localized. Each dislocation has a no net chirality, and the Weyl nodes on the $k_y$-axis do not form chiral modes.  (c) The spatially-resolved $k_x$ momentum density response of a Weyl node quadrupole semimetal to a pair of screw dislocations with opposite Burgers' vectors $b_x=\pm a_x$ located at $(y,z)=(20a_y, (20\pm 10)a_z)$ with the background gauge field $A_x=2.5\times 10^{-4}\hbar/ea_x$ and $\mathcal{Q}_{xx}=\pi^2/(2a_x^2)$. (d) Numerically calculated dependence of the $k_x$ momentum density localized on a screw dislocation with Burgers' vector $b_x=1$ as a function of the background gauge field $A_x$, using the same model as in (c). }
    \label{fig:3D_quad}
\end{figure*}

Finally, we will discuss some aspects of the crystalline response of 3D Weyl semimetals with gapless Weyl nodes forming a quadrupole pattern. Some of these responses were recently discussed in Refs.~\onlinecite{dbh2021,gioia2021, hirsbrunner2023}, and here we consider some of the responses in more microscopic detail and compare directly with lattice model calculations.

Recall from Sec.~\ref{sec:3dsemimetalresponses} the response action
\begin{equation*}
    S_{WQ}=\frac{e\mathcal{Q}_{\alpha\beta}}{8\pi^2}\int \mathfrak{e}^\alpha\wedge d\mathfrak{e}^\beta\wedge A.
\end{equation*} The bulk linear response implied by Eq.~(\ref{Eq:SWQResp}) is
\begin{equation}
\begin{split}
    \mathcal{J}_{\alpha}^{\mu} &= \frac{e}{8\pi^2}\varepsilon^{\mu\nu\rho\sigma}\mathcal{Q}_{\alpha\beta} \mathfrak{e}^{\beta}_\nu \partial_\rho A_\sigma\\
    &-\frac{e}{4\pi^2} \varepsilon^{\mu\nu\rho\sigma}\mathcal{Q}_{\alpha\beta} A_\nu \partial_\rho\mathfrak{e}^{\beta}_\sigma,
\end{split}
\label{Eq:QuadCurrents}
\end{equation} 
\begin{equation}
    j^{\mu}=-\frac{e}{8\pi^2}\epsilon^{\mu\nu\rho\sigma}Q_{\alpha\beta}\mathfrak{e}_\nu^{\alpha}\partial_\rho \mathfrak{e}_\sigma^{\beta}.\label{Eq:QuadChargeCurrents}
\end{equation} We also note that both of these currents can be anomalous when subjected to certain gauge field configurations:
\begin{gather}
\partial_\mu\mathcal{J}_{\alpha}^{\mu} = -\frac{e}{8\pi^2}\varepsilon^{\mu\nu\rho\sigma}\mathcal{Q}_{\alpha\beta} \partial_\mu\mathfrak{e}^{\beta}_\nu \partial_\rho A_\sigma, \\
\partial_\mu j^{\mu}=-\frac{e}{8\pi^2}\epsilon^{\mu\nu\rho\sigma}Q_{\alpha\beta}\partial_\mu\mathfrak{e}_\nu^{\alpha}\partial_\rho \mathfrak{e}_\sigma^{\beta}.
\end{gather}

Now let us consider several different phenomena associated to these response equations in the context of a lattice model introduced in Ref.~\onlinecite{dbh2021}:
\begin{equation}
\begin{split}
    H(\textbf{k})&=\sin k_x \sin k_y \Gamma^x+\sin k_z \Gamma^y\\
    &+\left(m+t(\cos k_x+\cos k_y+\cos k_z) \right)\Gamma^z.
    \label{Eq:QuadModel}
\end{split}
\end{equation}
Without any geometric deformations, the semimetal phase of our model with a Weyl node quadrupole has two nodes of one chirality at ${\bf{k}}=(\pm K,0,0)$ and two of the opposite chirality at $(0,\pm K,0).$ Thus the gapped, 2D $k_y k_z$ planes parameterized by $k_x$ will have a non-vanishing Chern number $C$ for $-K<k_x<0$ and a non-vanishing Chern number $-C$ for $0<k_x<K$ where $C=\pm 1.$ Similar statements can be made about the $k_x k_z$ planes. Without loss of generality let us choose the nodes on the $k_x$-axis to have positive chirality such that $\mathcal{Q}_{xx}>0$ and $C=+1.$ For our model this also implies that $\mathcal{Q}_{yy}<0$ and the non-vanishing $k_x k_z$ Chern number planes have a negative Chern number for $k_y<0$ and positive Chern number for $k_y>0.$ For example, in our model we can generate a configuration with this structure using $m=-2$, $t=1.$ 

\subsubsection{Response to flux and dislocation lines}
We will begin by studying the momentum density bound to magnetic flux and charge density bound to dislocations.  These two responses, some aspects of which are described in Ref.~\cite{dbh2021} (see also Refs.~\onlinecite{gioia2021, hirsbrunner2023}),  are the most straightforward because they are essentially bulk responses and do not generate anomalous currents, i.e., the RHS of the anomalous conservation laws above will vanish. Our model has $\mathcal{Q}_{xx}=-\mathcal{Q}_{yy}\neq 0,$ and the responses generated by these two coefficients give two separate sets of terms in the response action. Hence, for simplicity we consider only the $\mathcal{Q}_{xx}$ responses for now. 

Let us first microscopically calculate the expected response to inserting a magnetic flux or a screw dislocation and compare with the response theory. First, consider inserting a thin magnetic flux line along the $x$-direction having flux $\Phi$ localized at, say $(y,z)=(0,0).$  This flux will generate a Hall effect from each of the non-trivial $k_y k_z$ Chern planes. The total charge bound to the flux line will vanish because there are equal and opposite contributions from $k_x<0$ and $k_x>0.$ However, threading the flux will build up a non-vanishing $k_x$-momentum since planes with opposite $k_x$-momentum have opposite Chern number. The total momentum (spatial integral of momentum density) driven to the flux line by the Hall effect at each $k_x$ momentum is
\begin{equation}
   \Delta P_x=-\frac{\Phi}{\Phi_0}\frac{L_x}{2\pi}\int_{-\pi}^{\pi}C(k_x) \hbar k_x dk_x=\frac{\Phi}{\Phi_0}\frac{\hbar K^2 L_x}{2\pi},
\end{equation} where  the Chern number $C(k_x)$ is the piecewise-constant function across the $k_x$ BZ described above , and $\Phi_0=h/e$ is the quantum of magnetic flux. Using the fact that $\mathcal{Q}_{xx}=2K^2$ and dividing by the volume we find the momentum density
\begin{equation}
    \mathcal{J}_x^0=\frac{e\mathcal{Q}_{xx}}{8\pi^2}B_x.
\end{equation} This is the same result coming from the first term in Eq.~\ref{Eq:QuadCurrents} when $\mathfrak{e}_x^x=1.$

Next let us calculate the charge response to inserting dislocations. Consider a screw dislocation with Burgers vector component $b^x$ associated to a translation gauge field configuration $\mathcal{B}_x^x\equiv \partial_y \mathfrak{e}_z^x-\partial_z \mathfrak{e}_y^x =b^x\delta(y)\delta(z).$ From Eqs.~\ref{Eq:QuadCurrents} and \ref{Eq:QuadChargeCurrents} we see that both the momentum and charge currents have responses to dislocations, and we will first calculate the charge response. Heuristically the dislocation is like a $U(1)$ gauge flux that couples to momentum instead of electric charge, so the dislocation couples to $k_x$ momentum because it has a non-vanishing $b^x$. Hence each $k_y k_z$-plane having non-vanishing Chern number (and non-vanishing $k_x$) will generate a Hall response, but with a magnitude proportional to its $k_x$ charge. Indeed, each plane sees an effective flux $\Phi(k_x)=\frac{k_x b^x}{2\pi}\Phi_0.$ Hence, the total charge bound to the dislocation will be
\begin{equation}
   \Delta Q=\frac{e L_x}{2\pi}\int_{-\pi}^{\pi}\frac{k_x b_x}{2\pi}C(k_x)dk_x=-\frac{e b_x \mathcal{Q}_{xx}}{8\pi^2}L_x.\label{eq:chargedislocationquad}
\end{equation} This matches Eq.~\ref{Eq:QuadChargeCurrents}, again after setting $\mathfrak{e}_x^x=1$ (see also Refs.~\onlinecite{gioia2021, dbh2021, hirsbrunner2023}).

 Now we consider the momentum response to a dislocation, i.e., a momentum density bound to the dislocation when $A_x$ is non-vanishing (this comes from the second term in Eq.~\ref{Eq:QuadCurrents}). First we can compute the amount of momentum bound to a dislocation when $A_x=0$ by adding the contributions of each Chern plane:
\begin{equation}
\begin{aligned}
\Delta P_x &= \frac{L_x}{2\pi}\int_{-\pi}^{\pi} \frac{k_x b_x}{2\pi} C(k_x) \hbar k_x dk_x \\
&= \frac{L_x b_x \hbar}{4\pi^2}\left(\int_0^K k_x^2 dk_x-\int_{-K}^0 k_x^2 dk_x\right) \\
&= 0.
\end{aligned}
\end{equation} We note that this calculation is similar to Eq.~\ref{eq:chargedislocationquad} except with an additional factor of the ``momentum-charge" $\hbar k_x$ in the integrand. Now if we turn on an $A_x$ such that $k_x\to k_x+\tfrac{e}{\hbar}A_x,$ we can repeat the calculation to find
\begin{equation}
\begin{aligned}
\Delta P_x\vert_{A_x} &=\frac{L_x b_x \hbar}{4\pi^2}\left(\int_{-\tfrac{eA_x}{\hbar}}^{K-\tfrac{e A_x}{\hbar}} k_x^2 dk_x-\int_{-K-\tfrac{eA_x}{\hbar}}^{-\tfrac{eA_x}{\hbar}} k_x^2 dk_x\right) \\ 
&=-\frac{e L_x b_x 2K^2}{4\pi^2}A_x.
\end{aligned}
\end{equation} The final result yields
\begin{equation}
\mathcal{J}_x^0=-\frac{e \mathcal{Q}_{xx} A_x}{4\pi^2}\mathcal{B}_x^x,
\end{equation} which matches Eq.~\ref{Eq:QuadCurrents} and our numerical calculations in Figs.~\ref{fig:3D_quad}(c) and (d). For the numerics we inserted a pair of screw dislocations with burgers vectors $b^x=\pm a_x$ in the presence of a constant background gauge potential $A_x$. The resulting $k_{x}$ momentum density of the ground state as a function of the $y$ and $z$ lattice coordinates is shown in Fig.~\ref{fig:3D_quad} (c). Furthermore, the dependence of this momentum density on $A_x$ reproduces the expected response coefficient, as shown in Fig.~\ref{fig:3D_quad} (d).

\begin{figure}
    \centering
    \includegraphics[width=0.5\textwidth]{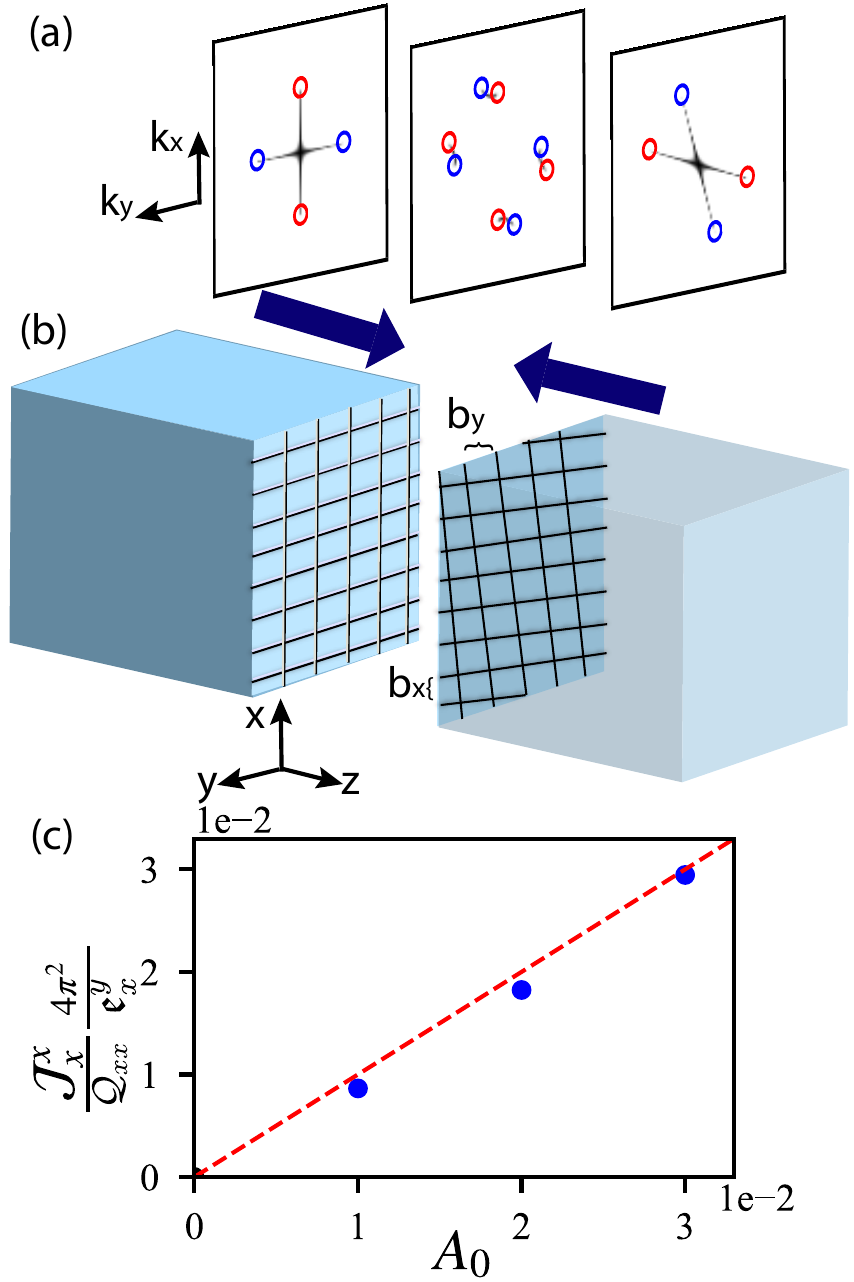}
    \caption{(a) The three panels show numerically calculated Fermi arcs in (left) the surface BZ of the un-deformed geometry, (right) the surface BZ of the deformed geometry with $\mathfrak{e}_y^x$ and $e_x^y$ are non-vanishing, and (center) the BZ of the interface formed gluing the deformed and un-deformed geometries together. See the caption in Fig.~\ref{fig:twistedinterface} for comments about the color guides on the open circles representing the surface BZ projections of the bulk Weyl nodes. (b) Illustrations of (left) un-deformed and (right) deformed geometries. (c) The numerically calculated momentum current localized at interface between deformed and un-deformed geometries as a function of the chemical potential shift $A_0$.}
    \label{fig:twistedinterfacequad}
\end{figure}

\subsubsection{Response of a deformed interface}
Next let us consider an interface between an un-deformed geometry and a geometry having a non-vanishing background $\mathfrak{e}_x^y$ and $\mathfrak{e}_y^x$ as shown in Fig.~\ref{fig:twistedinterfacequad}(b). To be explicit, let the interface between the two geometries occur as a function of $z$ at $z=0.$ On the surface of the un-deformed system we numerically calculated the characteristic (rank-2) Fermi arc structure as shown in the left surface-BZ panel in Fig.~\ref{fig:twistedinterfacequad}(a). For our deformed geometry we show the modified bulk Weyl node quadrupole and Fermi arcs when $\mathfrak{e}_x^y=\mathfrak{e}_y^x\neq 0$ in the right surface-BZ panel in Fig.~\ref{fig:twistedinterfacequad}(a). 

From these figures we see that the Weyl node quadrupole moment $\mathcal{Q}^{(R)}_{ab}$ on the deformed side is modified from the quadrupole moment $\mathcal{Q}_{ab}^{(L)}$ on the un-deformed side. Explicitly, we can compute:
\begin{equation}
\begin{aligned}
\mathcal{Q}_{xx}^{(R)}&=(\mathfrak{e}_x^x)^2 \mathcal{Q}_{xx}^{(L)}+2 \mathfrak{e}_x^x \mathfrak{e}_x^y \mathcal{Q}_{xy}^{(L)} +(\mathfrak{e}_x^y)^2 \mathcal{Q}_{yy}^{(L)} \\ 
\mathcal{Q}_{xy}^{(R)}&=\mathfrak{e}_x^x \mathfrak{e}_y^x \mathcal{Q}_{xx}^{(L)}+\mathfrak{e}_x^y \mathfrak{e}_y^y \mathcal{Q}_{yy}^{(L)}+ (\mathfrak{e}_x^x \mathfrak{e}_y^y +\mathfrak{e}_x^y \mathfrak{e}_y^x )\mathcal{Q}_{xy}^{(L)}\\
\mathcal{Q}_{yy}^{(R)}&=(\mathfrak{e}_y^x)^2 \mathcal{Q}_{xx}^{(L)}+2 \mathfrak{e}_y^x \mathfrak{e}_y^y \mathcal{Q}_{xy}^{(L)} +(\mathfrak{e}_y^y)^2 \mathcal{Q}_{yy}^{(L)},
\label{eq:deformquad}
\end{aligned}
\end{equation}
i.e., $\mathcal{Q}_{ij}^{(R)}=\mathfrak{e}_i^a Q_{ab}^{(L)}\mathfrak{e}_j^b.$ For our model and geometry we can make the simplifications $\mathfrak{e}_x^x=1=\mathfrak{e}_y^y, \mathfrak{e}_x^y=\mathfrak{e}_y^x, \mathcal{Q}_{xy}^{(L)}=0$, and  $\mathcal{Q}_{xx}^{(L)}=2K^2=-\mathcal{Q}_{yy}^{(L)}.$ Substituting these relations into Eq.~\ref{eq:deformquad} yields
 \begin{equation}
     \mathcal{Q}_{xx}^{(R)}=-\mathcal{Q}_{yy}^{(R)}=2K^2(1-(\mathfrak{e}_x^y)^2),
 \end{equation}
 and $Q_{xy}^{(R)}=0.$ Alternatively, we can see this result from the locations of the deformed Weyl nodes which will sit at $(K,K\mathfrak{e}_y^x,0)_{+}$, $(-K,-K\mathfrak{e}_y^x,0)_{+}$, $(K\mathfrak{e}_x^y,K,0)_{-}$, and $(-K\mathfrak{e}_x^y,-K,0)_{-}$ (where the subscripts $\pm$ encode the chirality for our choice of model parameters).

 Since the Weyl node quadrupole moments on the two sides of the interface are different, we expect gluing the two sides together will leave behind a signature at the interface. Indeed, from the middle surface-BZ panel in Fig.~\ref{fig:twistedinterfacequad}(a) we see gapless Fermi arcs that remain at the interface and stretch between the unmodified and modified projected locations of the bulk Weyl nodes. From Eqs.~\ref{Eq:QuadCurrents},~\ref{Eq:QuadChargeCurrents} we see there should be responses
 \begin{gather*}
     \mathcal{J}^x_x=-\frac{e}{4\pi^2}\mathcal{Q}_{xx}A_0\partial_z \mathfrak{e}_y^x,
     ,\quad
     \mathcal{J}^y_y=-\frac{e}{4\pi^2}\mathcal{Q}_{yy}A_0\partial_z \mathfrak{e}_x^y, \\ 
     j^0=\frac{e}{8\pi^2}\left(\mathcal{Q}_{xx}\mathfrak{e}_x^x\partial_z \mathfrak{e}_y^x-\mathcal{Q}_{yy}\mathfrak{e}_y^y\partial_z\mathfrak{e}_x^y\right)=\frac{e\mathcal{Q}_{xx}}{4\pi^2}\partial_z \mathfrak{e}_x^y,
 \end{gather*}
 where in the last equality we substituted in the relations that are specific to our model and interface geometry, which we stated above. 
 
 We confirmed the momentum and charge responses numerically, in particular the $\mathcal{J}_x^x$ response shown in Fig.~\ref{fig:twistedinterfacequad}(c), and we also provide microscopic analytic arguments here. The momentum currents both follow the same logic, so let us consider only $\mathcal{J}_x^x$ for now. From the center surface-BZ panel in Fig.~\ref{fig:twistedinterfacequad}(a) we see remnant Fermi arcs. If we increase $A_0,$ each linearly-dispersing point on the Fermi arc will have an excess charge density $\delta n({\bf{k}})=\frac{e A_0}{2\pi\hbar |v_F({\bf{k}})|}$ where $v_F({\bf{k}})$ is the Fermi velocity at the Fermi arc located at ${\bf{k}}$ in the surface-BZ. Hence, the contribution to the $k_x$ momentum current of such a point on the Fermi arc is $\mathcal{J}_x^x({\bf{k}})=\hbar k_x v_{F}({\bf{k}}) \delta n({\bf{k}}).$ For our model and geometry, the contributions to the $\mathcal{J}_x^x$ current that are linear in the deformations of $\mathfrak{e}_i^a$ arise from the Fermi arcs stretching between $(K,0)\to(K,K\mathfrak{e}_y^x)$ and $(-K,0)\to(-K,-K\mathfrak{e}_y^x).$ Each of these arcs has a fixed value $k_x=\pm K$ and each arc has an opposite Fermi velocity. Hence
\begin{equation}
\begin{aligned}
\mathcal{J}_x^x&=\hbar K v_F(K,k_y) \delta n \frac{K\mathfrak{e}_y^x}{2\pi}\nonumber \\
&+\hbar (-K) v_F(-K,k_y) \delta n \frac{K\mathfrak{e}_y^x}{2\pi}\nonumber\\
&=\frac{e \mathcal{Q}_{xx}^{(L)}\mathfrak{e}_y^x A_0}{4\pi^2}{\rm{sgn}}(v_F),
\end{aligned}
\end{equation}
 where $K\mathfrak{e}_y^x/2\pi$ counts the density of states on the Fermi arc in the $k_y$ direction, ${\rm{sgn}}(v_F)$ is sign of the velocity on the $k_x=+K$ arc, and the un-deformed $\mathcal{Q}_{xx}^{(L)}=2K^2.$ This result matches the prediction from the response theory and matches the numerical results in Fig.~\ref{fig:twistedinterfacequad}(c). 

  The calculation of the charge density $j^0$ at the interface is simpler since it comes from the bulk response to a translation magnetic field. At the interface there is a non-vanishing $\mathcal{B}_x^x=-\partial_z \mathfrak{e}_y^x$ and $\mathcal{B}_y^y=\partial_z \mathfrak{e}_x^y.$ Since the $k_y k_z$-planes and $k_x k_z$-planes have non-vanishing Chern numbers, they yield a density response similar to what we found on the dislocation line in Eq.~\ref{eq:chargedislocationquad}. Each $k_x$ state sees an effective magnetic flux $\Phi(k_x)=-\frac{k_x b^x}{2\pi}\Phi_0,$ and similarly for each $k_y$ state $\Phi(k_y)=\frac{k_y b^y}{2\pi}\Phi_0,$ where $b^{x}=\int dy \mathfrak{e}_y^x\vert_{z>0}$ and $b^y=\int dx \mathfrak{e}_x^y\vert_{z>0}$ are the Burgers vectors obtained when integrating across the entire periodic $y$- and $x$-directions respectively. Hence the total charge at the interface is
 \begin{equation}
 \begin{aligned}
     \Delta Q&=-\frac{2e L_x}{2\pi}\int_0^K \frac{k_x b^x}{2\pi} dk_x+\frac{2e L_y}{2\pi}\int_0^K \frac{k_y b^y}{2\pi} dk_y \\
     &=\frac{e}{8\pi^2}\left(-\mathcal{Q}_{xx}b^x L_x+ \mathcal{Q}_{yy}b^y L_y\right) \\
     &=-\frac{e\mathcal{Q}_{xx}b^x L_x}{4\pi^2},
 \end{aligned}
 \end{equation}
 where the leading factors of two in the first line account for identical contributions from the interval $k_x\in [-K,0],$ and in the last equation we used $\mathcal{Q}_{xx}=-\mathcal{Q}_{yy}$ and $L_x b^x=L_y b^y$ since $\mathfrak{e}_x^y=\mathfrak{e}_y^x.$ This final result matches  Eq.~\ref{Eq:QuadChargeCurrents}.

\section{Conclusion}
\label{sec:conclusion}
In this article we have presented a framework of explicit connections between a wide-ranging family of topological response theories from 0D to 3D.  Using this framework, we have shown how the coefficients for these response theories, most of which are well-known in insulators, can be obtained for topological semimetals.  This has allowed us to provide careful derivations and characterizations of mixed crystalline-electromagnetic responses of semimetallic  and insulating systems in various spatial dimensions. Finally, we have provided an extensive set of microscopic lattice calculations and numerical confirmations affirming that our predicted field theory responses do indeed arise in tight binding lattice models.
With the advent of topological quantum chemistry~\cite{vergniory2017graph, bradlyn2017topological, cano2018building, cano2018topology, bradlyn2018band, elcoro2021magnetic}, thousands of crystalline topological insulators and semimetals have been identified, but many open questions persist about how to probe their topological features. This work provides insight into how the topology in some of these materials may be probed and characterized, i.e., by combining geometric/strain distortions and electromagnetic responses.   

There is a growing body of work studying the mixed crystalline-electromagnetic responses of Weyl semimetals with dipole and quadrupole arrangements of nodes~\cite{dbh2021,gioia2021,hirsbrunner2023,Hiroaki2016,Kodama2019,Huang19,Huang20a,Huang20b,Liang20,pikulin2016chiral,you2016,dislocation_defect_2020,laurila_torsional_2020,chongwang2,jakko,gao_chiral_2021,amitani_torsion_2023,nissinen2018tetrads,nissinen2019elasticity,ferreiros2019mixed,chu2023chiral}, that indicate a broad interest in these topics.
Our work serves two major purposes in the context of this previous literature: (i) we identified several aspects of mixed crystalline-electromagnetic responses that have not yet been addressed in earlier work, and (ii) we synthesized aspects of the existing literature to present a unified description of these responses in terms of the momentum-space multipole moments of the nodal configurations, and to provide new intuition in previously studied responses. While prior work has examined the mixed crystalline-electromagnetic response of two-dimensional Dirac node dipole semimetals~\cite{RamamurthyPatterns,chongwangvishwanath}, we have advanced this understanding by identifying a Wilson loop correction the response coefficient that raises a subtle question about the connection between the charge polarization and the mixed-crystalline-electromagnetic response. Additionally, the Dirac node quadrupole semimetal has not been previously discussed, making our work the first study of its properties and mixed crystalline-electromagnetic responses. Furthermore, our model of a nodal line quadrupole semimetal and its corresponding response theory are new to the literature as well.

The results of this work point in many possible directions for future work. First, finding experimental realizations of the proposed topological responses in solid state or metamaterial systems is an exciting prospect. Rank-2 chiral fermions, which have an anomaly compensated by the bulk response of a Weyl quadrupole semimetal~\cite{dbh2021}, were realized in a recent experiment on non-Hermitian topo-electric circuit metamaterials~\cite{zhu2023higher}. In that platform, the mixed crystalline-electromagnetic response generates a momentum-resolved non-Hermitian skin effect that was observed in the experiment. Topo-electric circuits, along with other metamaterials and solid state platforms are promising arenas in which the many mixed crystalline-electromagnetic responses we discuss in this paper could be realized. Other extensions of this work include the consideration of additional crystalline gauge fields as was done in, e.g., Refs.~\onlinecite{manjunath2020classification, barkeshli2021,zhang2022fractional,gioia2021,may2022crystalline,may2023topological}. Some of us are also working on extending the nodal, higher-multipole responses to interacting systems and non-equilibrium systems where, in the latter, one can have mixed energy-momentum multipole moments. Studying the leading nodal dipole moments has already led to a rich set of phenomena, and the higher moments provide a large hierarchy of phenomena that can be explored in current experiments.


\begin{thebibliography}{100}%
\makeatletter
\providecommand \@ifxundefined [1]{%
 \@ifx{#1\undefined}
}%
\providecommand \@ifnum [1]{%
 \ifnum #1\expandafter \@firstoftwo
 \else \expandafter \@secondoftwo
 \fi
}%
\providecommand \@ifx [1]{%
 \ifx #1\expandafter \@firstoftwo
 \else \expandafter \@secondoftwo
 \fi
}%
\providecommand \natexlab [1]{#1}%
\providecommand \enquote  [1]{``#1''}%
\providecommand \bibnamefont  [1]{#1}%
\providecommand \bibfnamefont [1]{#1}%
\providecommand \citenamefont [1]{#1}%
\providecommand \href@noop [0]{\@secondoftwo}%
\providecommand \href [0]{\begingroup \@sanitize@url \@href}%
\providecommand \@href[1]{\@@startlink{#1}\@@href}%
\providecommand \@@href[1]{\endgroup#1\@@endlink}%
\providecommand \@sanitize@url [0]{\catcode `\\12\catcode `\$12\catcode
  `\&12\catcode `\#12\catcode `\^12\catcode `\_12\catcode `\%12\relax}%
\providecommand \@@startlink[1]{}%
\providecommand \@@endlink[0]{}%
\providecommand \url  [0]{\begingroup\@sanitize@url \@url }%
\providecommand \@url [1]{\endgroup\@href {#1}{\urlprefix }}%
\providecommand \urlprefix  [0]{URL }%
\providecommand \Eprint [0]{\href }%
\providecommand \doibase [0]{https://doi.org/}%
\providecommand \selectlanguage [0]{\@gobble}%
\providecommand \bibinfo  [0]{\@secondoftwo}%
\providecommand \bibfield  [0]{\@secondoftwo}%
\providecommand \translation [1]{[#1]}%
\providecommand \BibitemOpen [0]{}%
\providecommand \bibitemStop [0]{}%
\providecommand \bibitemNoStop [0]{.\EOS\space}%
\providecommand \EOS [0]{\spacefactor3000\relax}%
\providecommand \BibitemShut  [1]{\csname bibitem#1\endcsname}%
\let\auto@bib@innerbib\@empty
\bibitem [{\citenamefont {Halperin}(1982)}]{iqhe}%
  \BibitemOpen
  \bibfield  {author} {\bibinfo {author} {\bibfnamefont {B.~I.}\ \bibnamefont
  {Halperin}},\ }\bibfield  {title} {\bibinfo {title} {Quantized hall
  conductance, current-carrying edge states, and the existence of extended
  states in a two-dimensional disordered potential},\ }\href
  {https://doi.org/10.1103/PhysRevB.25.2185} {\bibfield  {journal} {\bibinfo
  {journal} {Phys. Rev. B}\ }\textbf {\bibinfo {volume} {25}},\ \bibinfo
  {pages} {2185} (\bibinfo {year} {1982})}\BibitemShut {NoStop}%
\bibitem [{\citenamefont {Laughlin}(1981)}]{laughlin}%
  \BibitemOpen
  \bibfield  {author} {\bibinfo {author} {\bibfnamefont {R.~B.}\ \bibnamefont
  {Laughlin}},\ }\bibfield  {title} {\bibinfo {title} {Quantized hall
  conductivity in two dimensions},\ }\href
  {https://doi.org/10.1103/PhysRevB.23.5632} {\bibfield  {journal} {\bibinfo
  {journal} {Phys. Rev. B}\ }\textbf {\bibinfo {volume} {23}},\ \bibinfo
  {pages} {5632} (\bibinfo {year} {1981})}\BibitemShut {NoStop}%
\bibitem [{\citenamefont {Thouless}\ \emph {et~al.}(1982)\citenamefont
  {Thouless}, \citenamefont {Kohmoto}, \citenamefont {Nightingale},\ and\
  \citenamefont {den Nijs}}]{tknn}%
  \BibitemOpen
  \bibfield  {author} {\bibinfo {author} {\bibfnamefont {D.~J.}\ \bibnamefont
  {Thouless}}, \bibinfo {author} {\bibfnamefont {M.}~\bibnamefont {Kohmoto}},
  \bibinfo {author} {\bibfnamefont {M.~P.}\ \bibnamefont {Nightingale}},\ and\
  \bibinfo {author} {\bibfnamefont {M.}~\bibnamefont {den Nijs}},\ }\bibfield
  {title} {\bibinfo {title} {Quantized hall conductance in a two-dimensional
  periodic potential},\ }\href {https://doi.org/10.1103/PhysRevLett.49.405}
  {\bibfield  {journal} {\bibinfo  {journal} {Phys. Rev. Lett.}\ }\textbf
  {\bibinfo {volume} {49}},\ \bibinfo {pages} {405} (\bibinfo {year}
  {1982})}\BibitemShut {NoStop}%
\bibitem [{\citenamefont {Wilczek}(1987)}]{wilczek}%
  \BibitemOpen
  \bibfield  {author} {\bibinfo {author} {\bibfnamefont {F.}~\bibnamefont
  {Wilczek}},\ }\bibfield  {title} {\bibinfo {title} {Two applications of axion
  electrodynamics},\ }\href {https://doi.org/10.1103/PhysRevLett.58.1799}
  {\bibfield  {journal} {\bibinfo  {journal} {Phys. Rev. Lett.}\ }\textbf
  {\bibinfo {volume} {58}},\ \bibinfo {pages} {1799} (\bibinfo {year}
  {1987})}\BibitemShut {NoStop}%
\bibitem [{\citenamefont {Qi}\ \emph {et~al.}(2008)\citenamefont {Qi},
  \citenamefont {Hughes},\ and\ \citenamefont {Zhang}}]{qi2008}%
  \BibitemOpen
  \bibfield  {author} {\bibinfo {author} {\bibfnamefont {X.-L.}\ \bibnamefont
  {Qi}}, \bibinfo {author} {\bibfnamefont {T.~L.}\ \bibnamefont {Hughes}},\
  and\ \bibinfo {author} {\bibfnamefont {S.-C.}\ \bibnamefont {Zhang}},\
  }\bibfield  {title} {\bibinfo {title} {Topological field theory of
  time-reversal invariant insulators},\ }\href
  {https://doi.org/10.1103/PhysRevB.78.195424} {\bibfield  {journal} {\bibinfo
  {journal} {Phys. Rev. B}\ }\textbf {\bibinfo {volume} {78}},\ \bibinfo
  {pages} {195424} (\bibinfo {year} {2008})}\BibitemShut {NoStop}%
\bibitem [{\citenamefont {Luttinger}(1964)}]{luttinger195x}%
  \BibitemOpen
  \bibfield  {author} {\bibinfo {author} {\bibfnamefont {J.~M.}\ \bibnamefont
  {Luttinger}},\ }\bibfield  {title} {\bibinfo {title} {Theory of thermal
  transport coefficients},\ }\href {https://doi.org/10.1103/PhysRev.135.A1505}
  {\bibfield  {journal} {\bibinfo  {journal} {Phys. Rev.}\ }\textbf {\bibinfo
  {volume} {135}},\ \bibinfo {pages} {A1505} (\bibinfo {year}
  {1964})}\BibitemShut {NoStop}%
\bibitem [{\citenamefont {Kapustin}\ and\ \citenamefont
  {Spodyneiko}(2020)}]{chiralcentralcharge}%
  \BibitemOpen
  \bibfield  {author} {\bibinfo {author} {\bibfnamefont {A.}~\bibnamefont
  {Kapustin}}\ and\ \bibinfo {author} {\bibfnamefont {L.}~\bibnamefont
  {Spodyneiko}},\ }\bibfield  {title} {\bibinfo {title} {Thermal hall
  conductance and a relative topological invariant of gapped two-dimensional
  systems},\ }\href {https://doi.org/10.1103/PhysRevB.101.045137} {\bibfield
  {journal} {\bibinfo  {journal} {Phys. Rev. B}\ }\textbf {\bibinfo {volume}
  {101}},\ \bibinfo {pages} {045137} (\bibinfo {year} {2020})}\BibitemShut
  {NoStop}%
\bibitem [{\citenamefont {Avron}\ \emph {et~al.}(1995)\citenamefont {Avron},
  \citenamefont {Seiler},\ and\ \citenamefont {Zograf}}]{avron1995}%
  \BibitemOpen
  \bibfield  {author} {\bibinfo {author} {\bibfnamefont {J.~E.}\ \bibnamefont
  {Avron}}, \bibinfo {author} {\bibfnamefont {R.}~\bibnamefont {Seiler}},\ and\
  \bibinfo {author} {\bibfnamefont {P.~G.}\ \bibnamefont {Zograf}},\ }\bibfield
   {title} {\bibinfo {title} {Viscosity of quantum hall fluids},\ }\href
  {https://doi.org/10.1103/PhysRevLett.75.697} {\bibfield  {journal} {\bibinfo
  {journal} {Phys. Rev. Lett.}\ }\textbf {\bibinfo {volume} {75}},\ \bibinfo
  {pages} {697} (\bibinfo {year} {1995})}\BibitemShut {NoStop}%
\bibitem [{\citenamefont {Read}(2009)}]{read2008}%
  \BibitemOpen
  \bibfield  {author} {\bibinfo {author} {\bibfnamefont {N.}~\bibnamefont
  {Read}},\ }\bibfield  {title} {\bibinfo {title} {Non-abelian adiabatic
  statistics and hall viscosity in quantum hall states and ${p}_{x}+i{p}_{y}$
  paired superfluids},\ }\href {https://doi.org/10.1103/PhysRevB.79.045308}
  {\bibfield  {journal} {\bibinfo  {journal} {Phys. Rev. B}\ }\textbf {\bibinfo
  {volume} {79}},\ \bibinfo {pages} {045308} (\bibinfo {year}
  {2009})}\BibitemShut {NoStop}%
\bibitem [{\citenamefont {Sherafati}\ \emph {et~al.}(2016)\citenamefont
  {Sherafati}, \citenamefont {Principi},\ and\ \citenamefont
  {Vignale}}]{vignaleXXX}%
  \BibitemOpen
  \bibfield  {author} {\bibinfo {author} {\bibfnamefont {M.}~\bibnamefont
  {Sherafati}}, \bibinfo {author} {\bibfnamefont {A.}~\bibnamefont
  {Principi}},\ and\ \bibinfo {author} {\bibfnamefont {G.}~\bibnamefont
  {Vignale}},\ }\bibfield  {title} {\bibinfo {title} {Hall viscosity and
  electromagnetic response of electrons in graphene},\ }\href
  {https://doi.org/10.1103/PhysRevB.94.125427} {\bibfield  {journal} {\bibinfo
  {journal} {Phys. Rev. B}\ }\textbf {\bibinfo {volume} {94}},\ \bibinfo
  {pages} {125427} (\bibinfo {year} {2016})}\BibitemShut {NoStop}%
\bibitem [{\citenamefont {Hughes}\ \emph
  {et~al.}(2011{\natexlab{a}})\citenamefont {Hughes}, \citenamefont {Leigh},\
  and\ \citenamefont {Fradkin}}]{hughesleighfradkin}%
  \BibitemOpen
  \bibfield  {author} {\bibinfo {author} {\bibfnamefont {T.~L.}\ \bibnamefont
  {Hughes}}, \bibinfo {author} {\bibfnamefont {R.~G.}\ \bibnamefont {Leigh}},\
  and\ \bibinfo {author} {\bibfnamefont {E.}~\bibnamefont {Fradkin}},\
  }\bibfield  {title} {\bibinfo {title} {Torsional response and dissipationless
  viscosity in topological insulators},\ }\href
  {https://doi.org/10.1103/PhysRevLett.107.075502} {\bibfield  {journal}
  {\bibinfo  {journal} {Phys. Rev. Lett.}\ }\textbf {\bibinfo {volume} {107}},\
  \bibinfo {pages} {075502} (\bibinfo {year} {2011}{\natexlab{a}})}\BibitemShut
  {NoStop}%
\bibitem [{\citenamefont {You}\ \emph {et~al.}(2016)\citenamefont {You},
  \citenamefont {Cho},\ and\ \citenamefont {Hughes}}]{you2016}%
  \BibitemOpen
  \bibfield  {author} {\bibinfo {author} {\bibfnamefont {Y.}~\bibnamefont
  {You}}, \bibinfo {author} {\bibfnamefont {G.~Y.}\ \bibnamefont {Cho}},\ and\
  \bibinfo {author} {\bibfnamefont {T.~L.}\ \bibnamefont {Hughes}},\ }\bibfield
   {title} {\bibinfo {title} {Response properties of axion insulators and weyl
  semimetals driven by screw dislocations and dynamical axion strings},\
  }\href@noop {} {\bibfield  {journal} {\bibinfo  {journal} {Physical Review
  B}\ }\textbf {\bibinfo {volume} {94}},\ \bibinfo {pages} {085102} (\bibinfo
  {year} {2016})}\BibitemShut {NoStop}%
\bibitem [{\citenamefont {Sumiyoshi}\ and\ \citenamefont
  {Fujimoto}(2016)}]{Hiroaki2016}%
  \BibitemOpen
  \bibfield  {author} {\bibinfo {author} {\bibfnamefont {H.}~\bibnamefont
  {Sumiyoshi}}\ and\ \bibinfo {author} {\bibfnamefont {S.}~\bibnamefont
  {Fujimoto}},\ }\bibfield  {title} {\bibinfo {title} {Torsional chiral
  magnetic effect in a weyl semimetal with a topological defect},\ }\href
  {https://doi.org/10.1103/PhysRevLett.116.166601} {\bibfield  {journal}
  {\bibinfo  {journal} {Phys. Rev. Lett.}\ }\textbf {\bibinfo {volume} {116}},\
  \bibinfo {pages} {166601} (\bibinfo {year} {2016})}\BibitemShut {NoStop}%
\bibitem [{\citenamefont {Teo}\ and\ \citenamefont
  {Hughes}(2017)}]{teoTopologicalDefectsSymmetryProtected2017a}%
  \BibitemOpen
  \bibfield  {author} {\bibinfo {author} {\bibfnamefont {J.~C.}\ \bibnamefont
  {Teo}}\ and\ \bibinfo {author} {\bibfnamefont {T.~L.}\ \bibnamefont
  {Hughes}},\ }\bibfield  {title} {\bibinfo {title} {Topological {{Defects}} in
  {{Symmetry-Protected Topological Phases}}},\ }\href
  {https://doi.org/10.1146/annurev-conmatphys-031016-025154} {\bibfield
  {journal} {\bibinfo  {journal} {Annual Review of Condensed Matter Physics}\
  }\textbf {\bibinfo {volume} {8}},\ \bibinfo {pages} {211} (\bibinfo {year}
  {2017})}\BibitemShut {NoStop}%
\bibitem [{\citenamefont {Soto-Garrido}\ \emph {et~al.}(2020)\citenamefont
  {Soto-Garrido}, \citenamefont {Mu\~noz},\ and\ \citenamefont {Juri\ifmmode
  \check{c}\else \v{c}\fi{}i\ifmmode~\acute{c}\else
  \'{c}\fi{}}}]{dislocation_defect_2020}%
  \BibitemOpen
  \bibfield  {author} {\bibinfo {author} {\bibfnamefont {R.}~\bibnamefont
  {Soto-Garrido}}, \bibinfo {author} {\bibfnamefont {E.}~\bibnamefont
  {Mu\~noz}},\ and\ \bibinfo {author} {\bibfnamefont {V.}~\bibnamefont
  {Juri\ifmmode \check{c}\else \v{c}\fi{}i\ifmmode~\acute{c}\else
  \'{c}\fi{}}},\ }\bibfield  {title} {\bibinfo {title} {Dislocation defect as a
  bulk probe of monopole charge of multi-weyl semimetals},\ }\href
  {https://doi.org/10.1103/PhysRevResearch.2.012043} {\bibfield  {journal}
  {\bibinfo  {journal} {Phys. Rev. Res.}\ }\textbf {\bibinfo {volume} {2}},\
  \bibinfo {pages} {012043} (\bibinfo {year} {2020})}\BibitemShut {NoStop}%
\bibitem [{\citenamefont {Laurila}\ and\ \citenamefont
  {Nissinen}(2020)}]{laurila_torsional_2020}%
  \BibitemOpen
  \bibfield  {author} {\bibinfo {author} {\bibfnamefont {S.}~\bibnamefont
  {Laurila}}\ and\ \bibinfo {author} {\bibfnamefont {J.}~\bibnamefont
  {Nissinen}},\ }\bibfield  {title} {\bibinfo {title} {Torsional landau levels
  and geometric anomalies in condensed matter weyl systems},\ }\href
  {https://doi.org/10.1103/PhysRevB.102.235163} {\bibfield  {journal} {\bibinfo
   {journal} {Phys. Rev. B}\ }\textbf {\bibinfo {volume} {102}},\ \bibinfo
  {pages} {235163} (\bibinfo {year} {2020})}\BibitemShut {NoStop}%
\bibitem [{\citenamefont {Dubinkin}\ \emph {et~al.}(2021)\citenamefont
  {Dubinkin}, \citenamefont {Burnell},\ and\ \citenamefont {Hughes}}]{dbh2021}%
  \BibitemOpen
  \bibfield  {author} {\bibinfo {author} {\bibfnamefont {O.}~\bibnamefont
  {Dubinkin}}, \bibinfo {author} {\bibfnamefont {F.}~\bibnamefont {Burnell}},\
  and\ \bibinfo {author} {\bibfnamefont {T.~L.}\ \bibnamefont {Hughes}},\
  }\bibfield  {title} {\bibinfo {title} {Higher rank chiral fermions in 3d weyl
  semimetals},\ }\href@noop {} {\bibfield  {journal} {\bibinfo  {journal}
  {arXiv preprint arXiv:2102.08959}\ } (\bibinfo {year} {2021})}\BibitemShut
  {NoStop}%
\bibitem [{\citenamefont {Gioia}\ \emph {et~al.}(2021)\citenamefont {Gioia},
  \citenamefont {Wang},\ and\ \citenamefont {Burkov}}]{gioia2021}%
  \BibitemOpen
  \bibfield  {author} {\bibinfo {author} {\bibfnamefont {L.}~\bibnamefont
  {Gioia}}, \bibinfo {author} {\bibfnamefont {C.}~\bibnamefont {Wang}},\ and\
  \bibinfo {author} {\bibfnamefont {A.}~\bibnamefont {Burkov}},\ }\bibfield
  {title} {\bibinfo {title} {Unquantized anomalies in topological semimetals},\
  }\href {https://doi.org/10.1103/PhysRevResearch.3.043067} {\bibfield
  {journal} {\bibinfo  {journal} {Physical Review Research}\ }\textbf {\bibinfo
  {volume} {3}},\ \bibinfo {pages} {043067} (\bibinfo {year}
  {2021})}\BibitemShut {NoStop}%
\bibitem [{\citenamefont {Wang}\ \emph {et~al.}(2021)\citenamefont {Wang},
  \citenamefont {Hickey}, \citenamefont {Ying},\ and\ \citenamefont
  {Burkov}}]{chongwang2}%
  \BibitemOpen
  \bibfield  {author} {\bibinfo {author} {\bibfnamefont {C.}~\bibnamefont
  {Wang}}, \bibinfo {author} {\bibfnamefont {A.}~\bibnamefont {Hickey}},
  \bibinfo {author} {\bibfnamefont {X.}~\bibnamefont {Ying}},\ and\ \bibinfo
  {author} {\bibfnamefont {A.~A.}\ \bibnamefont {Burkov}},\ }\bibfield  {title}
  {\bibinfo {title} {Emergent anomalies and generalized luttinger theorems in
  metals and semimetals},\ }\href {https://doi.org/10.1103/PhysRevB.104.235113}
  {\bibfield  {journal} {\bibinfo  {journal} {Phys. Rev. B}\ }\textbf {\bibinfo
  {volume} {104}},\ \bibinfo {pages} {235113} (\bibinfo {year}
  {2021})}\BibitemShut {NoStop}%
\bibitem [{\citenamefont {Nissinen}\ \emph {et~al.}(2021)\citenamefont
  {Nissinen}, \citenamefont {Heikkil\"a},\ and\ \citenamefont
  {Volovik}}]{jakko}%
  \BibitemOpen
  \bibfield  {author} {\bibinfo {author} {\bibfnamefont {J.}~\bibnamefont
  {Nissinen}}, \bibinfo {author} {\bibfnamefont {T.~T.}\ \bibnamefont
  {Heikkil\"a}},\ and\ \bibinfo {author} {\bibfnamefont {G.~E.}\ \bibnamefont
  {Volovik}},\ }\bibfield  {title} {\bibinfo {title} {Topological polarization,
  dual invariants, and surface flat bands in crystalline insulators},\ }\href
  {https://doi.org/10.1103/PhysRevB.103.245115} {\bibfield  {journal} {\bibinfo
   {journal} {Phys. Rev. B}\ }\textbf {\bibinfo {volume} {103}},\ \bibinfo
  {pages} {245115} (\bibinfo {year} {2021})}\BibitemShut {NoStop}%
\bibitem [{\citenamefont {Gao}\ \emph {et~al.}(2021)\citenamefont {Gao},
  \citenamefont {Kaushik}, \citenamefont {Kharzeev},\ and\ \citenamefont
  {Philip}}]{gao_chiral_2021}%
  \BibitemOpen
  \bibfield  {author} {\bibinfo {author} {\bibfnamefont {L.-L.}\ \bibnamefont
  {Gao}}, \bibinfo {author} {\bibfnamefont {S.}~\bibnamefont {Kaushik}},
  \bibinfo {author} {\bibfnamefont {D.~E.}\ \bibnamefont {Kharzeev}},\ and\
  \bibinfo {author} {\bibfnamefont {E.~J.}\ \bibnamefont {Philip}},\ }\bibfield
   {title} {\bibinfo {title} {Chiral kinetic theory of anomalous transport
  induced by torsion},\ }\href {https://doi.org/10.1103/PhysRevB.104.064307}
  {\bibfield  {journal} {\bibinfo  {journal} {Phys. Rev. B}\ }\textbf {\bibinfo
  {volume} {104}},\ \bibinfo {pages} {064307} (\bibinfo {year}
  {2021})}\BibitemShut {NoStop}%
\bibitem [{\citenamefont {You}\ \emph {et~al.}(2022)\citenamefont {You},
  \citenamefont {Bibo}, \citenamefont {Pollmann},\ and\ \citenamefont
  {Hughes}}]{you_fracton_2022}%
  \BibitemOpen
  \bibfield  {author} {\bibinfo {author} {\bibfnamefont {Y.}~\bibnamefont
  {You}}, \bibinfo {author} {\bibfnamefont {J.}~\bibnamefont {Bibo}}, \bibinfo
  {author} {\bibfnamefont {F.}~\bibnamefont {Pollmann}},\ and\ \bibinfo
  {author} {\bibfnamefont {T.~L.}\ \bibnamefont {Hughes}},\ }\bibfield  {title}
  {\bibinfo {title} {Fracton critical point at a higher-order topological phase
  transition},\ }\href {https://doi.org/10.1103/PhysRevB.106.235130} {\bibfield
   {journal} {\bibinfo  {journal} {Physical Review B}\ }\textbf {\bibinfo
  {volume} {106}},\ \bibinfo {pages} {235130} (\bibinfo {year}
  {2022})}\BibitemShut {NoStop}%
\bibitem [{\citenamefont {Amitani}\ and\ \citenamefont
  {Nishida}(2023)}]{amitani_torsion_2023}%
  \BibitemOpen
  \bibfield  {author} {\bibinfo {author} {\bibfnamefont {T.}~\bibnamefont
  {Amitani}}\ and\ \bibinfo {author} {\bibfnamefont {Y.}~\bibnamefont
  {Nishida}},\ }\bibfield  {title} {\bibinfo {title} {Torsion-induced chiral
  magnetic current in equilibrium},\ }\href
  {https://doi.org/10.1016/j.aop.2022.169181} {\bibfield  {journal} {\bibinfo
  {journal} {Annals of Physics}\ }\textbf {\bibinfo {volume} {448}},\ \bibinfo
  {pages} {169181} (\bibinfo {year} {2023})}\BibitemShut {NoStop}%
\bibitem [{\citenamefont {Hirsbrunner}\ \emph {et~al.}(2023)\citenamefont
  {Hirsbrunner}, \citenamefont {Gray},\ and\ \citenamefont
  {Hughes}}]{hirsbrunner2023}%
  \BibitemOpen
  \bibfield  {author} {\bibinfo {author} {\bibfnamefont {M.~R.}\ \bibnamefont
  {Hirsbrunner}}, \bibinfo {author} {\bibfnamefont {A.~D.}\ \bibnamefont
  {Gray}},\ and\ \bibinfo {author} {\bibfnamefont {T.~L.}\ \bibnamefont
  {Hughes}},\ }\href
  {https://doi.org/https://doi.org/10.48550/arXiv.2308.05796} {\bibinfo {title}
  {Crystalline-electromagnetic responses of higher order topological
  semimetals}} (\bibinfo {year} {2023}),\ \Eprint
  {https://arxiv.org/abs/2308.05796} {2308.05796 [cond-mat.mes-hall]}
  \BibitemShut {NoStop}%
\bibitem [{\citenamefont {Zak}(1989)}]{zak}%
  \BibitemOpen
  \bibfield  {author} {\bibinfo {author} {\bibfnamefont {J.}~\bibnamefont
  {Zak}},\ }\bibfield  {title} {\bibinfo {title} {Berry's phase for energy
  bands in solids},\ }\href {https://doi.org/10.1103/PhysRevLett.62.2747}
  {\bibfield  {journal} {\bibinfo  {journal} {Phys. Rev. Lett.}\ }\textbf
  {\bibinfo {volume} {62}},\ \bibinfo {pages} {2747} (\bibinfo {year}
  {1989})}\BibitemShut {NoStop}%
\bibitem [{\citenamefont {Vanderbilt}\ and\ \citenamefont
  {King-Smith}(1993)}]{vanderbilt1993}%
  \BibitemOpen
  \bibfield  {author} {\bibinfo {author} {\bibfnamefont {D.}~\bibnamefont
  {Vanderbilt}}\ and\ \bibinfo {author} {\bibfnamefont {R.~D.}\ \bibnamefont
  {King-Smith}},\ }\bibfield  {title} {\bibinfo {title} {Electric polarization
  as a bulk quantity and its relation to surface charge},\ }\href
  {https://doi.org/10.1103/PhysRevB.48.4442} {\bibfield  {journal} {\bibinfo
  {journal} {Phys. Rev. B}\ }\textbf {\bibinfo {volume} {48}},\ \bibinfo
  {pages} {4442} (\bibinfo {year} {1993})}\BibitemShut {NoStop}%
\bibitem [{\citenamefont {Benalcazar}\ \emph
  {et~al.}(2017{\natexlab{a}})\citenamefont {Benalcazar}, \citenamefont
  {Bernevig},\ and\ \citenamefont {Hughes}}]{bbh}%
  \BibitemOpen
  \bibfield  {author} {\bibinfo {author} {\bibfnamefont {W.~A.}\ \bibnamefont
  {Benalcazar}}, \bibinfo {author} {\bibfnamefont {B.~A.}\ \bibnamefont
  {Bernevig}},\ and\ \bibinfo {author} {\bibfnamefont {T.~L.}\ \bibnamefont
  {Hughes}},\ }\bibfield  {title} {\bibinfo {title} {Quantized electric
  multipole insulators},\ }\href {https://doi.org/10.1126/science.aah6442}
  {\bibfield  {journal} {\bibinfo  {journal} {Science}\ }\textbf {\bibinfo
  {volume} {357}},\ \bibinfo {pages} {61} (\bibinfo {year}
  {2017}{\natexlab{a}})},\ \Eprint
  {https://arxiv.org/abs/http://science.sciencemag.org/content}
  {http://science.sciencemag.org/content} \BibitemShut {NoStop}%
\bibitem [{\citenamefont {Essin}\ \emph {et~al.}(2009)\citenamefont {Essin},
  \citenamefont {Moore},\ and\ \citenamefont {Vanderbilt}}]{essin2009}%
  \BibitemOpen
  \bibfield  {author} {\bibinfo {author} {\bibfnamefont {A.~M.}\ \bibnamefont
  {Essin}}, \bibinfo {author} {\bibfnamefont {J.~E.}\ \bibnamefont {Moore}},\
  and\ \bibinfo {author} {\bibfnamefont {D.}~\bibnamefont {Vanderbilt}},\
  }\bibfield  {title} {\bibinfo {title} {Magnetoelectric polarizability and
  axion electrodynamics in crystalline insulators},\ }\href
  {https://doi.org/10.1103/PhysRevLett.102.146805} {\bibfield  {journal}
  {\bibinfo  {journal} {Phys. Rev. Lett.}\ }\textbf {\bibinfo {volume} {102}},\
  \bibinfo {pages} {146805} (\bibinfo {year} {2009})}\BibitemShut {NoStop}%
\bibitem [{\citenamefont {Armitage}\ and\ \citenamefont
  {Wu}(2019)}]{wuarmitage}%
  \BibitemOpen
  \bibfield  {author} {\bibinfo {author} {\bibfnamefont {N.~P.}\ \bibnamefont
  {Armitage}}\ and\ \bibinfo {author} {\bibfnamefont {L.}~\bibnamefont {Wu}},\
  }\bibfield  {title} {\bibinfo {title} {{On the matter of topological
  insulators as magnetoelectrics}},\ }\href
  {https://doi.org/10.21468/SciPostPhys.6.4.046} {\bibfield  {journal}
  {\bibinfo  {journal} {SciPost Phys.}\ }\textbf {\bibinfo {volume} {6}},\
  \bibinfo {pages} {46} (\bibinfo {year} {2019})}\BibitemShut {NoStop}%
\bibitem [{\citenamefont {Haldane}(2004)}]{haldaneAHE}%
  \BibitemOpen
  \bibfield  {author} {\bibinfo {author} {\bibfnamefont {F.~D.~M.}\
  \bibnamefont {Haldane}},\ }\bibfield  {title} {\bibinfo {title} {Berry
  curvature on the fermi surface: Anomalous hall effect as a topological
  fermi-liquid property},\ }\href
  {https://doi.org/10.1103/PhysRevLett.93.206602} {\bibfield  {journal}
  {\bibinfo  {journal} {Phys. Rev. Lett.}\ }\textbf {\bibinfo {volume} {93}},\
  \bibinfo {pages} {206602} (\bibinfo {year} {2004})}\BibitemShut {NoStop}%
\bibitem [{\citenamefont {Burkov}\ and\ \citenamefont
  {Balents}(2011)}]{burkov2011A}%
  \BibitemOpen
  \bibfield  {author} {\bibinfo {author} {\bibfnamefont {A.}~\bibnamefont
  {Burkov}}\ and\ \bibinfo {author} {\bibfnamefont {L.}~\bibnamefont
  {Balents}},\ }\bibfield  {title} {\bibinfo {title} {Weyl semimetal in a
  topological insulator multilayer},\ }\href@noop {} {\bibfield  {journal}
  {\bibinfo  {journal} {Physical review letters}\ }\textbf {\bibinfo {volume}
  {107}},\ \bibinfo {pages} {127205} (\bibinfo {year} {2011})}\BibitemShut
  {NoStop}%
\bibitem [{\citenamefont {Wan}\ \emph {et~al.}(2011)\citenamefont {Wan},
  \citenamefont {Turner}, \citenamefont {Vishwanath},\ and\ \citenamefont
  {Savrasov}}]{Wan2011}%
  \BibitemOpen
  \bibfield  {author} {\bibinfo {author} {\bibfnamefont {X.}~\bibnamefont
  {Wan}}, \bibinfo {author} {\bibfnamefont {A.~M.}\ \bibnamefont {Turner}},
  \bibinfo {author} {\bibfnamefont {A.}~\bibnamefont {Vishwanath}},\ and\
  \bibinfo {author} {\bibfnamefont {S.~Y.}\ \bibnamefont {Savrasov}},\
  }\bibfield  {title} {\bibinfo {title} {Topological semimetal and fermi-arc
  surface states in the electronic structure of pyrochlore iridates},\ }\href
  {https://doi.org/10.1103/PhysRevB.83.205101} {\bibfield  {journal} {\bibinfo
  {journal} {Phys. Rev. B}\ }\textbf {\bibinfo {volume} {83}},\ \bibinfo
  {pages} {205101} (\bibinfo {year} {2011})}\BibitemShut {NoStop}%
\bibitem [{\citenamefont {Zyuzin}\ and\ \citenamefont
  {Burkov}(2012)}]{zyuzin2012}%
  \BibitemOpen
  \bibfield  {author} {\bibinfo {author} {\bibfnamefont {A.~A.}\ \bibnamefont
  {Zyuzin}}\ and\ \bibinfo {author} {\bibfnamefont {A.~A.}\ \bibnamefont
  {Burkov}},\ }\bibfield  {title} {\bibinfo {title} {Topological response in
  weyl semimetals and the chiral anomaly},\ }\href
  {https://doi.org/10.1103/PhysRevB.86.115133} {\bibfield  {journal} {\bibinfo
  {journal} {Phys. Rev. B}\ }\textbf {\bibinfo {volume} {86}},\ \bibinfo
  {pages} {115133} (\bibinfo {year} {2012})}\BibitemShut {NoStop}%
\bibitem [{\citenamefont {Ramamurthy}\ and\ \citenamefont
  {Hughes}(2015)}]{RamamurthyPatterns}%
  \BibitemOpen
  \bibfield  {author} {\bibinfo {author} {\bibfnamefont {S.~T.}\ \bibnamefont
  {Ramamurthy}}\ and\ \bibinfo {author} {\bibfnamefont {T.~L.}\ \bibnamefont
  {Hughes}},\ }\bibfield  {title} {\bibinfo {title} {Patterns of
  electromagnetic response in topological semimetals},\ }\href
  {https://doi.org/10.1103/PhysRevB.92.085105} {\bibfield  {journal} {\bibinfo
  {journal} {Phys. Rev. B}\ }\textbf {\bibinfo {volume} {92}},\ \bibinfo
  {pages} {085105} (\bibinfo {year} {2015})}\BibitemShut {NoStop}%
\bibitem [{\citenamefont {Ramamurthy}\ and\ \citenamefont
  {Hughes}(2017)}]{ramamurthylinenode}%
  \BibitemOpen
  \bibfield  {author} {\bibinfo {author} {\bibfnamefont {S.~T.}\ \bibnamefont
  {Ramamurthy}}\ and\ \bibinfo {author} {\bibfnamefont {T.~L.}\ \bibnamefont
  {Hughes}},\ }\bibfield  {title} {\bibinfo {title} {Quasitopological
  electromagnetic response of line-node semimetals},\ }\href
  {https://doi.org/10.1103/PhysRevB.95.075138} {\bibfield  {journal} {\bibinfo
  {journal} {Phys. Rev. B}\ }\textbf {\bibinfo {volume} {95}},\ \bibinfo
  {pages} {075138} (\bibinfo {year} {2017})}\BibitemShut {NoStop}%
\bibitem [{\citenamefont {Halperin}(1987)}]{halperin1987}%
  \BibitemOpen
  \bibfield  {author} {\bibinfo {author} {\bibfnamefont {B.~I.}\ \bibnamefont
  {Halperin}},\ }\bibfield  {title} {\bibinfo {title} {Possible states for a
  three-dimensional electron gas in a strong magnetic field},\ }\href@noop {}
  {\bibfield  {journal} {\bibinfo  {journal} {Jpn. J. Appl. Phys. Suppl.}\
  }\textbf {\bibinfo {volume} {26}},\ \bibinfo {pages} {1913} (\bibinfo {year}
  {1987})}\BibitemShut {NoStop}%
\bibitem [{\citenamefont {Fu}\ \emph {et~al.}(2007)\citenamefont {Fu},
  \citenamefont {Kane},\ and\ \citenamefont {Mele}}]{fu2007topological}%
  \BibitemOpen
  \bibfield  {author} {\bibinfo {author} {\bibfnamefont {L.}~\bibnamefont
  {Fu}}, \bibinfo {author} {\bibfnamefont {C.~L.}\ \bibnamefont {Kane}},\ and\
  \bibinfo {author} {\bibfnamefont {E.~J.}\ \bibnamefont {Mele}},\ }\bibfield
  {title} {\bibinfo {title} {Topological insulators in three dimensions},\
  }\href@noop {} {\bibfield  {journal} {\bibinfo  {journal} {Phys. Rev. Lett.}\
  }\textbf {\bibinfo {volume} {98}},\ \bibinfo {pages} {106803} (\bibinfo
  {year} {2007})}\BibitemShut {NoStop}%
\bibitem [{\citenamefont {Ran}\ \emph {et~al.}(2009)\citenamefont {Ran},
  \citenamefont {Zhang},\ and\ \citenamefont {Vishwanath}}]{ran2009}%
  \BibitemOpen
  \bibfield  {author} {\bibinfo {author} {\bibfnamefont {Y.}~\bibnamefont
  {Ran}}, \bibinfo {author} {\bibfnamefont {Y.}~\bibnamefont {Zhang}},\ and\
  \bibinfo {author} {\bibfnamefont {A.}~\bibnamefont {Vishwanath}},\ }\bibfield
   {title} {\bibinfo {title} {One-dimensional topologically protected modes in
  topological insulators with lattice dislocations},\ }\href
  {https://doi.org/10.1038/nphys1220} {\bibfield  {journal} {\bibinfo
  {journal} {Nature Physics}\ }\textbf {\bibinfo {volume} {5}},\ \bibinfo
  {pages} {298} (\bibinfo {year} {2009})}\BibitemShut {NoStop}%
\bibitem [{\citenamefont {Zhu}\ \emph {et~al.}(2023)\citenamefont {Zhu},
  \citenamefont {Sun}, \citenamefont {Hughes},\ and\ \citenamefont
  {Bahl}}]{zhu2023higher}%
  \BibitemOpen
  \bibfield  {author} {\bibinfo {author} {\bibfnamefont {P.}~\bibnamefont
  {Zhu}}, \bibinfo {author} {\bibfnamefont {X.-Q.}\ \bibnamefont {Sun}},
  \bibinfo {author} {\bibfnamefont {T.~L.}\ \bibnamefont {Hughes}},\ and\
  \bibinfo {author} {\bibfnamefont {G.}~\bibnamefont {Bahl}},\ }\bibfield
  {title} {\bibinfo {title} {Higher rank chirality and non-hermitian skin
  effect in a topolectrical circuit},\ }\href
  {https://doi.org/https://doi.org/10.1038/s41467-023-36130-x} {\bibfield
  {journal} {\bibinfo  {journal} {Nature communications}\ }\textbf {\bibinfo
  {volume} {14}},\ \bibinfo {pages} {720} (\bibinfo {year} {2023})}\BibitemShut
  {NoStop}%
\bibitem [{\citenamefont {Thouless}(1983)}]{thoulesspumping}%
  \BibitemOpen
  \bibfield  {author} {\bibinfo {author} {\bibfnamefont {D.~J.}\ \bibnamefont
  {Thouless}},\ }\bibfield  {title} {\bibinfo {title} {Quantization of particle
  transport},\ }\href {https://doi.org/10.1103/PhysRevB.27.6083} {\bibfield
  {journal} {\bibinfo  {journal} {Phys. Rev. B}\ }\textbf {\bibinfo {volume}
  {27}},\ \bibinfo {pages} {6083} (\bibinfo {year} {1983})}\BibitemShut
  {NoStop}%
\bibitem [{\citenamefont {Landau}\ \emph {et~al.}(1986)\citenamefont {Landau},
  \citenamefont {Pitaevskii}, \citenamefont {Lifshitz},\ and\ \citenamefont
  {Kosevich}}]{landaulifshitzelasticity}%
  \BibitemOpen
  \bibfield  {author} {\bibinfo {author} {\bibfnamefont {L.~D.}\ \bibnamefont
  {Landau}}, \bibinfo {author} {\bibfnamefont {L.~P.}\ \bibnamefont
  {Pitaevskii}}, \bibinfo {author} {\bibfnamefont {E.~M.}\ \bibnamefont
  {Lifshitz}},\ and\ \bibinfo {author} {\bibfnamefont {A.~M.}\ \bibnamefont
  {Kosevich}},\ }\href@noop {} {\emph {\bibinfo {title} {Theory of
  Elasticity}}},\ \bibinfo {edition} {3rd}\ ed.\ (\bibinfo  {publisher}
  {Butterworth-Heinemann},\ \bibinfo {year} {1986})\BibitemShut {NoStop}%
\bibitem [{\citenamefont {Katanaev}\ and\ \citenamefont
  {Volovich}(1992)}]{KATANAEV1992}%
  \BibitemOpen
  \bibfield  {author} {\bibinfo {author} {\bibfnamefont {M.}~\bibnamefont
  {Katanaev}}\ and\ \bibinfo {author} {\bibfnamefont {I.}~\bibnamefont
  {Volovich}},\ }\bibfield  {title} {\bibinfo {title} {Theory of defects in
  solids and three-dimensional gravity},\ }\href
  {https://doi.org/https://doi.org/10.1016/0003-4916(52)90040-7} {\bibfield
  {journal} {\bibinfo  {journal} {Annals of Physics}\ }\textbf {\bibinfo
  {volume} {216}},\ \bibinfo {pages} {1} (\bibinfo {year} {1992})}\BibitemShut
  {NoStop}%
\bibitem [{\citenamefont {Vozmediano}\ \emph {et~al.}(2010)\citenamefont
  {Vozmediano}, \citenamefont {Katsnelson},\ and\ \citenamefont
  {Guinea}}]{Vozmediano2010}%
  \BibitemOpen
  \bibfield  {author} {\bibinfo {author} {\bibfnamefont {M.}~\bibnamefont
  {Vozmediano}}, \bibinfo {author} {\bibfnamefont {M.}~\bibnamefont
  {Katsnelson}},\ and\ \bibinfo {author} {\bibfnamefont {F.}~\bibnamefont
  {Guinea}},\ }\bibfield  {title} {\bibinfo {title} {Gauge fields in
  graphene},\ }\href
  {https://doi.org/https://doi.org/10.1016/j.physrep.2010.07.003} {\bibfield
  {journal} {\bibinfo  {journal} {Physics Reports}\ }\textbf {\bibinfo {volume}
  {496}},\ \bibinfo {pages} {109} (\bibinfo {year} {2010})}\BibitemShut
  {NoStop}%
\bibitem [{\citenamefont {Guinea}\ \emph {et~al.}(2010)\citenamefont {Guinea},
  \citenamefont {Katsnelson},\ and\ \citenamefont {Geim}}]{Guinea2010}%
  \BibitemOpen
  \bibfield  {author} {\bibinfo {author} {\bibfnamefont {F.}~\bibnamefont
  {Guinea}}, \bibinfo {author} {\bibfnamefont {M.~I.}\ \bibnamefont
  {Katsnelson}},\ and\ \bibinfo {author} {\bibfnamefont {A.~K.}\ \bibnamefont
  {Geim}},\ }\bibfield  {title} {\bibinfo {title} {Energy gaps and a zero-field
  quantum hall effect in graphene by strain engineering},\ }\href
  {https://doi.org/10.1038/nphys1420} {\bibfield  {journal} {\bibinfo
  {journal} {Nature Physics}\ }\textbf {\bibinfo {volume} {6}},\ \bibinfo
  {pages} {30} (\bibinfo {year} {2010})}\BibitemShut {NoStop}%
\bibitem [{\citenamefont {Levy}\ \emph {et~al.}(2010)\citenamefont {Levy},
  \citenamefont {Burke}, \citenamefont {Meaker}, \citenamefont {Panlasigui},
  \citenamefont {Zettl}, \citenamefont {Guinea}, \citenamefont {Neto},\ and\
  \citenamefont {Crommie}}]{Levy2010}%
  \BibitemOpen
  \bibfield  {author} {\bibinfo {author} {\bibfnamefont {N.}~\bibnamefont
  {Levy}}, \bibinfo {author} {\bibfnamefont {S.}~\bibnamefont {Burke}},
  \bibinfo {author} {\bibfnamefont {K.}~\bibnamefont {Meaker}}, \bibinfo
  {author} {\bibfnamefont {M.}~\bibnamefont {Panlasigui}}, \bibinfo {author}
  {\bibfnamefont {A.}~\bibnamefont {Zettl}}, \bibinfo {author} {\bibfnamefont
  {F.}~\bibnamefont {Guinea}}, \bibinfo {author} {\bibfnamefont {A.~C.}\
  \bibnamefont {Neto}},\ and\ \bibinfo {author} {\bibfnamefont {M.~F.}\
  \bibnamefont {Crommie}},\ }\bibfield  {title} {\bibinfo {title}
  {Strain-induced pseudo--magnetic fields greater than 300 tesla in graphene
  nanobubbles},\ }\href {https://doi.org/10.1126/science.1191700} {\bibfield
  {journal} {\bibinfo  {journal} {Science}\ }\textbf {\bibinfo {volume}
  {329}},\ \bibinfo {pages} {544} (\bibinfo {year} {2010})}\BibitemShut
  {NoStop}%
\bibitem [{\citenamefont {Rachel}\ \emph {et~al.}(2016)\citenamefont {Rachel},
  \citenamefont {Fritz},\ and\ \citenamefont {Vojta}}]{Rachel2016}%
  \BibitemOpen
  \bibfield  {author} {\bibinfo {author} {\bibfnamefont {S.}~\bibnamefont
  {Rachel}}, \bibinfo {author} {\bibfnamefont {L.}~\bibnamefont {Fritz}},\ and\
  \bibinfo {author} {\bibfnamefont {M.}~\bibnamefont {Vojta}},\ }\bibfield
  {title} {\bibinfo {title} {Landau levels of majorana fermions in a spin
  liquid},\ }\href {https://doi.org/10.1103/PhysRevLett.116.167201} {\bibfield
  {journal} {\bibinfo  {journal} {Phys. Rev. Lett.}\ }\textbf {\bibinfo
  {volume} {116}},\ \bibinfo {pages} {167201} (\bibinfo {year}
  {2016})}\BibitemShut {NoStop}%
\bibitem [{\citenamefont {Cortijo}\ \emph {et~al.}(2015)\citenamefont
  {Cortijo}, \citenamefont {Ferreir\'os}, \citenamefont {Landsteiner},\ and\
  \citenamefont {Vozmediano}}]{Cortijo2016}%
  \BibitemOpen
  \bibfield  {author} {\bibinfo {author} {\bibfnamefont {A.}~\bibnamefont
  {Cortijo}}, \bibinfo {author} {\bibfnamefont {Y.}~\bibnamefont
  {Ferreir\'os}}, \bibinfo {author} {\bibfnamefont {K.}~\bibnamefont
  {Landsteiner}},\ and\ \bibinfo {author} {\bibfnamefont {M.~A.~H.}\
  \bibnamefont {Vozmediano}},\ }\bibfield  {title} {\bibinfo {title} {Elastic
  gauge fields in weyl semimetals},\ }\href
  {https://doi.org/10.1103/PhysRevLett.115.177202} {\bibfield  {journal}
  {\bibinfo  {journal} {Phys. Rev. Lett.}\ }\textbf {\bibinfo {volume} {115}},\
  \bibinfo {pages} {177202} (\bibinfo {year} {2015})}\BibitemShut {NoStop}%
\bibitem [{\citenamefont {Pikulin}\ \emph
  {et~al.}(2016{\natexlab{a}})\citenamefont {Pikulin}, \citenamefont {Chen},\
  and\ \citenamefont {Franz}}]{Pikulin2016}%
  \BibitemOpen
  \bibfield  {author} {\bibinfo {author} {\bibfnamefont {D.~I.}\ \bibnamefont
  {Pikulin}}, \bibinfo {author} {\bibfnamefont {A.}~\bibnamefont {Chen}},\ and\
  \bibinfo {author} {\bibfnamefont {M.}~\bibnamefont {Franz}},\ }\bibfield
  {title} {\bibinfo {title} {Chiral anomaly from strain-induced gauge fields in
  dirac and weyl semimetals},\ }\href
  {https://doi.org/10.1103/PhysRevX.6.041021} {\bibfield  {journal} {\bibinfo
  {journal} {Phys. Rev. X}\ }\textbf {\bibinfo {volume} {6}},\ \bibinfo {pages}
  {041021} (\bibinfo {year} {2016}{\natexlab{a}})}\BibitemShut {NoStop}%
\bibitem [{\citenamefont {Grushin}\ \emph {et~al.}(2016)\citenamefont
  {Grushin}, \citenamefont {Venderbos}, \citenamefont {Vishwanath},\ and\
  \citenamefont {Ilan}}]{Grushin2016}%
  \BibitemOpen
  \bibfield  {author} {\bibinfo {author} {\bibfnamefont {A.~G.}\ \bibnamefont
  {Grushin}}, \bibinfo {author} {\bibfnamefont {J.~W.~F.}\ \bibnamefont
  {Venderbos}}, \bibinfo {author} {\bibfnamefont {A.}~\bibnamefont
  {Vishwanath}},\ and\ \bibinfo {author} {\bibfnamefont {R.}~\bibnamefont
  {Ilan}},\ }\bibfield  {title} {\bibinfo {title} {Inhomogeneous weyl and dirac
  semimetals: Transport in axial magnetic fields and fermi arc surface states
  from pseudo-landau levels},\ }\href
  {https://doi.org/10.1103/PhysRevX.6.041046} {\bibfield  {journal} {\bibinfo
  {journal} {Phys. Rev. X}\ }\textbf {\bibinfo {volume} {6}},\ \bibinfo {pages}
  {041046} (\bibinfo {year} {2016})}\BibitemShut {NoStop}%
\bibitem [{\citenamefont {Matsushita}\ \emph {et~al.}(2020)\citenamefont
  {Matsushita}, \citenamefont {Fujimoto},\ and\ \citenamefont
  {Schnyder}}]{Matsushita2020}%
  \BibitemOpen
  \bibfield  {author} {\bibinfo {author} {\bibfnamefont {T.}~\bibnamefont
  {Matsushita}}, \bibinfo {author} {\bibfnamefont {S.}~\bibnamefont
  {Fujimoto}},\ and\ \bibinfo {author} {\bibfnamefont {A.~P.}\ \bibnamefont
  {Schnyder}},\ }\bibfield  {title} {\bibinfo {title} {Topological
  piezoelectric effect and parity anomaly in nodal line semimetals},\ }\href
  {https://doi.org/10.1103/PhysRevResearch.2.043311} {\bibfield  {journal}
  {\bibinfo  {journal} {Phys. Rev. Res.}\ }\textbf {\bibinfo {volume} {2}},\
  \bibinfo {pages} {043311} (\bibinfo {year} {2020})}\BibitemShut {NoStop}%
\bibitem [{\citenamefont {Thorngren}\ and\ \citenamefont
  {Else}(2018)}]{thorngren2018}%
  \BibitemOpen
  \bibfield  {author} {\bibinfo {author} {\bibfnamefont {R.}~\bibnamefont
  {Thorngren}}\ and\ \bibinfo {author} {\bibfnamefont {D.~V.}\ \bibnamefont
  {Else}},\ }\bibfield  {title} {\bibinfo {title} {Gauging spatial symmetries
  and the classification of topological crystalline phases},\ }\href@noop {}
  {\bibfield  {journal} {\bibinfo  {journal} {Phys. Rev. X}\ }\textbf {\bibinfo
  {volume} {8}},\ \bibinfo {pages} {011040} (\bibinfo {year}
  {2018})}\BibitemShut {NoStop}%
\bibitem [{\citenamefont {Song}\ \emph {et~al.}(2021)\citenamefont {Song},
  \citenamefont {He}, \citenamefont {Vishwanath},\ and\ \citenamefont
  {Wang}}]{chongwangvishwanath}%
  \BibitemOpen
  \bibfield  {author} {\bibinfo {author} {\bibfnamefont {X.-Y.}\ \bibnamefont
  {Song}}, \bibinfo {author} {\bibfnamefont {Y.-C.}\ \bibnamefont {He}},
  \bibinfo {author} {\bibfnamefont {A.}~\bibnamefont {Vishwanath}},\ and\
  \bibinfo {author} {\bibfnamefont {C.}~\bibnamefont {Wang}},\ }\bibfield
  {title} {\bibinfo {title} {Electric polarization as a nonquantized
  topological response and boundary luttinger theorem},\ }\href
  {https://doi.org/10.1103/PhysRevResearch.3.023011} {\bibfield  {journal}
  {\bibinfo  {journal} {Phys. Rev. Research}\ }\textbf {\bibinfo {volume}
  {3}},\ \bibinfo {pages} {023011} (\bibinfo {year} {2021})}\BibitemShut
  {NoStop}%
\bibitem [{\citenamefont {Manjunath}\ and\ \citenamefont
  {Barkeshli}(2021)}]{barkeshli2021}%
  \BibitemOpen
  \bibfield  {author} {\bibinfo {author} {\bibfnamefont {N.}~\bibnamefont
  {Manjunath}}\ and\ \bibinfo {author} {\bibfnamefont {M.}~\bibnamefont
  {Barkeshli}},\ }\bibfield  {title} {\bibinfo {title} {Crystalline gauge
  fields and quantized discrete geometric response for abelian topological
  phases with lattice symmetry},\ }\href
  {https://doi.org/10.1103/PhysRevResearch.3.013040} {\bibfield  {journal}
  {\bibinfo  {journal} {Phys. Rev. Research}\ }\textbf {\bibinfo {volume}
  {3}},\ \bibinfo {pages} {013040} (\bibinfo {year} {2021})}\BibitemShut
  {NoStop}%
\bibitem [{\citenamefont {Kaluza}(1921)}]{kaluza}%
  \BibitemOpen
  \bibfield  {author} {\bibinfo {author} {\bibfnamefont {T.}~\bibnamefont
  {Kaluza}},\ }\bibfield  {title} {\bibinfo {title} {{Zum Unit\"atsproblem der
  Physik}},\ }\href {https://doi.org/10.1142/S0218271818700017} {\bibfield
  {journal} {\bibinfo  {journal} {Sitzungsber. Preuss. Akad. Wiss. Berlin
  (Math. Phys. )}\ }\textbf {\bibinfo {volume} {1921}},\ \bibinfo {pages} {966}
  (\bibinfo {year} {1921})},\ \Eprint {https://arxiv.org/abs/1803.08616}
  {arXiv:1803.08616 [physics.hist-ph]} \BibitemShut {NoStop}%
\bibitem [{\citenamefont {Klein}(1926)}]{klein}%
  \BibitemOpen
  \bibfield  {author} {\bibinfo {author} {\bibfnamefont {O.}~\bibnamefont
  {Klein}},\ }\bibfield  {title} {\bibinfo {title} {{Quantum Theory and
  Five-Dimensional Theory of Relativity. (In German and English)}},\ }\href
  {https://doi.org/10.1007/BF01397481} {\bibfield  {journal} {\bibinfo
  {journal} {Z. Phys.}\ }\textbf {\bibinfo {volume} {37}},\ \bibinfo {pages}
  {895} (\bibinfo {year} {1926})}\BibitemShut {NoStop}%
\bibitem [{Note1()}]{Note1}%
  \BibitemOpen
  \bibinfo {note} {As mentioned, this analogy is precise for gapped systems.
  For gapless systems the analogy predicts the correct form of the action, but
  does not uniquely determine the coefficient.}\BibitemShut {Stop}%
\bibitem [{\citenamefont {Su}\ \emph {et~al.}(1979)\citenamefont {Su},
  \citenamefont {Schrieffer},\ and\ \citenamefont {Heeger}}]{su1979}%
  \BibitemOpen
  \bibfield  {author} {\bibinfo {author} {\bibfnamefont {W.~P.}\ \bibnamefont
  {Su}}, \bibinfo {author} {\bibfnamefont {J.~R.}\ \bibnamefont {Schrieffer}},\
  and\ \bibinfo {author} {\bibfnamefont {A.~J.}\ \bibnamefont {Heeger}},\
  }\bibfield  {title} {\bibinfo {title} {Solitons in polyacetylene},\ }\href
  {https://doi.org/10.1103/PhysRevLett.42.1698} {\bibfield  {journal} {\bibinfo
   {journal} {Phys. Rev. Lett.}\ }\textbf {\bibinfo {volume} {42}},\ \bibinfo
  {pages} {1698} (\bibinfo {year} {1979})}\BibitemShut {NoStop}%
\bibitem [{\citenamefont {Benalcazar}\ \emph
  {et~al.}(2017{\natexlab{b}})\citenamefont {Benalcazar}, \citenamefont
  {Bernevig},\ and\ \citenamefont {Hughes}}]{benalcazar2017}%
  \BibitemOpen
  \bibfield  {author} {\bibinfo {author} {\bibfnamefont {W.~A.}\ \bibnamefont
  {Benalcazar}}, \bibinfo {author} {\bibfnamefont {B.~A.}\ \bibnamefont
  {Bernevig}},\ and\ \bibinfo {author} {\bibfnamefont {T.~L.}\ \bibnamefont
  {Hughes}},\ }\bibfield  {title} {\bibinfo {title} {Electric multipole
  moments, topological multipole moment pumping, and chiral hinge states in
  crystalline insulators},\ }\href {https://doi.org/10.1103/PhysRevB.96.245115}
  {\bibfield  {journal} {\bibinfo  {journal} {Phys. Rev. B}\ }\textbf {\bibinfo
  {volume} {96}},\ \bibinfo {pages} {245115} (\bibinfo {year}
  {2017}{\natexlab{b}})}\BibitemShut {NoStop}%
\bibitem [{\citenamefont {Parrikar}\ \emph {et~al.}(2014)\citenamefont
  {Parrikar}, \citenamefont {Hughes},\ and\ \citenamefont {Leigh}}]{parrikar2}%
  \BibitemOpen
  \bibfield  {author} {\bibinfo {author} {\bibfnamefont {O.}~\bibnamefont
  {Parrikar}}, \bibinfo {author} {\bibfnamefont {T.~L.}\ \bibnamefont
  {Hughes}},\ and\ \bibinfo {author} {\bibfnamefont {R.~G.}\ \bibnamefont
  {Leigh}},\ }\bibfield  {title} {\bibinfo {title} {Torsion, parity-odd
  response, and anomalies in topological states},\ }\href
  {https://doi.org/10.1103/PhysRevD.90.105004} {\bibfield  {journal} {\bibinfo
  {journal} {Phys. Rev. D}\ }\textbf {\bibinfo {volume} {90}},\ \bibinfo
  {pages} {105004} (\bibinfo {year} {2014})}\BibitemShut {NoStop}%
\bibitem [{\citenamefont {Pikulin}\ \emph
  {et~al.}(2016{\natexlab{b}})\citenamefont {Pikulin}, \citenamefont {Chen},\
  and\ \citenamefont {Franz}}]{pikulin2016chiral}%
  \BibitemOpen
  \bibfield  {author} {\bibinfo {author} {\bibfnamefont {D.}~\bibnamefont
  {Pikulin}}, \bibinfo {author} {\bibfnamefont {A.}~\bibnamefont {Chen}},\ and\
  \bibinfo {author} {\bibfnamefont {M.}~\bibnamefont {Franz}},\ }\bibfield
  {title} {\bibinfo {title} {Chiral anomaly from strain-induced gauge fields in
  dirac and weyl semimetals},\ }\href@noop {} {\bibfield  {journal} {\bibinfo
  {journal} {Physical Review X}\ }\textbf {\bibinfo {volume} {6}},\ \bibinfo
  {pages} {041021} (\bibinfo {year} {2016}{\natexlab{b}})}\BibitemShut
  {NoStop}%
\bibitem [{\citenamefont {Hughes}\ \emph {et~al.}(2013)\citenamefont {Hughes},
  \citenamefont {Leigh},\ and\ \citenamefont {Parrikar}}]{parrikar1}%
  \BibitemOpen
  \bibfield  {author} {\bibinfo {author} {\bibfnamefont {T.~L.}\ \bibnamefont
  {Hughes}}, \bibinfo {author} {\bibfnamefont {R.~G.}\ \bibnamefont {Leigh}},\
  and\ \bibinfo {author} {\bibfnamefont {O.}~\bibnamefont {Parrikar}},\
  }\bibfield  {title} {\bibinfo {title} {Torsional anomalies, hall viscosity,
  and bulk-boundary correspondence in topological states},\ }\href
  {https://doi.org/10.1103/PhysRevD.88.025040} {\bibfield  {journal} {\bibinfo
  {journal} {Phys. Rev. D}\ }\textbf {\bibinfo {volume} {88}},\ \bibinfo
  {pages} {025040} (\bibinfo {year} {2013})}\BibitemShut {NoStop}%
\bibitem [{\citenamefont {Bradlyn}\ and\ \citenamefont
  {Read}(2015)}]{bradlyn2015low}%
  \BibitemOpen
  \bibfield  {author} {\bibinfo {author} {\bibfnamefont {B.}~\bibnamefont
  {Bradlyn}}\ and\ \bibinfo {author} {\bibfnamefont {N.}~\bibnamefont {Read}},\
  }\bibfield  {title} {\bibinfo {title} {Low-energy effective theory in the
  bulk for transport in a topological phase},\ }\href@noop {} {\bibfield
  {journal} {\bibinfo  {journal} {Phys. Rev. B}\ }\textbf {\bibinfo {volume}
  {91}},\ \bibinfo {pages} {125303} (\bibinfo {year} {2015})}\BibitemShut
  {NoStop}%
\bibitem [{\citenamefont {Rao}\ and\ \citenamefont
  {Bradlyn}(2020)}]{bradlynrao}%
  \BibitemOpen
  \bibfield  {author} {\bibinfo {author} {\bibfnamefont {P.}~\bibnamefont
  {Rao}}\ and\ \bibinfo {author} {\bibfnamefont {B.}~\bibnamefont {Bradlyn}},\
  }\bibfield  {title} {\bibinfo {title} {Hall viscosity in quantum systems with
  discrete symmetry: Point group and lattice anisotropy},\ }\href
  {https://doi.org/10.1103/PhysRevX.10.021005} {\bibfield  {journal} {\bibinfo
  {journal} {Phys. Rev. X}\ }\textbf {\bibinfo {volume} {10}},\ \bibinfo
  {pages} {021005} (\bibinfo {year} {2020})}\BibitemShut {NoStop}%
\bibitem [{\citenamefont {Wang}\ and\ \citenamefont {Zhang}(2012)}]{Zhong2012}%
  \BibitemOpen
  \bibfield  {author} {\bibinfo {author} {\bibfnamefont {Z.}~\bibnamefont
  {Wang}}\ and\ \bibinfo {author} {\bibfnamefont {S.-C.}\ \bibnamefont
  {Zhang}},\ }\bibfield  {title} {\bibinfo {title} {Simplified topological
  invariants for interacting insulators},\ }\href
  {https://doi.org/10.1103/PhysRevX.2.031008} {\bibfield  {journal} {\bibinfo
  {journal} {Phys. Rev. X}\ }\textbf {\bibinfo {volume} {2}},\ \bibinfo {pages}
  {031008} (\bibinfo {year} {2012})}\BibitemShut {NoStop}%
\bibitem [{Note2()}]{Note2}%
  \BibitemOpen
  \bibinfo {note} {As shown in the Appendix, this anomaly has two
  contributions. One comes from the low-energy currents that contribute with a
  factor of $1/2\pi $ and a second from a change of system-size for a ground
  state carrying a non-vanishing momentum density with a factor of $-1/4\pi
  $.}\BibitemShut {Stop}%
\bibitem [{\citenamefont {Hughes}\ \emph
  {et~al.}(2011{\natexlab{b}})\citenamefont {Hughes}, \citenamefont {Prodan},\
  and\ \citenamefont {Bernevig}}]{hughesinversion2011}%
  \BibitemOpen
  \bibfield  {author} {\bibinfo {author} {\bibfnamefont {T.~L.}\ \bibnamefont
  {Hughes}}, \bibinfo {author} {\bibfnamefont {E.}~\bibnamefont {Prodan}},\
  and\ \bibinfo {author} {\bibfnamefont {B.~A.}\ \bibnamefont {Bernevig}},\
  }\bibfield  {title} {\bibinfo {title} {Inversion-symmetric topological
  insulators},\ }\href {https://doi.org/10.1103/PhysRevB.83.245132} {\bibfield
  {journal} {\bibinfo  {journal} {Phys. Rev. B}\ }\textbf {\bibinfo {volume}
  {83}},\ \bibinfo {pages} {245132} (\bibinfo {year}
  {2011}{\natexlab{b}})}\BibitemShut {NoStop}%
\bibitem [{\citenamefont {Turner}\ \emph {et~al.}(2012)\citenamefont {Turner},
  \citenamefont {Zhang}, \citenamefont {Mong},\ and\ \citenamefont
  {Vishwanath}}]{turner2012inversion}%
  \BibitemOpen
  \bibfield  {author} {\bibinfo {author} {\bibfnamefont {A.~M.}\ \bibnamefont
  {Turner}}, \bibinfo {author} {\bibfnamefont {Y.}~\bibnamefont {Zhang}},
  \bibinfo {author} {\bibfnamefont {R.~S.~K.}\ \bibnamefont {Mong}},\ and\
  \bibinfo {author} {\bibfnamefont {A.}~\bibnamefont {Vishwanath}},\ }\bibfield
   {title} {\bibinfo {title} {Quantized response and topology of magnetic
  insulators with inversion symmetry},\ }\href
  {https://doi.org/10.1103/PhysRevB.85.165120} {\bibfield  {journal} {\bibinfo
  {journal} {Phys. Rev. B}\ }\textbf {\bibinfo {volume} {85}},\ \bibinfo
  {pages} {165120} (\bibinfo {year} {2012})}\BibitemShut {NoStop}%
\bibitem [{Note3()}]{Note3}%
  \BibitemOpen
  \bibinfo {note} {Even though there is a $\protect \mathfrak {e}^\alpha $
  wedge product with a lower-dimensional action, it is not transverse to the
  lower-dimensional action since $\protect \mathcal {Q}_{\alpha \beta }$ is
  symmetric. For example, there will be terms where, say, $\protect \mathfrak
  {e}^x$ couples to $d\protect \mathfrak {e}^x$, which cannot be interpreted as
  a conventional stacked action.}\BibitemShut {Stop}%
\bibitem [{Note4()}]{Note4}%
  \BibitemOpen
  \bibinfo {note} {Alternatively we can assume the fields $A_w, e^\alpha _w$
  are locked to their ground state values and thus have vanishing derivatives
  in all directions.}\BibitemShut {Stop}%
\bibitem [{\citenamefont {Lin}\ and\ \citenamefont {Hughes}(2018)}]{linHOTSM}%
  \BibitemOpen
  \bibfield  {author} {\bibinfo {author} {\bibfnamefont {M.}~\bibnamefont
  {Lin}}\ and\ \bibinfo {author} {\bibfnamefont {T.~L.}\ \bibnamefont
  {Hughes}},\ }\bibfield  {title} {\bibinfo {title} {Topological quadrupolar
  semimetals},\ }\href {https://doi.org/10.1103/PhysRevB.98.241103} {\bibfield
  {journal} {\bibinfo  {journal} {Phys. Rev. B}\ }\textbf {\bibinfo {volume}
  {98}},\ \bibinfo {pages} {241103} (\bibinfo {year} {2018})}\BibitemShut
  {NoStop}%
\bibitem [{\citenamefont {Wieder}\ \emph {et~al.}(2020)\citenamefont {Wieder},
  \citenamefont {Wang}, \citenamefont {Cano}, \citenamefont {Dai},
  \citenamefont {Schoop}, \citenamefont {Bradlyn},\ and\ \citenamefont
  {Bernevig}}]{wieder2020strong}%
  \BibitemOpen
  \bibfield  {author} {\bibinfo {author} {\bibfnamefont {B.~J.}\ \bibnamefont
  {Wieder}}, \bibinfo {author} {\bibfnamefont {Z.}~\bibnamefont {Wang}},
  \bibinfo {author} {\bibfnamefont {J.}~\bibnamefont {Cano}}, \bibinfo {author}
  {\bibfnamefont {X.}~\bibnamefont {Dai}}, \bibinfo {author} {\bibfnamefont
  {L.~M.}\ \bibnamefont {Schoop}}, \bibinfo {author} {\bibfnamefont
  {B.}~\bibnamefont {Bradlyn}},\ and\ \bibinfo {author} {\bibfnamefont {B.~A.}\
  \bibnamefont {Bernevig}},\ }\bibfield  {title} {\bibinfo {title} {Strong and
  fragile topological dirac semimetals with higher-order fermi arcs},\
  }\href@noop {} {\bibfield  {journal} {\bibinfo  {journal} {Nat. Comms.}\
  }\textbf {\bibinfo {volume} {11}},\ \bibinfo {pages} {627} (\bibinfo {year}
  {2020})}\BibitemShut {NoStop}%
\bibitem [{\citenamefont {Chang}\ and\ \citenamefont
  {Niu}(1995)}]{chang1995berry}%
  \BibitemOpen
  \bibfield  {author} {\bibinfo {author} {\bibfnamefont {M.-C.}\ \bibnamefont
  {Chang}}\ and\ \bibinfo {author} {\bibfnamefont {Q.}~\bibnamefont {Niu}},\
  }\bibfield  {title} {\bibinfo {title} {Berry phase, hyperorbits, and the
  hofstadter spectrum},\ }\href@noop {} {\bibfield  {journal} {\bibinfo
  {journal} {Physical review letters}\ }\textbf {\bibinfo {volume} {75}},\
  \bibinfo {pages} {1348} (\bibinfo {year} {1995})}\BibitemShut {NoStop}%
\bibitem [{\citenamefont {Chang}\ and\ \citenamefont
  {Niu}(1996)}]{chang1996berry}%
  \BibitemOpen
  \bibfield  {author} {\bibinfo {author} {\bibfnamefont {M.-C.}\ \bibnamefont
  {Chang}}\ and\ \bibinfo {author} {\bibfnamefont {Q.}~\bibnamefont {Niu}},\
  }\bibfield  {title} {\bibinfo {title} {Berry phase, hyperorbits, and the
  hofstadter spectrum: Semiclassical dynamics in magnetic bloch bands},\
  }\href@noop {} {\bibfield  {journal} {\bibinfo  {journal} {Physical Review
  B}\ }\textbf {\bibinfo {volume} {53}},\ \bibinfo {pages} {7010} (\bibinfo
  {year} {1996})}\BibitemShut {NoStop}%
\bibitem [{Note5()}]{Note5}%
  \BibitemOpen
  \bibinfo {note} {We comment that even though the Chern-Simons term for the
  translation gauge fields shares some properties with the electromagnetic
  Chern-Simons term, there is a key distinction: The translation gauge fields
  have a constant background. This allows the Dirac node quadrupole system to
  have a static momentum polarization, whereas the electromagnetic Chern Simons
  term in a Chern insulator would predict generating an electric polarization
  as one tunes the vector potential.}\BibitemShut {Stop}%
\bibitem [{\citenamefont {Burkov}(2018)}]{BurkovReview18}%
  \BibitemOpen
  \bibfield  {author} {\bibinfo {author} {\bibfnamefont {A.}~\bibnamefont
  {Burkov}},\ }\bibfield  {title} {\bibinfo {title} {Weyl metals},\ }\href
  {https://doi.org/10.1146/annurev-conmatphys-033117-054129} {\bibfield
  {journal} {\bibinfo  {journal} {Annual Review of Condensed Matter Physics}\
  }\textbf {\bibinfo {volume} {9}},\ \bibinfo {pages} {359} (\bibinfo {year}
  {2018})},\ \Eprint
  {https://arxiv.org/abs/https://doi.org/10.1146/annurev-conmatphys-033117-054129}
  {https://doi.org/10.1146/annurev-conmatphys-033117-054129} \BibitemShut
  {NoStop}%
\bibitem [{\citenamefont {Kodama}\ and\ \citenamefont
  {Takane}(2019)}]{Kodama2019}%
  \BibitemOpen
  \bibfield  {author} {\bibinfo {author} {\bibfnamefont {K.}~\bibnamefont
  {Kodama}}\ and\ \bibinfo {author} {\bibfnamefont {Y.}~\bibnamefont
  {Takane}},\ }\bibfield  {title} {\bibinfo {title} {Persistent current due to
  a screw dislocation in weyl semimetals: Role of one-dimensional chiral
  states},\ }\href {https://doi.org/10.7566/JPSJ.88.054715} {\bibfield
  {journal} {\bibinfo  {journal} {Journal of the Physical Society of Japan}\
  }\textbf {\bibinfo {volume} {88}},\ \bibinfo {pages} {054715} (\bibinfo
  {year} {2019})},\ \Eprint
  {https://arxiv.org/abs/https://doi.org/10.7566/JPSJ.88.054715}
  {https://doi.org/10.7566/JPSJ.88.054715} \BibitemShut {NoStop}%
\bibitem [{\citenamefont {Huang}\ \emph {et~al.}(2019)\citenamefont {Huang},
  \citenamefont {Li}, \citenamefont {Zhou},\ and\ \citenamefont
  {Zhang}}]{Huang19}%
  \BibitemOpen
  \bibfield  {author} {\bibinfo {author} {\bibfnamefont {Z.-M.}\ \bibnamefont
  {Huang}}, \bibinfo {author} {\bibfnamefont {L.}~\bibnamefont {Li}}, \bibinfo
  {author} {\bibfnamefont {J.}~\bibnamefont {Zhou}},\ and\ \bibinfo {author}
  {\bibfnamefont {H.-H.}\ \bibnamefont {Zhang}},\ }\bibfield  {title} {\bibinfo
  {title} {Torsional response and liouville anomaly in weyl semimetals with
  dislocations},\ }\href {https://doi.org/10.1103/PhysRevB.99.155152}
  {\bibfield  {journal} {\bibinfo  {journal} {Phys. Rev. B}\ }\textbf {\bibinfo
  {volume} {99}},\ \bibinfo {pages} {155152} (\bibinfo {year}
  {2019})}\BibitemShut {NoStop}%
\bibitem [{\citenamefont {Huang}\ \emph {et~al.}(2020)\citenamefont {Huang},
  \citenamefont {Han},\ and\ \citenamefont {Stone}}]{Huang20a}%
  \BibitemOpen
  \bibfield  {author} {\bibinfo {author} {\bibfnamefont {Z.-M.}\ \bibnamefont
  {Huang}}, \bibinfo {author} {\bibfnamefont {B.}~\bibnamefont {Han}},\ and\
  \bibinfo {author} {\bibfnamefont {M.}~\bibnamefont {Stone}},\ }\bibfield
  {title} {\bibinfo {title} {Nieh-yan anomaly: Torsional landau levels, central
  charge, and anomalous thermal hall effect},\ }\href
  {https://doi.org/10.1103/PhysRevB.101.125201} {\bibfield  {journal} {\bibinfo
   {journal} {Phys. Rev. B}\ }\textbf {\bibinfo {volume} {101}},\ \bibinfo
  {pages} {125201} (\bibinfo {year} {2020})}\BibitemShut {NoStop}%
\bibitem [{\citenamefont {Huang}\ and\ \citenamefont {Han}(2020)}]{Huang20b}%
  \BibitemOpen
  \bibfield  {author} {\bibinfo {author} {\bibfnamefont {Z.-M.}\ \bibnamefont
  {Huang}}\ and\ \bibinfo {author} {\bibfnamefont {B.}~\bibnamefont {Han}},\
  }\bibfield  {title} {\bibinfo {title} {Torsional anomalies and
  bulk-dislocation correspondence in weyl systems},\ }\href
  {https://arxiv.org/abs/2003.04853} {\bibfield  {journal} {\bibinfo  {journal}
  {arXiv preprint: 2003.04853}\ } (\bibinfo {year} {2020})}\BibitemShut
  {NoStop}%
\bibitem [{\citenamefont {Liang}\ and\ \citenamefont {Ojanen}(2020)}]{Liang20}%
  \BibitemOpen
  \bibfield  {author} {\bibinfo {author} {\bibfnamefont {L.}~\bibnamefont
  {Liang}}\ and\ \bibinfo {author} {\bibfnamefont {T.}~\bibnamefont {Ojanen}},\
  }\bibfield  {title} {\bibinfo {title} {Topological magnetotorsional effect in
  weyl semimetals},\ }\href {https://doi.org/10.1103/PhysRevResearch.2.022016}
  {\bibfield  {journal} {\bibinfo  {journal} {Phys. Rev. Research}\ }\textbf
  {\bibinfo {volume} {2}},\ \bibinfo {pages} {022016} (\bibinfo {year}
  {2020})}\BibitemShut {NoStop}%
\bibitem [{\citenamefont {Nissinen}\ and\ \citenamefont
  {Volovik}(2020)}]{Nissinen20}%
  \BibitemOpen
  \bibfield  {author} {\bibinfo {author} {\bibfnamefont {J.}~\bibnamefont
  {Nissinen}}\ and\ \bibinfo {author} {\bibfnamefont {G.~E.}\ \bibnamefont
  {Volovik}},\ }\bibfield  {title} {\bibinfo {title} {Thermal nieh-yan anomaly
  in weyl superfluids},\ }\href
  {https://doi.org/10.1103/PhysRevResearch.2.033269} {\bibfield  {journal}
  {\bibinfo  {journal} {Phys. Rev. Research}\ }\textbf {\bibinfo {volume}
  {2}},\ \bibinfo {pages} {033269} (\bibinfo {year} {2020})}\BibitemShut
  {NoStop}%
\bibitem [{\citenamefont {Nissinen}\ and\ \citenamefont
  {Volovik}(2018)}]{nissinen2018tetrads}%
  \BibitemOpen
  \bibfield  {author} {\bibinfo {author} {\bibfnamefont {J.}~\bibnamefont
  {Nissinen}}\ and\ \bibinfo {author} {\bibfnamefont {G.}~\bibnamefont
  {Volovik}},\ }\bibfield  {title} {\bibinfo {title} {Tetrads in solids: from
  elasticity theory to topological quantum hall systems and weyl fermions},\
  }\href@noop {} {\bibfield  {journal} {\bibinfo  {journal} {Journal of
  Experimental and Theoretical Physics}\ }\textbf {\bibinfo {volume} {127}},\
  \bibinfo {pages} {948} (\bibinfo {year} {2018})}\BibitemShut {NoStop}%
\bibitem [{\citenamefont {Nissinen}\ and\ \citenamefont
  {Volovik}(2019)}]{nissinen2019elasticity}%
  \BibitemOpen
  \bibfield  {author} {\bibinfo {author} {\bibfnamefont {J.}~\bibnamefont
  {Nissinen}}\ and\ \bibinfo {author} {\bibfnamefont {G.}~\bibnamefont
  {Volovik}},\ }\bibfield  {title} {\bibinfo {title} {Elasticity tetrads, mixed
  axial-gravitational anomalies, and (3+ 1)-d quantum hall effect},\
  }\href@noop {} {\bibfield  {journal} {\bibinfo  {journal} {Physical Review
  Research}\ }\textbf {\bibinfo {volume} {1}},\ \bibinfo {pages} {023007}
  (\bibinfo {year} {2019})}\BibitemShut {NoStop}%
\bibitem [{\citenamefont {Ferreiros}\ \emph {et~al.}(2019)\citenamefont
  {Ferreiros}, \citenamefont {Kedem}, \citenamefont {Bergholtz},\ and\
  \citenamefont {Bardarson}}]{ferreiros2019mixed}%
  \BibitemOpen
  \bibfield  {author} {\bibinfo {author} {\bibfnamefont {Y.}~\bibnamefont
  {Ferreiros}}, \bibinfo {author} {\bibfnamefont {Y.}~\bibnamefont {Kedem}},
  \bibinfo {author} {\bibfnamefont {E.~J.}\ \bibnamefont {Bergholtz}},\ and\
  \bibinfo {author} {\bibfnamefont {J.~H.}\ \bibnamefont {Bardarson}},\
  }\bibfield  {title} {\bibinfo {title} {Mixed axial-torsional anomaly in weyl
  semimetals},\ }\href@noop {} {\bibfield  {journal} {\bibinfo  {journal}
  {Physical review letters}\ }\textbf {\bibinfo {volume} {122}},\ \bibinfo
  {pages} {056601} (\bibinfo {year} {2019})}\BibitemShut {NoStop}%
\bibitem [{\citenamefont {Chu}\ and\ \citenamefont
  {Miao}(2023)}]{chu2023chiral}%
  \BibitemOpen
  \bibfield  {author} {\bibinfo {author} {\bibfnamefont {C.-S.}\ \bibnamefont
  {Chu}}\ and\ \bibinfo {author} {\bibfnamefont {R.-X.}\ \bibnamefont {Miao}},\
  }\bibfield  {title} {\bibinfo {title} {Chiral current induced by torsional
  weyl anomaly},\ }\href@noop {} {\bibfield  {journal} {\bibinfo  {journal}
  {Physical Review B}\ }\textbf {\bibinfo {volume} {107}},\ \bibinfo {pages}
  {205410} (\bibinfo {year} {2023})}\BibitemShut {NoStop}%
\bibitem [{Note6()}]{Note6}%
  \BibitemOpen
  \bibinfo {note} {While the coefficient of the response in Eq.~\ref
  {Eq:WeylDipoleCurrents} is half the size of our numerical and analytic
  result, our calculations inherently determine the \protect \emph {covariant}
  anomaly of the interface Fermi arc states which receives inflow from the bulk
  term in Eq.~\ref {Eq:WeylDipoleCurrents}, i.e., inflow from a boundary term
  of the same magnitude, hence doubling the result~\cite
  {callanharvey,naculich1988,stone2012,parrikar1}.}\BibitemShut {Stop}%
\bibitem [{\citenamefont {Callan}\ and\ \citenamefont
  {Harvey}(1985)}]{callanharvey}%
  \BibitemOpen
  \bibfield  {author} {\bibinfo {author} {\bibfnamefont {C.}~\bibnamefont
  {Callan}}\ and\ \bibinfo {author} {\bibfnamefont {J.}~\bibnamefont
  {Harvey}},\ }\bibfield  {title} {\bibinfo {title} {Anomalies and fermion zero
  modes on strings and domain walls},\ }\href
  {https://doi.org/https://doi.org/10.1016/0550-3213(85)90489-4} {\bibfield
  {journal} {\bibinfo  {journal} {Nuclear Physics B}\ }\textbf {\bibinfo
  {volume} {250}},\ \bibinfo {pages} {427} (\bibinfo {year}
  {1985})}\BibitemShut {NoStop}%
\bibitem [{\citenamefont {Naculich}(1988)}]{naculich1988}%
  \BibitemOpen
  \bibfield  {author} {\bibinfo {author} {\bibfnamefont {S.~G.}\ \bibnamefont
  {Naculich}},\ }\bibfield  {title} {\bibinfo {title} {Axionic strings:
  covariant anomalies and bosonization of chiral zero modes},\ }\href@noop {}
  {\bibfield  {journal} {\bibinfo  {journal} {Nucl. Phys. B}\ }\textbf
  {\bibinfo {volume} {296}},\ \bibinfo {pages} {837} (\bibinfo {year}
  {1988})}\BibitemShut {NoStop}%
\bibitem [{\citenamefont {Stone}(2012)}]{stone2012}%
  \BibitemOpen
  \bibfield  {author} {\bibinfo {author} {\bibfnamefont {M.}~\bibnamefont
  {Stone}},\ }\bibfield  {title} {\bibinfo {title} {Gravitational anomalies and
  thermal hall effect in topological insulators},\ }\href@noop {} {\bibfield
  {journal} {\bibinfo  {journal} {Phys. Rev. B}\ }\textbf {\bibinfo {volume}
  {85}},\ \bibinfo {pages} {184503} (\bibinfo {year} {2012})}\BibitemShut
  {NoStop}%
\bibitem [{\citenamefont {Vergniory}\ \emph {et~al.}(2017)\citenamefont
  {Vergniory}, \citenamefont {Elcoro}, \citenamefont {Wang}, \citenamefont
  {Cano}, \citenamefont {Felser}, \citenamefont {Aroyo}, \citenamefont
  {Bernevig},\ and\ \citenamefont {Bradlyn}}]{vergniory2017graph}%
  \BibitemOpen
  \bibfield  {author} {\bibinfo {author} {\bibfnamefont {M.}~\bibnamefont
  {Vergniory}}, \bibinfo {author} {\bibfnamefont {L.}~\bibnamefont {Elcoro}},
  \bibinfo {author} {\bibfnamefont {Z.}~\bibnamefont {Wang}}, \bibinfo {author}
  {\bibfnamefont {J.}~\bibnamefont {Cano}}, \bibinfo {author} {\bibfnamefont
  {C.}~\bibnamefont {Felser}}, \bibinfo {author} {\bibfnamefont
  {M.}~\bibnamefont {Aroyo}}, \bibinfo {author} {\bibfnamefont {B.~A.}\
  \bibnamefont {Bernevig}},\ and\ \bibinfo {author} {\bibfnamefont
  {B.}~\bibnamefont {Bradlyn}},\ }\bibfield  {title} {\bibinfo {title} {Graph
  theory data for topological quantum chemistry},\ }\href@noop {} {\bibfield
  {journal} {\bibinfo  {journal} {Physical Review E}\ }\textbf {\bibinfo
  {volume} {96}},\ \bibinfo {pages} {023310} (\bibinfo {year}
  {2017})}\BibitemShut {NoStop}%
\bibitem [{\citenamefont {Bradlyn}\ \emph {et~al.}(2017)\citenamefont
  {Bradlyn}, \citenamefont {Elcoro}, \citenamefont {Cano}, \citenamefont
  {Vergniory}, \citenamefont {Wang}, \citenamefont {Felser}, \citenamefont
  {Aroyo},\ and\ \citenamefont {Bernevig}}]{bradlyn2017topological}%
  \BibitemOpen
  \bibfield  {author} {\bibinfo {author} {\bibfnamefont {B.}~\bibnamefont
  {Bradlyn}}, \bibinfo {author} {\bibfnamefont {L.}~\bibnamefont {Elcoro}},
  \bibinfo {author} {\bibfnamefont {J.}~\bibnamefont {Cano}}, \bibinfo {author}
  {\bibfnamefont {M.~G.}\ \bibnamefont {Vergniory}}, \bibinfo {author}
  {\bibfnamefont {Z.}~\bibnamefont {Wang}}, \bibinfo {author} {\bibfnamefont
  {C.}~\bibnamefont {Felser}}, \bibinfo {author} {\bibfnamefont {M.~I.}\
  \bibnamefont {Aroyo}},\ and\ \bibinfo {author} {\bibfnamefont {B.~A.}\
  \bibnamefont {Bernevig}},\ }\bibfield  {title} {\bibinfo {title} {Topological
  quantum chemistry},\ }\href@noop {} {\bibfield  {journal} {\bibinfo
  {journal} {Nature}\ }\textbf {\bibinfo {volume} {547}},\ \bibinfo {pages}
  {298} (\bibinfo {year} {2017})}\BibitemShut {NoStop}%
\bibitem [{\citenamefont {Cano}\ \emph
  {et~al.}(2018{\natexlab{a}})\citenamefont {Cano}, \citenamefont {Bradlyn},
  \citenamefont {Wang}, \citenamefont {Elcoro}, \citenamefont {Vergniory},
  \citenamefont {Felser}, \citenamefont {Aroyo},\ and\ \citenamefont
  {Bernevig}}]{cano2018building}%
  \BibitemOpen
  \bibfield  {author} {\bibinfo {author} {\bibfnamefont {J.}~\bibnamefont
  {Cano}}, \bibinfo {author} {\bibfnamefont {B.}~\bibnamefont {Bradlyn}},
  \bibinfo {author} {\bibfnamefont {Z.}~\bibnamefont {Wang}}, \bibinfo {author}
  {\bibfnamefont {L.}~\bibnamefont {Elcoro}}, \bibinfo {author} {\bibfnamefont
  {M.~G.}\ \bibnamefont {Vergniory}}, \bibinfo {author} {\bibfnamefont
  {C.}~\bibnamefont {Felser}}, \bibinfo {author} {\bibfnamefont {M.~I.}\
  \bibnamefont {Aroyo}},\ and\ \bibinfo {author} {\bibfnamefont {B.~A.}\
  \bibnamefont {Bernevig}},\ }\bibfield  {title} {\bibinfo {title} {Building
  blocks of topological quantum chemistry: Elementary band representations},\
  }\href@noop {} {\bibfield  {journal} {\bibinfo  {journal} {Physical Review
  B}\ }\textbf {\bibinfo {volume} {97}},\ \bibinfo {pages} {035139} (\bibinfo
  {year} {2018}{\natexlab{a}})}\BibitemShut {NoStop}%
\bibitem [{\citenamefont {Cano}\ \emph
  {et~al.}(2018{\natexlab{b}})\citenamefont {Cano}, \citenamefont {Bradlyn},
  \citenamefont {Wang}, \citenamefont {Elcoro}, \citenamefont {Vergniory},
  \citenamefont {Felser}, \citenamefont {Aroyo},\ and\ \citenamefont
  {Bernevig}}]{cano2018topology}%
  \BibitemOpen
  \bibfield  {author} {\bibinfo {author} {\bibfnamefont {J.}~\bibnamefont
  {Cano}}, \bibinfo {author} {\bibfnamefont {B.}~\bibnamefont {Bradlyn}},
  \bibinfo {author} {\bibfnamefont {Z.}~\bibnamefont {Wang}}, \bibinfo {author}
  {\bibfnamefont {L.}~\bibnamefont {Elcoro}}, \bibinfo {author} {\bibfnamefont
  {M.~G.}\ \bibnamefont {Vergniory}}, \bibinfo {author} {\bibfnamefont
  {C.}~\bibnamefont {Felser}}, \bibinfo {author} {\bibfnamefont {M.~I.}\
  \bibnamefont {Aroyo}},\ and\ \bibinfo {author} {\bibfnamefont {B.~A.}\
  \bibnamefont {Bernevig}},\ }\bibfield  {title} {\bibinfo {title} {Topology of
  disconnected elementary band representations},\ }\href
  {https://doi.org/10.1103/PhysRevLett.120.266401} {\bibfield  {journal}
  {\bibinfo  {journal} {Phys. Rev. Lett.}\ }\textbf {\bibinfo {volume} {120}},\
  \bibinfo {pages} {266401} (\bibinfo {year} {2018}{\natexlab{b}})}\BibitemShut
  {NoStop}%
\bibitem [{\citenamefont {Bradlyn}\ \emph {et~al.}(2018)\citenamefont
  {Bradlyn}, \citenamefont {Elcoro}, \citenamefont {Vergniory}, \citenamefont
  {Cano}, \citenamefont {Wang}, \citenamefont {Felser}, \citenamefont {Aroyo},\
  and\ \citenamefont {Bernevig}}]{bradlyn2018band}%
  \BibitemOpen
  \bibfield  {author} {\bibinfo {author} {\bibfnamefont {B.}~\bibnamefont
  {Bradlyn}}, \bibinfo {author} {\bibfnamefont {L.}~\bibnamefont {Elcoro}},
  \bibinfo {author} {\bibfnamefont {M.~G.}\ \bibnamefont {Vergniory}}, \bibinfo
  {author} {\bibfnamefont {J.}~\bibnamefont {Cano}}, \bibinfo {author}
  {\bibfnamefont {Z.}~\bibnamefont {Wang}}, \bibinfo {author} {\bibfnamefont
  {C.}~\bibnamefont {Felser}}, \bibinfo {author} {\bibfnamefont {M.~I.}\
  \bibnamefont {Aroyo}},\ and\ \bibinfo {author} {\bibfnamefont {B.~A.}\
  \bibnamefont {Bernevig}},\ }\bibfield  {title} {\bibinfo {title} {Band
  connectivity for topological quantum chemistry: Band structures as a graph
  theory problem},\ }\href {https://doi.org/10.1103/PhysRevB.97.035138}
  {\bibfield  {journal} {\bibinfo  {journal} {Phys. Rev. B}\ }\textbf {\bibinfo
  {volume} {97}},\ \bibinfo {pages} {035138} (\bibinfo {year}
  {2018})}\BibitemShut {NoStop}%
\bibitem [{\citenamefont {Elcoro}\ \emph {et~al.}(2021)\citenamefont {Elcoro},
  \citenamefont {Wieder}, \citenamefont {Song}, \citenamefont {Xu},
  \citenamefont {Bradlyn},\ and\ \citenamefont
  {Bernevig}}]{elcoro2021magnetic}%
  \BibitemOpen
  \bibfield  {author} {\bibinfo {author} {\bibfnamefont {L.}~\bibnamefont
  {Elcoro}}, \bibinfo {author} {\bibfnamefont {B.~J.}\ \bibnamefont {Wieder}},
  \bibinfo {author} {\bibfnamefont {Z.}~\bibnamefont {Song}}, \bibinfo {author}
  {\bibfnamefont {Y.}~\bibnamefont {Xu}}, \bibinfo {author} {\bibfnamefont
  {B.}~\bibnamefont {Bradlyn}},\ and\ \bibinfo {author} {\bibfnamefont {B.~A.}\
  \bibnamefont {Bernevig}},\ }\bibfield  {title} {\bibinfo {title} {Magnetic
  topological quantum chemistry},\ }\href@noop {} {\bibfield  {journal}
  {\bibinfo  {journal} {Nature communications}\ }\textbf {\bibinfo {volume}
  {12}},\ \bibinfo {pages} {5965} (\bibinfo {year} {2021})}\BibitemShut
  {NoStop}%
\bibitem [{\citenamefont {Manjunath}\ and\ \citenamefont
  {Barkeshli}(2020)}]{manjunath2020classification}%
  \BibitemOpen
  \bibfield  {author} {\bibinfo {author} {\bibfnamefont {N.}~\bibnamefont
  {Manjunath}}\ and\ \bibinfo {author} {\bibfnamefont {M.}~\bibnamefont
  {Barkeshli}},\ }\bibfield  {title} {\bibinfo {title} {Classification of
  fractional quantum hall states with spatial symmetries},\ }\href@noop {}
  {\bibfield  {journal} {\bibinfo  {journal} {arXiv preprint arXiv:2012.11603}\
  } (\bibinfo {year} {2020})}\BibitemShut {NoStop}%
\bibitem [{\citenamefont {Zhang}\ \emph {et~al.}(2022)\citenamefont {Zhang},
  \citenamefont {Manjunath}, \citenamefont {Nambiar},\ and\ \citenamefont
  {Barkeshli}}]{zhang2022fractional}%
  \BibitemOpen
  \bibfield  {author} {\bibinfo {author} {\bibfnamefont {Y.}~\bibnamefont
  {Zhang}}, \bibinfo {author} {\bibfnamefont {N.}~\bibnamefont {Manjunath}},
  \bibinfo {author} {\bibfnamefont {G.}~\bibnamefont {Nambiar}},\ and\ \bibinfo
  {author} {\bibfnamefont {M.}~\bibnamefont {Barkeshli}},\ }\bibfield  {title}
  {\bibinfo {title} {Fractional disclination charge and discrete shift in the
  hofstadter butterfly},\ }\href
  {https://doi.org/10.1103/PhysRevLett.129.275301} {\bibfield  {journal}
  {\bibinfo  {journal} {Phys. Rev. Lett.}\ }\textbf {\bibinfo {volume} {129}},\
  \bibinfo {pages} {275301} (\bibinfo {year} {2022})}\BibitemShut {NoStop}%
\bibitem [{\citenamefont {May-Mann}\ and\ \citenamefont
  {Hughes}(2022)}]{may2022crystalline}%
  \BibitemOpen
  \bibfield  {author} {\bibinfo {author} {\bibfnamefont {J.}~\bibnamefont
  {May-Mann}}\ and\ \bibinfo {author} {\bibfnamefont {T.~L.}\ \bibnamefont
  {Hughes}},\ }\bibfield  {title} {\bibinfo {title} {Crystalline responses for
  rotation-invariant higher-order topological insulators},\ }\href@noop {}
  {\bibfield  {journal} {\bibinfo  {journal} {Phys. Rev. B}\ }\textbf {\bibinfo
  {volume} {106}},\ \bibinfo {pages} {L241113} (\bibinfo {year}
  {2022})}\BibitemShut {NoStop}%
\bibitem [{\citenamefont {May-Mann}\ \emph {et~al.}(2023)\citenamefont
  {May-Mann}, \citenamefont {Hirsbrunner}, \citenamefont {Cao},\ and\
  \citenamefont {Hughes}}]{may2023topological}%
  \BibitemOpen
  \bibfield  {author} {\bibinfo {author} {\bibfnamefont {J.}~\bibnamefont
  {May-Mann}}, \bibinfo {author} {\bibfnamefont {M.~R.}\ \bibnamefont
  {Hirsbrunner}}, \bibinfo {author} {\bibfnamefont {X.}~\bibnamefont {Cao}},\
  and\ \bibinfo {author} {\bibfnamefont {T.~L.}\ \bibnamefont {Hughes}},\
  }\bibfield  {title} {\bibinfo {title} {Topological field theories of
  three-dimensional rotation symmetric insulators: Coupling curvature and
  electromagnetism},\ }\href@noop {} {\bibfield  {journal} {\bibinfo  {journal}
  {Phys. Rev. B}\ }\textbf {\bibinfo {volume} {107}},\ \bibinfo {pages}
  {205149} (\bibinfo {year} {2023})}\BibitemShut {NoStop}%
\bibitem [{\citenamefont {Nielsen}\ and\ \citenamefont
  {Ninomiya}(1981)}]{Nielsen81}%
  \BibitemOpen
  \bibfield  {author} {\bibinfo {author} {\bibfnamefont {H.}~\bibnamefont
  {Nielsen}}\ and\ \bibinfo {author} {\bibfnamefont {M.}~\bibnamefont
  {Ninomiya}},\ }\bibfield  {title} {\bibinfo {title} {A no-go theorem for
  regularizing chiral fermions},\ }\href
  {https://doi.org/https://doi.org/10.1016/0370-2693(81)91026-1} {\bibfield
  {journal} {\bibinfo  {journal} {Physics Letters B}\ }\textbf {\bibinfo
  {volume} {105}},\ \bibinfo {pages} {219} (\bibinfo {year}
  {1981})}\BibitemShut {NoStop}%
\end{thebibliography}

%

\appendix

\section{Translation gauge fields derived from the teleparallel prescription}
\label{app:tp_derivation}
In this section we provide a derivation of the translation gauge field $\mathfrak{e}^\mu_\nu$ and its' coupling prescription that follow directly from gauging the translational symmetry group, which can be done in a similar fashion to gauging the ordinary electromagnetic $U(1)$ symmetry. 
Consider a translation transformation 
\begin{equation}
    r^\mu\to r^\mu+a^\mu,
\end{equation}
which is generated by corresponding operators $\hat{P}_\mu=-i\hbar\partial_\mu$. Under such an infinitesimal translation the wave function changes by $\delta\psi = ia^\mu(\hat{P}_\mu/\hbar)\psi$.  Promoting the transformation to a local one, $a^\mu\to a^\mu (r)$, we find that the derivative of $\psi$ does not transform covariantly anymore:
\begin{equation}
    \hbar\delta(\partial_\nu\psi)=ia^\mu(r)\partial_\nu(\hat{P}_\mu\psi)+i\hat{P}_\mu\psi\partial_\nu a^\mu(r).
\end{equation}
We can compensate the second term by introducing an additional gauge potential $B^\mu_\nu$ that obeys the gauge transformation rules $B^\mu_\nu\to B^\mu_\nu - \partial_\nu a^\mu(r)$. This allows us to define a covariant derivative:
\begin{equation}
    D_\nu\psi=\partial_\nu\psi+iB^\mu_\nu (\hat{P}_\mu/\hbar)\psi.
    \label{eqn:tp_cov_deriv}
\end{equation}
Now it is straightforward to check that the covariant derivative transforms as expected:
\begin{equation}
    \hbar\delta(D_\nu\psi)=ia^\mu(t,x)D_\nu(\hat{P}_\mu\psi)
\end{equation}
We can re-express the partial derivative in Eq.~\ref{eqn:tp_cov_deriv} as a momentum operator to write down:
\begin{equation}
    D_\nu\psi = i\mathfrak{e}^\mu_\nu\hat{P}_\mu/\hbar
\end{equation}
where $\mathfrak{e}^\mu_\nu = \delta^\mu_\nu+B^\mu_\nu$ is a translation gauge field that inherits  its gauge transformations from the gauge potential $B^\mu_\nu$:
\begin{equation}
    \mathfrak{e}^\mu_\nu\to \mathfrak{e}^\mu_\nu-\partial_\nu a^\mu(r).
\end{equation}

\section{Gradient expansion}
\label{app:grad_exp}
In this appendix we would like to do a quick review of the gradient expansion procedure. As we will be interested in responses involving both electromagnetic and translation gauge fields, we need to consider how the electron wave vector gets shifted in the presence of spatially-varying gauge fields $A_\mu(r)$ and $\mathfrak{e}^\lambda_\mu(r)=\delta^\lambda_\mu+B^\lambda_\mu(r)$ (see Appendix~\ref{app:tp_derivation}):
\begin{equation}
    k_\mu\to k_\mu + \frac{e}{\hbar}A_\mu(r)+ k_\lambda B^\lambda_\mu(r).
\end{equation}
For small gauge fields, we can obtain a simple form of the resulting single-particle Green's functions performing a Taylor expansion:
\begin{equation}
\begin{split}
    &G_0(k)^{-1}\to G_{AB}^{-1}(k,r)=G_{0}^{-1}\left(k_\mu + \frac{e}{\hbar}A_\mu(r)+ k_\lambda B^\lambda_\mu(r)\right)\\
    &\approx G_{0}^{-1}(k)+\frac{e}{\hbar}A_\mu(r)\frac{\partial G_{0}^{-1}}{\partial k_\mu}+k_\lambda B^\lambda_\mu(r)\frac{\partial G_{0}^{-1}}{\partial k_\mu}(k)+...\\
    &\approx G_{0}^{-1}(k) +\left(\frac{e}{\hbar}A_\mu(r)+k_\lambda \mathfrak{e}^\lambda_\mu(r) - k_\mu\right)\frac{\partial G_{0}^{-1}}{\partial k_\mu}(k)+...
\end{split}
\end{equation}
We then follow the standard procedure to derive the effective action:
\begin{equation}
\begin{aligned}
    \frac{i}{\hbar}S&=\log\left(\frac{Z_{AB}}{Z_0}\right)=\log\left(\frac{\text{Det} G_{AB}^{-1}}{\text{Det} G_{0}^{-1}}\right) \\
    &\approx\text{Tr}\log\left(I+G_0\Sigma\right)
\end{aligned}
\end{equation}
where
\begin{equation}
\begin{split}
    \Sigma =& \left(\frac{e}{\hbar}A_\mu(r)+k_\lambda \mathfrak{e}^\lambda_\mu(r) - k_\mu\right)\frac{\partial G_{0}^{-1}}{\partial k_\mu}(k)+...
\end{split}
\end{equation}
Expanding the trace of logarithm we get
\begin{equation}
\begin{aligned}
    \frac{i}{\hbar}S &\approx \text{Tr}\log\left(I+G_0\Sigma\right) \\
    &\approx \text{Tr}\left(G_0\Sigma\right)-\frac{1}{2}\text{Tr}\left(G_0\Sigma G_0\Sigma\right) \\
    &+\frac{1}{3}\text{Tr}\left(G_0\Sigma G_0\Sigma G_0\Sigma\right)-...
\end{aligned}
\label{eqn:app_trlog_exp}
\end{equation}
The RHS of this equation is a sum of integrals over the entire phase space and the products under functional traces are convolutions. Therefore we need to use the Moyal product formula, expanding each $G_0\Sigma$ term as:
\begin{equation}
    G_0\star \Sigma \approx G_0\Sigma + \frac{i}{2}\{G_0,\Sigma\}+...
\end{equation}
where $\star$ is the Moyal product operator and $\{\cdot,\cdot\}$ are the Poisson brackets for the $r^\mu$ and $k_\mu$ variables. 
The RHS of the last equation contains ordinary products of $G_0$ and $\Sigma$ that are subsequently integrated over the phase-space. For example, in $d$ space-time dimensions we get for the $\text{Tr}\left(G_0\Sigma \right)$ term in the 0-th order of the Moyal product expansion:
\begin{equation}
    \int d^dr\frac{d^dk}{(2\pi)^d}\left(\frac{e}{\hbar}A_\mu+k_\lambda B^\lambda_\mu \right)\text{tr}\left(G_0\frac{\partial G_0^{-1}}{\partial k_\mu}\right).
\end{equation}
where `tr' denotes the ordinary trace over orbital and spin degrees of freedom.

\section{Electric polarization as a Berry curvature dipole}
\label{app:pol_berry_dipole}
Let us consider the expression for the polarization of a 2D system with a single filled band,
\begin{equation}
\begin{split}
     P_e^y =\frac{e\Omega}{(2\pi)^2}i\int_{BZ} d^2\textbf{k}\langle u_\textbf{k}|\partial_{k_y} u_\textbf{k}\rangle \equiv \frac{e\Omega}{(2\pi)^2}i P,
\end{split}
\end{equation}
where $\Omega$ is the area of the unit cell.
We will rewrite the last integral denoted as '$P$' as a first moment of the Berry curvature, $\mathcal{F}^{xy}=i\langle \partial_{k_x} u_{\textbf{k}}|\partial_{k_y} u_{\textbf{k}}\rangle-i\langle \partial_{k_y} u_{\textbf{k}}|\partial_{k_x} u_{\textbf{k}}\rangle$.
Consider the following quantity:
\begin{equation}
\begin{aligned}
        F&=-i\int_{BZ} d^2\textbf{k}\ k_x\mathcal{F}^{xy} \\
        &=\int_{BZ} d^2\textbf{k}\ k_x\langle\partial_{k_x} u_{\textbf{k}}|\partial_{k_y} u_{\textbf{k}}\rangle \\
        &-\int_{BZ} d^2\textbf{k}\ k_x\langle \partial_{k_y} u_{\textbf{k}}|\partial_{k_x} u_{\textbf{k}}\rangle,
\end{aligned}
\end{equation}
where we assume $\mathcal{F}^{xy}$ to be smooth and integrable in the Brillouin zone spanning $k_x\in[-\pi/a_x,\pi/a_x)$ and $k_y\in[-\pi/a_y,\pi/a_y)$. Clearly, the integrand jumps in value at the $k_x = \pi/a_x$ boundary of the Brillouin zone and so we will treat the $k_x$ direction as open.
Integrating by parts with respect to $k_x$ we find:
\begin{equation}
\begin{aligned}
    F& = -i\frac{4\pi}{a_x}\oint dk_y \mathcal{A}^y(k_x = \pi/a_x, k_y)  \\
    &- \int_{BZ} d^2\textbf{k}\big(\langle u_{\textbf{k}}|\partial_{k_y}u_{\textbf{k}}\rangle - \langle \partial_{k_y} u_{\textbf{k}}|u_{\textbf{k}}\rangle\big)\\
    & - \int_{BZ} d^2\textbf{k}\ k_x \langle u_{\textbf{k}}|\partial_{k_x}\partial_{k_y}u_{\textbf{k}}\rangle \\
    & + \int_{BZ} d^2\textbf{k}\ k_x \langle \partial_{k_x}\partial_{k_y} u_{\textbf{k}}|u_{\textbf{k}}\rangle,
\end{aligned}
\end{equation}
where $\mathcal{A}^\mu(\textbf{k})=i\langle u_{\textbf{k}}|\partial_{k_\mu}u_\textbf{k}\rangle$ is the Berry connection.
The first term is proportional to a Wilson loop $W^y(k_x = \pi/a_x)$ along the $k_x = \pi/a_x$ line.
It is easy to recognize that the second term is twice the integral of interest $-2P$.
Integrating the third and fourth terms by parts with respect to $k_y$ we find:
\begin{equation}
    \begin{aligned}
        &-\int_{BZ} d^2\textbf{k}\ k_x \big(\langle u_{\textbf{k}}|\partial_{k_x}\partial_{k_y}u_{\textbf{k}}\rangle - \langle \partial_{k_x}\partial_{k_y} u_{\textbf{k}}|u_{\textbf{k}}\rangle)\\
        &=\int_{BZ} d^2\textbf{k}\ k_x\big(\langle \partial_{k_y} u_{\textbf{k}}|\partial_{k_x} u_{\textbf{k}}\rangle - \langle\partial_{k_x} u_{\textbf{k}}|\partial_{k_y} u_{\textbf{k}}\rangle\big)\\
        &=-i\int d^2\textbf{k}\ k_x \mathcal{F}^{yx} = - F
    \end{aligned}
\end{equation}
Summing up, we found:
\begin{equation*}
\begin{gathered}
    F = -i\frac{4\pi}{a_x}W^y -2P - F \\
    \Downarrow \\
    P = -F - i\frac{2\pi}{a_x}W^y
\end{gathered}
\end{equation*}
and the polarization is therefore given by
\begin{equation}
\begin{aligned}
    P_e^y &=-\frac{e\Omega}{(2\pi)^2}\int_{BZ} d^2\textbf{k}\ k_x\mathcal{F}^{xy} \\
    &+\frac{e a_y}{2\pi}\oint dk_y \mathcal{A}^y(k_x = \pi/a_x, k_y).
\end{aligned}
\end{equation}
Performing a similar calculation for $P_e^x$, we find the general formula
\begin{equation}
    P_e^i = \frac{e\Omega}{(2\pi)^2}\varepsilon^{ij}\int d^2\textbf{k}\ k_j\mathcal{F}^{xy} + e a_i W^i.
\end{equation}
In the case when the system has inversion symmetry, the Wilson loop taken along a high-symmetry line satisfies $W^i(k_j=\pi/a_j)=-W^i(k_j=\pi/a_j)$ and $W^i(k_j=0)=-W^i(k_j=0)$ for $i\neq j$ and we find that the non-quantized part of the polarization is accounted for entirely by the Berry curvature's dipole moment.

\section{Momentum polarization as a Berry curvature quadrupole}
\label{app:quad_pol}
Let us consider the following expression for the $k_x$ momentum polarization in the $\hat{y}$ direction of a 2D system with a single filled band:
\begin{equation}
\begin{split}
     P_{k_x}^y =\frac{\Omega}{(2\pi)^2}i\int_{BZ} d^2\textbf{k}\ k_x\langle u_\textbf{k}|\partial_{k_y} u_\textbf{k}\rangle \equiv \frac{\Omega}{(2\pi)^2}i Q,
\end{split}
\end{equation}
which is just a natural extension of the analogous expression for the charge polarization.
We can rewrite the integral denoted as `$Q$' as a second moment of the Berry curvature, as we now show.
Consider the following quantity:
\begin{equation}
\begin{aligned}
        F&=-\frac{i}{2}\int_{BZ} d^2\textbf{k}\ k^2_x\mathcal{F}^{xy} \\
        &=\frac{1}{2}\int_{BZ} d^2\textbf{k}\ k^2_x\langle\partial_{k_x} u_{\textbf{k}}|\partial_{k_y} u_{\textbf{k}}\rangle \\
        &- \frac{1}{2}\int_{BZ} d^2\textbf{k}\ k^2_x\langle \partial_{k_y} u_{\textbf{k}}|\partial_{k_x} u_{\textbf{k}}\rangle
\end{aligned}
\end{equation}
where we once again assume $\mathcal{F}^{xy}$ to be smooth and integrable in the Brillouin zone spanned by $k_x\in[-\pi/a_x,\pi/a_x)$ and $k_y\in[-\pi/a_y,\pi/a_y)$. 
Treating the $k_x$ direction of the BZ as open, we integrate by parts with respect to $k_x$ to find:
\begin{equation}
\begin{aligned}
    F &=- \int_{BZ} d^2\textbf{k}\ k_x\big(\langle u_{\textbf{k}}|\partial_{k_y}u_{\textbf{k}}\rangle - \langle \partial_{k_y} u_{\textbf{k}}|u_{\textbf{k}}\rangle\big)\\
    & - \frac{1}{2}\int_{BZ} d^2\textbf{k}\ k^2_x \langle u_{\textbf{k}}|\partial_{k_x}\partial_{k_y}u_{\textbf{k}}\rangle \\
    & + \frac{1}{2}\int_{BZ} d^2\textbf{k}\ k^2_x \langle \partial_{k_x}\partial_{k_y} u_{\textbf{k}}|u_{\textbf{k}}\rangle
\end{aligned}
\end{equation}
Note the absence of the Wilson loop contribution we found in the previous section, which is a result of the symmetric nature of the function $k_x^2$.
We see again that the first term is twice the integral of interest $-2Q$.
Integrating the third and fourth terms by parts with respect to $k_y$ we find:
\begin{equation}
    \begin{aligned}
        &-\frac{1}{2}\int_{BZ} d^2\textbf{k}\ k^2_x \big(\langle u_{\textbf{k}}|\partial_{k_x}\partial_{k_y}u_{\textbf{k}}\rangle - \langle \partial_{k_x}\partial_{k_y} u_{\textbf{k}}|u_{\textbf{k}}\rangle)\\
        &=\frac{1}{2}\int_{BZ} d^2\textbf{k}\ k^2_x\big(\langle \partial_{k_y} u_{\textbf{k}}|\partial_{k_x} u_{\textbf{k}}\rangle - \langle\partial_{k_x} u_{\textbf{k}}|\partial_{k_y} u_{\textbf{k}}\rangle\big)\\
        &=-\frac{i}{2}\int d^2\textbf{k}\ k^2_x \mathcal{F}^{yx} = \frac{i}{2}\int d^2\textbf{k}\ k^2_x \mathcal{F}^{xy} = - F
    \end{aligned}
\end{equation}
Summing up, we found:
\begin{equation}
\begin{gathered}
    F = -2Q - F \\
    \Downarrow \\
    Q = -F 
\end{gathered}
\end{equation}
and we found the polarization to be:
\begin{equation}
    P_{k_x}^y =-\frac{\Omega}{8\pi^2}\int_{BZ} d^2\textbf{k}\ k^2_x\mathcal{F}^{xy}.
\end{equation}
Performing a similar calculation for $P_{k_y}^x$, we find the general relation:
\begin{equation}
    P_{k_y}^x = \frac{\Omega}{8\pi^2}\int_{BZ} d^2\textbf{k}\ k^2_y\mathcal{F}^{xy}.
\end{equation}

\section{Responses for 1D systems}
\label{app:responsesin1d}
In this Appendix we will discuss responses of isolated one-dimensional metals having a fixed number of electrons $N_e$. For the cases we consider, the Fermi surface consists of an even integer number $N_F$ of Fermi points having chiralities $\chi_a={\rm{sgn}}\ v_a$ where $v_a$ is the Fermi velocity of the $a$-th Fermi point. From the Fermion doubling theorem~\cite{Nielsen81} the total chirality vanishes, $\chi=\sum_{a=1}^{N_F}\chi_a = 0.$ We wish to define three more quantities besides $\chi$ that will characterize 1D metals:
\begin{align}
    \mathcal{P}_x&=\sum_{a=1}^{N_F}\chi_a k_{F x}^{(a)}\\
    \mathcal{Q}_{xx}&=\sum_{a=1}^{N_F}\chi_a \left(k_{F x}^{(a)}\right)^2\\
    \mathcal{O}_{xxx}&=\sum_{a=1}^{N_F}\chi_a \left(k_{F x}^{(a)}\right)^3.
\end{align}
These three quantities represent the momentum space dipole, quadrupole, and octupole moments of the Fermi-points, respectively (see Fig.~\ref{fig:1dresponse}(a), (b), (c)). We could go beyond the octupole moment to any higher moment, but for brevity we stop at this order. Importantly, these momentum moments are related to the ground state properties of the metal. The total charge is proportional to the dipole moment,
\begin{equation}
    Q=\frac{eL}{2\pi}\mathcal{P}_x,
\end{equation}
the total momentum $\langle \hbar k_x\rangle$ is proportional to the quadrupole moment,
\begin{equation}
    P_x = \frac{1}{2}\frac{\hbar L}{2\pi}\mathcal{Q}_{xx},\label{eq:totalmomentumPx}
\end{equation}
the total momentum squared $\langle (\hbar k_x)^2\rangle$ is proportional to the octupole moment,
\begin{equation}
    P_{xx}=\frac{1}{3}\frac{\hbar^2 L}{2\pi}\mathcal{O}_{xxx},\label{eq:totalPxx}
\end{equation}
and so on for higher moments. From this we see that each of the momentum-space moments determines the density of higher and higher powers of momentum, starting at zeroth order where the charge is proportional to the momentum dipole.  There are two important caveats to note: (i) in order for the $n$-th moment and its associated physical quantity to be independent of the origin of the BZ, all lower moments must vanish, and (ii)  these results hold only up to constants independent of the set of $k_{Fx}^{(a)}$ which result from contributions from filled bands.

We now want to consider a family of anomalous responses to various gauge fields in 1D metals. We have already considered some of these anomalies in Sec.~\ref{sec:1dsemimetalderivation}, and we will go into more detail in this Appendix. To proceed, we will introduce a family of gauge fields $\mathfrak{e}, \mathfrak{e}^\alpha, \mathfrak{e}^{\alpha\beta}, \mathfrak{e}^{\alpha\beta\gamma}, \ldots.$ Each of these fields couples to charges that are powers of momentum. The field $\mathfrak{e}$ we identify with the  family of electromagnetic gauge field one-forms $\tfrac{e}{\hbar}A$ as it couples to zero powers of momentum. The field $\mathfrak{e}^\alpha$ is the translation gauge field we have extensively discussed, and it couples linearly to momentum $k_\alpha.$ In general the fields $\mathfrak{e}^{\alpha\beta\gamma\ldots \zeta}$ couple to the momentum charges $k_\alpha k_\beta k_\gamma\ldots k_{\zeta}.$ Since we consider momentum-space moments only up to the octupole moment $\mathcal{O}_{xxx},$ we will consider gauge fields only up to $\mathfrak{e}^{\alpha\beta\gamma}.$

Using these gauge fields we can consider the following set of actions
\begin{widetext}
\begin{align}
S_{\chi}=&\frac{e^2\chi}{2\pi\hbar}\int dx dt\, A_0 A_x\\
S_{\mathcal{P}}=&\frac{e}{2\pi}\int dx dt\, \mathcal{P}_x (\mathfrak{e}_0^x A_x-\mathfrak{e}_x^x A_0)\\
S_{\mathcal{Q}}=& \frac{\hbar}{2\pi}\int dx dt\, \mathcal{Q}_{xx}\left[\tfrac{1}{2} \mathfrak{e}_{0}^x\mathfrak{e}_{x}^x+ e (\mathfrak{e}_0^{xx} A_x-\mathfrak{e}_x^{xx} A_0)\right]\\
S_{\mathcal{O}}=&\frac{\hbar^2}{2\pi}\int dx dt\, \mathcal{O}_{xxx}\left[\tfrac{1}{3}(\mathfrak{e}_0^{x}\mathfrak{e}_x^{xx}-\mathfrak{e}_x^{x} \mathfrak{e}_0^{xx})+e (\mathfrak{e}_0^{xxx} A_x-\mathfrak{e}_x^{xxx} A_0)\right].
\end{align}
\end{widetext}
These actions capture two important phenomena associated to each of the momentum moments: (i) the connection to the assiociated ground state quantity, i.e., $Q, P_x,$ and $P_{xx}$, and (ii) the shift in $Q, P_x, P_{xx},$ and $P_{xxx}$ when an electric field is turned on.  As a first example let us consider $S_{\chi}.$ We can calculate the electromagnetic charge density and current to find $\rho=\frac{e^2\chi}{h}A_x$ and $j^x=\frac{e^2\chi}{h}A_0.$ If we use these results to calculate the conservation law we find
\begin{equation*}
    \partial_\mu j^\mu = \frac{e^2\chi}{h}E_x,
\end{equation*} which is just the usual $U(1)$ anomaly of a chiral fermion. The fact that $\chi=0$ for any lattice model has two immediate consequences: (i) the $U(1)$ charge anomaly above vanishes for lattice systems, and (ii) the momentum dipole moment $\mathcal{P}_x$ is well-defined and independent of the choice of momentum space origin. Just as for conventional electric or magnetic multipole moments, in order for the $n$-th moment to be well-defined, all of the lower moments must vanish. As such, the action $S_{\mathcal{Q}}$ is well-defined only if $\chi=\mathcal{P}_x=0.$ Similarly, for $S_{\mathcal{O}}$ to be well-defined we must have $\chi=\mathcal{P}_x=\mathcal{Q}_{xx}=0.$

Now let us consider each of the remaining actions in turn. We will begin with $S_{\mathcal{P}}.$ As mentioned in Sec.~\ref{sec:1dsemimetalderivation}, $\mathcal{P}_x$  is related~\cite{RamamurthyPatterns} to the charge density of a 1D metal via $\rho= -\frac{e}{2\pi} \mathcal{P}_x\mathfrak{e}_{x}^{x}$, and the momentum density via $\mathcal{J}^0_x=\frac{e}{2\pi}\mathcal{P}_x A_x.$ Assuming that our system is translation invariant, let us consider stretching our system via a time-dependent $\mathfrak{e}_x^x.$ During this process the total number of electrons cannot change. Working from the charge density we find
\begin{equation}
    \partial_t \rho = -\frac{e}{2\pi}\partial_{t}(\mathcal{P}_x\mathfrak{e}^x_x).\label{eq:app_dtdensity}
\end{equation}
Naively we are just changing $\mathfrak{e}^x_x,$ however if we stretch the system at fixed particle number the Fermi momenta will change inversely. Indeed we have $\partial_t \mathcal{P}_x=-\tfrac{\mathcal{P}_x}{\mathfrak{e}^x_x}\partial_t \mathfrak{e}_x^x.$ Inserting this into Eq.~\ref{eq:app_dtdensity} we find
\begin{equation}
    \partial_t \rho = -\frac{e}{2\pi}\left(-\mathfrak{e}_x^x\tfrac{\mathcal{P}_x}{\mathfrak{e}^x_x}\partial_t \mathfrak{e}_x^x+\mathcal{P}_x\partial_t \mathfrak{e}_x^x\right) = 0. 
\end{equation} Using this equation we find
\begin{equation}
    \Delta Q = \int dt \int dx\, \partial_t\rho =0
\end{equation} as we expect for a fixed number of electrons.  

To be self-contained, let us reiterate our argument from the main text. At a fixed particle number we know the total charge cannot change. Intuitively we might expect that the density should decrease if we stretch the system. However, the quantity $\rho$ above, which is defined as $\tfrac{\delta S}{\delta A_0}$ is not a scalar density. For general geometries the scalar charge density would be defined as
\begin{equation}
    \bar{\rho}=\frac{1}{\mathfrak{e}^x_x}\frac{\delta S}{\delta A_0}.
\end{equation} To calculate the total charge we would then use
\begin{equation}
    Q=\int dx\, \mathfrak{e}_x^x \bar{\rho}=\int dx\, \rho.
\end{equation} Indeed, the scalar charge density $\bar{\rho}$ will decrease as the system is stretched since $\partial_t \bar{\rho}\propto \partial_t \mathcal{P}_x,$ which decreases as the system size increases at fixed electron number.

Next, we can see that another consequence of a non-vanishing $\mathcal{P}_x$ is a mixed crystalline-electromagnetic anomaly. To illustrate this, let us consider the change in momentum-density in an applied electric field generated by a change in $A_x.$ We find
\begin{equation}
    \partial_t \mathcal{J}_0^x = \frac{e}{2\pi}\partial_t(\mathcal{P}_x A_x).
\end{equation} Unlike the previous case, when we turn on a non-vanishing $A_x$ the dipole $\mathcal{P}_x$ does not change. Hence we find the anomalous conservation law
    \begin{eqnarray}
    \partial_\mu \mathcal{J}^{\mu}_x&=&\frac{e \mathcal{P}_x }{2\pi}E_x.
\end{eqnarray} 

Moving on, let us discuss the action $S_{\mathcal{Q}}.$ To have a well-defined quadrupole moment $\mathcal{Q}_{xx}$ we need $\mathcal{P}_x=0.$ This scenario can happen non-trivially in systems with more than one occupied band near the Fermi-level, as shown in Fig.~\ref{fig:1dresponse}(b). As long as any perturbations we apply keep $\chi$ and $\mathcal{P}_x$ fixed to zero, then the phenomena associated to $\mathcal{Q}_{xx}$ are physically meaningful. From this action we can derive three separate conservation laws:
\begin{align}
    \partial_t\rho &= \frac{e}{2\pi}\partial_t(\mathcal{Q}_{xx}\mathfrak{e}_{x}^{xx})\\
    \partial_t\mathcal{J}^0_{x} &= \frac{\hbar}{4\pi}\partial_t(\mathcal{Q}_{xx}\mathfrak{e}_{x}^{x})\label{eq:app_momanomaly}\\
    \partial_t\mathcal{J}^0_{xx} &= -\frac{e}{2\pi}\partial_t(\mathcal{Q}_{xx}A_x),
\end{align}
where the quantities $P_x$ and $P_{xx}$ in Eqs.~\ref{eq:totalmomentumPx} and \ref{eq:totalPxx} are determined by $P_x=\int \mathcal{J}^0_x dx, P_{xx}=\int \mathcal{J}^0_{xx}dx.$ The first and third equations generate a kind of mixed anomaly, so let us discuss those first. For fixed electron number we know that $\partial_t \rho$ must vanish, which implies that $\partial_t \mathcal{Q}_{xx}=-\frac{Q_{xx}}{\mathfrak{e}_x^{xx}}\partial_t \mathfrak{e}_{x}^{xx}.$ Thus the first equation is simply $\partial_t\rho=0.$ For the third equation,  since changing $A_x$ while keeping $\chi=\mathcal{P}_x=0$ does not change $\mathcal{Q}_{xx}$, we have
\begin{equation}
    \partial_\mu \mathcal{J}^{\mu}_{xx}=\frac{e\mathcal{Q}_{xx}}{2\pi}E_x.
\end{equation} This implies that if we insert flux into the system, then the momentum quadrupole moment changes, i.e., the expectation value of the momentum squared in the resulting excited state changes while the total charge and momentum remain fixed. 

Returning to the middle equation, we consider the change in momentum as we stretch the system. Crucially we use the relationship $\partial_t \mathcal{Q}_{xx}=-2\frac{\mathcal{Q}_{xx}}{\mathfrak{e}_x^x}\partial_t \mathfrak{e}_x^x$ (heuristically, this comes from the fact that quadratic powers of momentum are proportional to $L^{-2})$. Inserting this in Eq.~\ref{eq:app_momanomaly} we find
\begin{equation*}
    \partial_t\mathcal{J}_x^0 =-\frac{\hbar\mathcal{Q}_{xx}}{2\pi}\partial_t\mathfrak{e}_x^x+\frac{\hbar\mathcal{Q}_{xx}}{4\pi}\partial_t\mathfrak{e}_x^x= -\frac{\hbar\mathcal{Q}_{xx}}{4\pi}\partial_t\mathfrak{e}_x^x.
\end{equation*}
We can interpret the first contribution in the middle section of the above equation as coming from the change in the Fermi points $k_{Fx}^{(a)}$ induced by changing $\mathfrak{e}_x^x.$ The second contribution arises from the existence of a non-vanishing ground state momentum density when the length of the system is changed. Note that while the coefficient of the final result is the same magnitude as Eq.~\ref{eq:app_momanomaly}, the sign is opposite. The full conservation law becomes
\begin{equation}
    \partial_\mu \mathcal{J}^\mu_{x}=\frac{\hbar\mathcal{Q}_{xx}}{4\pi}\mathcal{E}_x^x.
\end{equation}

Finally, if we have a scenario where $\chi, \mathcal{P}_x,$ and $\mathcal{Q}_{xx}$ are all vanishing, and remain vanishing after applying any gauge fields, then the phenomena associated to $\mathcal{O}_{xxx}$ become physically relevant. Such a scenario can exist in a 1D metal where four bands appear at the Fermi surface (see Fig.~\ref{fig:1dresponse}(c)). Just as above, let us consider the conservation laws we can derive from $S_{\mathcal{O}}:$
\begin{align}
\partial_t \rho =& -\frac{e\hbar^2 }{2\pi}\partial_t(\mathcal{O}_{xxx}\mathfrak{e}_x^{xxx})\\
\partial_t \mathcal{J}_x^0 =& \frac{\hbar^2}{6\pi}\partial_t(\mathcal{O}_{xxx}\mathfrak{e}_x^{xx})\\
\partial_t \mathcal{J}_{xx}^0 =&-\frac{\hbar^2}{6\pi}\partial_t(\mathcal{O}_{xxx}\mathfrak{e}_x^{x})\\
\partial_t \mathcal{J}_{xxx}^0 =& \frac{e\hbar^2}{2\pi}\partial_t(\mathcal{O}_{xxx}A_x).
\end{align}
We can use identical arguments as above to determine that $\partial_t\mathcal{O}_{xxx}=-\frac{\mathcal{O}_{xxx}}{\mathfrak{e}_{x}^{xxx}}\partial_t \mathfrak{e}_x^{xxx}$ so that the total charge remains fixed. Under a change of $A_x$ we have $\partial_t \mathcal{O}_{xxx}=0,$ and under a change in $\mathfrak{e}_x^x$ we can determine that $\partial_t\mathcal{O}_{xxx}=-3\frac{\mathcal{O}_{xxx}}{\mathfrak{e}_{x}^{x}}\partial_t \mathfrak{e}_x^{x}.$ Using these relationships we can reduce three of the conservation laws to find 
\begin{align}
\partial_\mu j^\mu =&0\\
\partial_\mu \mathcal{J}^{\mu}_{xx} =&-\frac{\hbar^2\mathcal{O}_{xxx}}{3\pi}\mathcal{E}_{x}^{x}\\
\partial_\mu\mathcal{J}^{\mu}_{xxx}  =& -\frac{e\hbar^2\mathcal{O}_{xxx}}{2\pi}E_x.
\end{align}
To get the final conservation law we need to determine how $\mathcal{O}_{xxx}$ changes when $\mathfrak{e}_{x}^{xx}$ changes. From counting powers of length we find $\partial_t \mathcal{O}_{xxx}=-\tfrac{3}{2}\frac{\mathcal{O}_{xxx}}{\mathfrak{e}_{x}^{xx}}\partial_t \mathfrak{e}_x^{xx}.$ Inserting this into the conservation law for $\rho^x$ generates
\begin{equation}
    \partial_\mu \mathcal{J}^{\mu}_x = \frac{\hbar^2\mathcal{O}_{xxx}}{12\pi}\mathcal{E}_{x}^{xx}.
\end{equation}

In summary, while the anomalous responses we have written in this section are formally correct, it is impossible to uniquely determine the coefficients $\mathcal{P}_x, \mathcal{Q}_{xx},$ or $\mathcal{O}_{xxx}$ unless all lower moments vanish (starting with the chirality $\chi$). Additionally, even if the lower moments are initially vanishing, turning on gauge fields may generate these moments anomalously and hence invalidate the higher moments.  We expect that under the assumptions of vanishing lower moments that the highest moment will generate the physical responses described above.

\end{document}